\documentclass[aps,prb,reprint,superscriptaddress,floatfix]{revtex4-2}
\usepackage{amsmath,amssymb}
\usepackage{graphicx}

\begin{document}

\title{Electron--phonon matrix elements for hybrid functionals in the
projector augmented-wave method: Eigenvalue projection, nondiagonal
supercells, and Wannier interpolation}

\author{Qian Wang}
\affiliation{Key Laboratory of Material Simulation Methods and Software of Ministry of Education, College of Physics, Jilin University, Changchun 130012, China}
\affiliation{State Key Laboratory of High Pressure and Superhard Materials, College of Physics, Jilin University, Changchun 130012, China}
\author{Yu Xie}
\email{xieyu@jlu.edu.cn}
\affiliation{Key Laboratory of Material Simulation Methods and Software of Ministry of Education, College of Physics, Jilin University, Changchun 130012, China}
\affiliation{Key Laboratory of Physics and Technology for Advanced Batteries of Ministry of Education, College of Physics, Jilin University, Changchun 130012, China}
\author{Yanming Ma}
\affiliation{Key Laboratory of Material Simulation Methods and Software of Ministry of Education, College of Physics, Jilin University, Changchun 130012, China}
\affiliation{State Key Laboratory of High Pressure and Superhard Materials, College of Physics, Jilin University, Changchun 130012, China}
\affiliation{International Center of Future Science, Jilin University, Changchun 130012, China}

\date{\today}

\begin{abstract}
We present a finite-displacement approach to electron--phonon matrix elements
that requires only eigenvalues and wave functions from the displaced
calculations, for both norm-conserving pseudopotentials and the projector
augmented-wave (PAW) method. Extending the eigenvalue-projectability approach,
previously formulated for norm-conserving pseudopotentials, we reconstruct the
displaced Hamiltonian from the displaced eigenvalues and from overlaps
evaluated in the PAW metric of each displaced structure. Analytic terms
constructed from the PAW dataset then yield the all-electron matrix element,
without derivatives of the local potential or of the PAW strengths. The
approach therefore applies in principle to any electronic-structure method
available in PAW. To reduce the supercell cost, we combine it with nondiagonal supercells
that sample the Brillouin zone equivalently to the primitive cell and recover
the full matrix elements by crystal symmetry. Implemented in the ARES-EPC
code, the method reproduces matrix elements from density-functional
perturbation theory (DFPT) and from VASP+phelel for seven materials with mean
absolute errors below 0.5~meV. Nondiagonal supercells reproduce the
diagonal-supercell results to the same accuracy and reduce the cost of the
displaced calculations by up to a factor of 23. Through an interface to EPW,
the norm-conserving matrix elements give an electron mobility of Si within
0.9\% of that from DFPT. The HSE06 hybrid functional increases the mobility of
Si by 13\% relative to the PBE functional, because the lighter conduction band
and the harder phonons outweigh the stronger coupling.
\end{abstract}

\maketitle

\section{Introduction}
\label{sec:introduction}

Electron--phonon coupling (EPC) governs carrier scattering, electrical and
thermal transport, phonon-mediated superconductivity, and phonon lifetimes
\cite{Giustino2017,Monserrat2018}. It also underlies polaron formation and the
temperature dependence of electronic and optical spectra
\cite{Devreese2009,Zacharias2016}. These properties are built from the EPC
matrix elements $g_{mn\nu}(\mathbf k,\mathbf q)$. Their first-principles
evaluation is limited in two ways: by the electronic-structure methods for
which the matrix elements can be obtained, and by the cost of sampling the
phonon wave vectors $\mathbf q$.

Density-functional perturbation theory (DFPT) is the standard route. It
provides phonons and EPC matrix elements at arbitrary $\mathbf q$ from the
linear response of the primitive cell
\cite{Baroni1987,Giannozzi1991,DeGironcoli1995,Baroni2001,Savrasov1994}.
The linear response, however, depends on the pseudopotential type and on
the electronic-structure method. DFPT has been extended to ultrasoft and
projector augmented-wave (PAW) pseudopotentials
\cite{DalCorso2001,Audouze2006,DalCorso2010}, but in these formulations the
EPC matrix element between pseudo states lacks symmetries of the
all-electron matrix element~\cite{Engel2020}. DFPT-based EPC calculations
therefore mostly use norm-conserving (NC) pseudopotentials~\cite{Lee2023}.
Semilocal density-functional theory (DFT) can underestimate the matrix
elements, and screened hybrid functionals and $GW$ have been shown to correct
them~\cite{Yin2013,Li2019GWPT}. Yet for hybrid functionals, EPC required a
density-matrix perturbation formulation~\cite{Yang2022}, and $GW$
perturbation theory its own many-body linear response~\cite{Li2019GWPT}.
Each further method calls for a similar derivation and implementation, which
hinders EPC calculations with advanced methods.

Finite-displacement calculations avoid method-specific linear response,
since they require only self-consistent calculations of supercells with
displaced atoms~\cite{Monserrat2018}. Within PAW, for example, stochastic
displacements have been used to obtain the phonon-induced renormalization of
band gaps~\cite{Karsai2018}. Such sampling approaches
obtain the renormalization directly from the eigenvalues of the displaced
structures, without the individual matrix elements. The full matrix
elements, in contrast, are usually obtained from finite differences of the
self-consistent potential, which reintroduces a dependence on the
Hamiltonian representation. The nonlocal part of the pseudopotential is
treated separately, so its derivative has to be worked out for every
pseudopotential type~\cite{Gunst2016,Chaput2019}. Beyond semilocal DFT, the
effective potential itself becomes a nonlocal, orbital- or
frequency-dependent operator, and its change can no longer be obtained by subtracting potentials on a
real-space grid~\cite{Poliukhin2026}.

Poliukhin \emph{et al.}\ avoided potential differences altogether with an
eigenvalue-projectability approach~\cite{Poliukhin2026}, referred to below as
eigenvalue projection. The displaced Hamiltonian is represented in the basis
of the undisplaced eigenstates, using only the displaced eigenvalues and the
overlaps between the two sets of eigenstates. Since the Hamiltonian itself
is never needed, the approach applies to any electronic-structure method, as
they demonstrated for hybrid functionals and $G_0W_0$. Their implementation
uses NC pseudopotentials, for which the overlaps are ordinary inner products;
for ultrasoft and PAW pseudopotentials, they noted, the scalar product
involves an overlap operator and requires an additional treatment.

The PAW method~\cite{Blochl1994,KresseJoubert1999} is widely used in
plane-wave calculations because it combines smooth pseudo wave functions with
access to the all-electron states near the nuclei. Eigenvalue projection in
PAW would allow EPC to be computed at any level of theory available in PAW
codes, but the extension raises two issues. First, the pseudo states solve
a generalized eigenvalue problem whose overlap operator depends on the atomic
positions, so the overlaps must be taken in the metric of the displaced
structure. Second, the reconstruction then gives only the derivative of the pseudo-Hamiltonian, whereas the all-electron matrix element
also contains the derivative of the PAW transformation, which moves with the
atoms. In the finite-displacement PAW formulation of Chaput \emph{et al.},
this matrix element is evaluated from the derivatives of the local potential
and of the nonlocal PAW strengths (the one-center coefficients of the
nonlocal part of the pseudo-Hamiltonian), supplemented by analytic projector and
partial-wave terms~\cite{Chaput2019}. This requires the code to supply the
potential derivatives; matrix elements obtained in this way have been
interfaced with EPW~\cite{Lee2023} for semilocal and
meta-generalized-gradient approximation (meta-GGA) functionals, with hybrid
functionals left for future work~\cite{Radevych2025}.

Independently of the level of theory, finite-displacement
calculations are limited by the supercell size. A phonon wave vector
$\mathbf q$ must be commensurate with the supercell, so finer $\mathbf q$
sampling requires larger supercells. Lloyd-Williams and Monserrat showed that
nondiagonal supercell matrices accommodate the same $\mathbf q$ points with
far fewer primitive cells, and used them for phonons and for the zero-point
renormalization of band gaps obtained from band-energy changes
\cite{LloydWilliams2015}. Using nondiagonal supercells for the full matrix
elements additionally requires that each supercell sample the electronic
Brillouin zone equivalently to the primitive cell, so that the contributions
of different supercells are consistent.

In this work, we extend eigenvalue projection to PAW by resolving both
issues. Generalized overlaps give the full derivative of the
pseudo-Hamiltonian without differentiating the local potential or the PAW
strengths, and the derivative of the PAW transformation is evaluated
analytically from the PAW dataset. The displaced calculations thus need to
provide only eigenvalues and pseudo wave functions, which a PAW code provides
for hybrid functionals as for semilocal ones. We
implement the method for diagonal and nondiagonal supercells in ARES-EPC, a
standalone code that prepares the supercell calculations and processes their
output. It is interfaced with Quantum ESPRESSO
\cite{Giannozzi2009,Giannozzi2017}, VASP
\cite{KresseHafner1993,KresseFurthmueller1996}, and ARES
\cite{Xue2024}. An interface to EPW~\cite{Lee2023} provides Wannier
interpolation and transport calculations.

We validate the method for seven materials, including metals,
semiconductors, and a high-pressure hydride. NC matrix elements are compared
with DFPT, PAW matrix elements with VASP+phelel
\cite{Chaput2019,Engel2022,Chaput2026}, and nondiagonal with diagonal PAW
supercells. In all three comparisons, the mean absolute errors (MAEs) are
below 0.5~meV, and the nondiagonal supercells reduce the cost of the displaced
calculations by up to a factor of 23. For Si, the electron mobility computed
with EPW from the NC matrix elements of ARES-EPC agrees with that from DFPT
to 0.9\%. Using nondiagonal supercells, we find that the HSE06 hybrid
functional~\cite{Heyd2003,Krukau2006} increases the mobility of Si by 13\%
relative to the Perdew--Burke--Ernzerhof (PBE) functional~\cite{Perdew1996}.
As a further application, we compute the Eliashberg spectral function of
CaH$_6$ at 150~GPa.

The paper is organized as follows. Section~\ref{sec:theory} formulates
eigenvalue projection with generalized overlaps, derives the all-electron
matrix element in PAW, and describes the use of nondiagonal supercells,
crystal symmetry, and Wannier interpolation. The implementation is outlined in Sec.~\ref{sec:workflow},
and the computational parameters are given in Sec.~\ref{sec:computational}.
Section~\ref{sec:results} assesses the accuracy and cost of the method
against DFPT, VASP+phelel, and diagonal supercells, and compares the
mobility of Si with that from DFPT. Section~\ref{sec:si-hse} presents the
mobility of Si with HSE06, and Sec.~\ref{sec:cah6} the application to
CaH$_6$. The approximations, the origin of the remaining
differences, the cost, and possible extensions are discussed in
Sec.~\ref{sec:discussion}, and
conclusions are drawn in Sec.~\ref{sec:conclusions}.
Appendix~\ref{app:plane-wave-overlaps} gives the plane-wave evaluation of
the overlaps and the completeness relation, Appendix~\ref{app:paw-corrections}
the analytic PAW terms, and Appendix~\ref{app:symmetry} the symmetry
transformation of the matrix elements.

\section{Theory}
\label{sec:theory}

\subsection{Electron--phonon matrix elements}
\label{sec:matrix-elements}

Let $\mathbf A=[\mathbf a_1,\mathbf a_2,\mathbf a_3]$ contain the primitive
lattice vectors as columns and $\mathbf B=2\pi\mathbf A^{-T}$ the reciprocal
ones. Wave vectors $\mathbf k$ and $\mathbf q$ are given in reduced
coordinates of $\mathbf B$, primitive cell $l$ is located at
$\mathbf A\mathbf n_l$ with $\mathbf n_l\in\mathbb Z^3$, $l=0$
($\mathbf n_0=0$) denotes the home cell, and atom $\kappa$ has
reduced position $\boldsymbol\tau_\kappa$; spin indices are suppressed. We
denote by $u_{l\kappa\alpha}$
the Cartesian displacement of atom $\kappa$ in cell $l$ along $\alpha$, and by
$\mathbf k'=[\mathbf k+\mathbf q]_1$ the final-state momentum, where
$[\cdot]_1$ reduces each component modulo one. The atomic-displacement
matrix elements are
\begin{equation}\label{eq:cartesian-definition}
\begin{aligned}
 {}[g_{\kappa\alpha}(\mathbf k,\mathbf q)]_{mn}
 &=\langle\psi_{m\mathbf k'}|\partial_{\kappa\alpha,\mathbf q}H^{\mathrm{AE}}|
   \psi_{n\mathbf k}\rangle,\\
 \partial_{\kappa\alpha,\mathbf q}
 &=\sum_l e^{2\pi i\mathbf q\cdot\mathbf n_l}
   \frac{\partial}{\partial u_{l\kappa\alpha}},
\end{aligned}
\end{equation}
where $H^{\mathrm{AE}}$ is the all-electron Kohn--Sham Hamiltonian and
$\psi_{n\mathbf k}$ are the all-electron Bloch states of the undisplaced
structure. Primes are used throughout for quantities that refer to the final
state. The mode-resolved matrix elements follow as
\begin{equation}\label{eq:mode-vertex}
 g_{mn\nu}(\mathbf k,\mathbf q)=
 \sum_{\kappa\alpha}
 \sqrt{\frac{\hbar}{2M_\kappa\omega_{\mathbf q\nu}}}\,
 e_{\kappa\alpha,\nu}(\mathbf q)\,
 [g_{\kappa\alpha}(\mathbf k,\mathbf q)]_{mn},
\end{equation}
with atomic masses $M_\kappa$, phonon frequencies $\omega_{\mathbf q\nu}$, and
eigenvectors $e_{\kappa\alpha,\nu}(\mathbf q)$ defined by the mode
displacement pattern
$u_{l\kappa\alpha}\propto M_\kappa^{-1/2}e_{\kappa\alpha,\nu}(\mathbf q)
e^{2\pi i\mathbf q\cdot\mathbf n_l}$.

In a finite-displacement calculation, Eq.~\eqref{eq:cartesian-definition}
is evaluated from the Hamiltonians $H_x$ of displaced configurations, where
$x$ specifies the displaced atom, the direction, and the sign of a
displacement of amplitude $\delta$, and $x=0$ denotes the undisplaced
structure. The reference states $\psi_{n\mathbf k}$ are the eigenstates of
the undisplaced primitive cell and are held fixed. For any displaced configuration, the
spectral representation $H_x=\sum_j\varepsilon^x_j|\Psi^x_j\rangle
\langle\Psi^x_j|$ gives
\begin{equation}\label{eq:nc-reconstruction}
 \langle\psi_{m\mathbf k'}|H_x|\psi_{n\mathbf k}\rangle
 =\sum_j\varepsilon^x_j
 \langle\psi_{m\mathbf k'}|\Psi^x_j\rangle
 \langle\Psi^x_j|\psi_{n\mathbf k}\rangle .
\end{equation}
This is the eigenvalue projection of Poliukhin
\emph{et al.}~\cite{Poliukhin2026}; the displaced Hamiltonian enters only
through its eigenvalues and eigenstates. Below we extend it to the generalized
eigenvalue problem of PAW and recover the all-electron matrix element.

\subsection{Eigenvalue projection with generalized overlaps}
\label{sec:generalized-projection}

In PAW, the pseudo states of configuration $x$ solve
\begin{equation}\label{eq:generalized-eigenproblem}
 \tilde H_x|\tilde\Psi^x_j\rangle
 =\varepsilon^x_j\tilde S_x|\tilde\Psi^x_j\rangle,
 \qquad
 \langle\tilde\Psi^x_h|\tilde S_x|\tilde\Psi^x_j\rangle=\delta_{hj},
\end{equation}
with a Hermitian, positive-definite overlap operator $\tilde S_x$. The
eigenstates obey the generalized completeness relation
$\sum_j|\tilde\Psi^x_j\rangle\langle\tilde\Psi^x_j|\tilde S_x=1$, which
yields the spectral representation
\begin{equation}\label{eq:paw-spectral-operator}
 \tilde H_x
 =\tilde H_x\sum_j|\tilde\Psi^x_j\rangle\langle\tilde\Psi^x_j|\tilde S_x
 =\sum_j\varepsilon^x_j\,
  \tilde S_x|\tilde\Psi^x_j\rangle\langle\tilde\Psi^x_j|\tilde S_x .
\end{equation}
Taking matrix elements between the fixed reference pseudo states, we define
the generalized overlaps
\begin{equation}\label{eq:generalized-overlap}
 U_x(j,n)=\langle\tilde\Psi^x_j|\tilde S_x|\tilde\psi_{n\mathbf k}\rangle,
 \qquad
 U'_x(j,m)=\langle\tilde\Psi^x_j|\tilde S_x|\tilde\psi_{m\mathbf k'}\rangle,
\end{equation}
and obtain the projected pseudo-Hamiltonian
\begin{equation}\label{eq:spectral-reconstruction}
 \mathcal H_x
 =U_x'^{\dagger}\,\mathcal E_x\,U_x,
 \qquad
 [\mathcal H_x]_{mn}
 =\langle\tilde\psi_{m\mathbf k'}|\tilde H_x|\tilde\psi_{n\mathbf k}\rangle,
\end{equation}
where $\mathcal E_x$ is diagonal with entries $\varepsilon^x_j$. For
$\tilde S_x=1$, Eqs.~\eqref{eq:generalized-overlap} and
\eqref{eq:spectral-reconstruction} reduce to the NC expression,
Eq.~\eqref{eq:nc-reconstruction}. The PAW overlap operator is
\begin{equation}\label{eq:paw-overlap-operator}
\begin{aligned}
 \tilde S_x
 &=1+\sum_{l\kappa\mu\eta}|\tilde p^x_{l\kappa\mu}\rangle
   (Q_\kappa)_{\mu\eta}\langle\tilde p^x_{l\kappa\eta}|,\\
 (Q_\kappa)_{\mu\eta}
 &=\langle\phi_{\kappa\mu}|\phi_{\kappa\eta}\rangle
  -\langle\tilde\phi_{\kappa\mu}|\tilde\phi_{\kappa\eta}\rangle,
\end{aligned}
\end{equation}
where $\tilde p^x_{l\kappa\mu}$ are the projectors at the displaced site
$(l,\kappa)$, $\phi_{\kappa\mu}$ and $\tilde\phi_{\kappa\mu}$ the all-electron
and pseudo partial waves, and $\mu,\eta$ label the partial-wave channels
$(\iota,\ell,m_\ell)$ of atom $\kappa$ (Appendix~\ref{app:plane-wave-overlaps}). Since
$\tilde S_x=\mathcal T_x^\dagger\mathcal T_x$ with the PAW transformation
$\mathcal T_x$, the overlaps in Eq.~\eqref{eq:generalized-overlap} equal
$\langle\mathcal T_x\tilde\Psi^x_j|\mathcal T_x\tilde\psi_{n\mathbf k}\rangle$:
the reference pseudo state is augmented at the displaced geometry. The
difference between $\mathcal T_x$ and the reference transformation
$\mathcal T_0$ is restored by the analytic terms of
Sec.~\ref{sec:paw-ae}. Plane-wave expressions for the overlaps are given in
Appendix~\ref{app:plane-wave-overlaps}.

Equation~\eqref{eq:spectral-reconstruction} holds when all displaced states
are retained. If only the first $N_s$ displaced states are retained,
completeness implies
\begin{equation}\label{eq:projectability}
 \lim_{N_s\to\infty}U_{x,N_s}'^{\dagger}U_{x,N_s}
 =\Sigma_x,
 \qquad
 [\Sigma_x]_{mn}=\langle\tilde\psi_{m\mathbf k'}|\tilde S_x|
 \tilde\psi_{n\mathbf k}\rangle,
\end{equation}
where $U_{x,N_s}$ retains the rows $j\le N_s$
(Appendix~\ref{app:completeness}). Unlike the NC case, $\Sigma_x$ differs from
the identity even for $\mathbf k'=\mathbf k$, because the reference states
are orthonormal in $\tilde S_0$ rather than in $\tilde S_x$. We define the
projectability residual as the largest absolute element of
$U_{x,N_s}'^{\dagger}U_{x,N_s}-\Sigma_x$ over the reference bands considered,
with $\Sigma_x$ evaluated independently from the projector coefficients. It
verifies the truncation and generalizes the projectability condition of
Ref.~\cite{Poliukhin2026} to the PAW metric.
A small residual implies that the displaced states into which the reference
states evolve, including complete degenerate multiplets, are retained. Each
omitted state has overlaps of order $\delta$ with opposite signs for
$\pm\delta$, so its eigenvalue-weighted contribution to $\mathcal H_x$ is of
order $\delta^2$ with an odd part of order $\delta^3$ and affects the
central-difference derivative only at order $\delta^2$.

\subsection{All-electron matrix element in PAW}
\label{sec:paw-ae}

In PAW, $\psi_{n\mathbf k}=\mathcal T_0\tilde\psi_{n\mathbf k}$ and
$\tilde H_x=\mathcal T_x^\dagger H^{\mathrm{AE}}_x\mathcal T_x$, with
\begin{equation}\label{eq:paw-transformation}
 \mathcal T_x=1+\sum_{l\kappa\mu}
 \bigl(|\phi^x_{l\kappa\mu}\rangle-|\tilde\phi^x_{l\kappa\mu}\rangle\bigr)
 \langle\tilde p^x_{l\kappa\mu}|
\end{equation}
\cite{Blochl1994,KresseJoubert1999}. We write
$\langle\tilde\psi'|\cdots|\tilde\psi\rangle$ for the matrix over final
bands at $\mathbf k'$ and initial bands at $\mathbf k$. Differentiating
$\tilde H$ at the reference geometry and using
$H^{\mathrm{AE}}_0\psi_{n\mathbf k}=\epsilon_{n\mathbf k}\psi_{n\mathbf k}$
gives~\cite{Chaput2019}
\begin{equation}\label{eq:ae-from-pseudo}
 g_{\kappa\alpha}
 =\langle\tilde\psi'|\partial_{\kappa\alpha,\mathbf q}\tilde H|\tilde\psi\rangle
 -E'X_{\kappa\alpha}-Y_{\kappa\alpha}E,
\end{equation}
where $E$ and $E'$ are diagonal matrices of the reference eigenvalues
$\epsilon_{n\mathbf k}$ and $\epsilon_{m\mathbf k'}$, and
\begin{equation}\label{eq:paw-xy}
\begin{aligned}
 X_{\kappa\alpha}
 &=\langle\tilde\psi'|\mathcal T_0^\dagger
   (\partial_{\kappa\alpha,\mathbf q}\mathcal T)|\tilde\psi\rangle,\\
 Y_{\kappa\alpha}
 &=\langle\tilde\psi'|(\partial_{\kappa\alpha,\mathbf q}\mathcal T^\dagger)
   \mathcal T_0|\tilde\psi\rangle .
\end{aligned}
\end{equation}
Here $\partial_{\kappa\alpha,\mathbf q}$ acts on $\mathcal T^\dagger$ itself,
i.e.,
$\partial_{\kappa\alpha,\mathbf q}\mathcal T^\dagger
=(\partial_{\kappa\alpha,-\mathbf q}\mathcal T)^\dagger$.
Equation~\eqref{eq:ae-from-pseudo} omits two terms in which
$H^{\mathrm{AE}}_0-\epsilon$ acts on the displacement derivatives of the
partial waves; they vanish for a complete PAW basis and correspond to Pulay
corrections in force calculations~\cite{Chaput2019}.

The first term of Eq.~\eqref{eq:ae-from-pseudo} is the spectral term
\begin{equation}\label{eq:spectral-term}
 g^{\mathrm H}_{\kappa\alpha}
 =\langle\tilde\psi'|\partial_{\kappa\alpha,\mathbf q}\tilde H|\tilde\psi\rangle .
\end{equation}
It is obtained from central differences of the matrices $\mathcal H_x$ of
Eq.~\eqref{eq:spectral-reconstruction} for displacements $\pm\delta$, with
the lattice sum in $\partial_{\kappa\alpha,\mathbf q}$ evaluated in
commensurate supercells (Sec.~\ref{sec:supercells}).

The remaining terms are evaluated analytically. We define the projector
coefficients of the reference states at the home-cell site and their
derivatives with respect to the projector position,
\begin{equation}\label{eq:projector-coefficients}
\begin{aligned}
 C_\kappa(\mu,n)&=\langle\tilde p_{0\kappa\mu}|\tilde\psi_{n\mathbf k}\rangle,
 &
 D_{\kappa\alpha}(\mu,n)&=\Bigl\langle
   \frac{\partial\tilde p_{0\kappa\mu}}{\partial u_{0\kappa\alpha}}
   \Bigm|\tilde\psi_{n\mathbf k}\Bigr\rangle,
\end{aligned}
\end{equation}
with $C'_\kappa$ and $D'_{\kappa\alpha}$ defined analogously for the final
states, and the partial-wave matrix
\begin{equation}\label{eq:paw-r}
 (R_{\kappa\alpha})_{\mu\eta}
 =\Bigl\langle\phi_{\kappa\mu}\Bigm|
   \frac{\partial\phi_{\kappa\eta}}{\partial u_{0\kappa\alpha}}\Bigr\rangle
 -\Bigl\langle\tilde\phi_{\kappa\mu}\Bigm|
   \frac{\partial\tilde\phi_{\kappa\eta}}{\partial u_{0\kappa\alpha}}\Bigr\rangle .
\end{equation}
Because $Q_\kappa$ is invariant under a rigid translation of the partial
waves, $R_{\kappa\alpha}+R_{\kappa\alpha}^\dagger=0$. The Bloch property of
the reference states reduces the lattice sum in
$\partial_{\kappa\alpha,\mathbf q}$ to the home-cell site: since
$\mathbf k'-\mathbf k-\mathbf q\in\mathbb Z^3$, the phase
$e^{2\pi i\mathbf q\cdot\mathbf n_l}$ cancels the Bloch phase of site $l$,
and each site contributes equally. Differentiating
Eq.~\eqref{eq:paw-transformation} and using the completeness of the pseudo
partial waves inside non-overlapping augmentation spheres,
$\sum_\mu|\tilde\phi_{\kappa\mu}\rangle\langle\tilde p_{0\kappa\mu}|=1$,
together with its derivative at fixed $\tilde\psi$~\cite{Chaput2019},
gives
\begin{equation}\label{eq:x-decomposition}
\begin{aligned}
 X_{\kappa\alpha}&=P_{\kappa\alpha}+Z_{\kappa\alpha},
 &
 Y_{\kappa\alpha}&=\bar P_{\kappa\alpha}-Z_{\kappa\alpha},
\end{aligned}
\end{equation}
with $P_{\kappa\alpha}=C_\kappa'^{\dagger}Q_\kappa D_{\kappa\alpha}$,
$\bar P_{\kappa\alpha}=D_{\kappa\alpha}'^{\dagger}Q_\kappa C_\kappa$, and
$Z_{\kappa\alpha}=C_\kappa'^{\dagger}R_{\kappa\alpha}C_\kappa$; the second
relation follows from the first by Hermitian conjugation, with
$\mathbf k\leftrightarrow\mathbf k'$ and $\mathbf q\to-\mathbf q$, and
$R^\dagger_{\kappa\alpha}=-R_{\kappa\alpha}$. Substituting
Eq.~\eqref{eq:x-decomposition} into Eq.~\eqref{eq:ae-from-pseudo} gives
\begin{equation}\label{eq:paw-corrections}
\begin{aligned}
 g_{\kappa\alpha}
 &=g^{\mathrm H}_{\kappa\alpha}+g^{\mathrm P}_{\kappa\alpha}
   +g^{\mathrm R}_{\kappa\alpha},\\
 g^{\mathrm P}_{\kappa\alpha}
 &=-E'P_{\kappa\alpha}-\bar P_{\kappa\alpha}E,\\
 g^{\mathrm R}_{\kappa\alpha}
 &=Z_{\kappa\alpha}E-E'Z_{\kappa\alpha},
\end{aligned}
\end{equation}
with the spectral term $g^{\mathrm H}_{\kappa\alpha}$ of
Eq.~\eqref{eq:spectral-term}. The term $g^{\mathrm R}$ coincides with the
partial-wave contribution of Ref.~\cite{Chaput2019}, whereas $g^{\mathrm P}$
contains only the part of its projector contribution proportional to
$\epsilon Q_\kappa$.
The remaining contributions of that work, from the local potential, the
nonlocal PAW strengths, and the part of the projector contribution with the
PAW strengths, are all contained in $g^{\mathrm H}$ and need not be
differentiated separately. Plane-wave expressions for $C$, $D$, and the
Bloch phases $e^{2\pi i\mathbf k\cdot\boldsymbol\tau_\kappa}$ they carry are
given in Appendix~\ref{app:paw-corrections}. In NC,
$\mathcal T=1$ and $g_{\kappa\alpha}=g^{\mathrm H}_{\kappa\alpha}$.

The all-electron matrix element of Eq.~\eqref{eq:cartesian-definition}
differs in general from the pseudo matrix element of the generalized
eigenproblem, $\tilde g_{\kappa\alpha}
=\langle\tilde\psi'|\partial_{\kappa\alpha,\mathbf q}\tilde H|\tilde\psi\rangle
-\langle\tilde\psi'|\partial_{\kappa\alpha,\mathbf q}\tilde S|\tilde\psi\rangle E$
\cite{Engel2020,Engel2022}. Since
$\partial_{\kappa\alpha,\mathbf q}\tilde S
=\partial_{\kappa\alpha,\mathbf q}(\mathcal T^\dagger\mathcal T)$ gives
$\langle\tilde\psi'|\partial_{\kappa\alpha,\mathbf q}\tilde S|\tilde\psi\rangle
=X_{\kappa\alpha}+Y_{\kappa\alpha}$,
\begin{equation}\label{eq:vertex-conventions}
 g_{\kappa\alpha}-\tilde g_{\kappa\alpha}
 =X_{\kappa\alpha}E-E'X_{\kappa\alpha},
\end{equation}
which vanishes between degenerate states but not in general. We report the all-electron
matrix element $g_{\kappa\alpha}$.

\subsection{Diagonal and nondiagonal supercells}
\label{sec:supercells}

The lattice sum in $\partial_{\kappa\alpha,\mathbf q}$ is evaluated in
supercells commensurate with $\mathbf q$. Supercell $p$ has lattice
vectors $\mathbf A_p=\mathbf A\mathbf M_p$, with an integer matrix
$\mathbf M_p$, $N_p=\det\mathbf M_p>0$ primitive cells, and reciprocal
vectors $\mathbf B_p=\mathbf B\mathbf M_p^{-T}$. It is commensurate with
$\mathbf q$ if $\mathbf M_p^T\mathbf q\in\mathbb Z^3$. Primitive momenta then
fold to a common supercell momentum,
\begin{equation}\label{eq:k-folding}
\begin{aligned}
 \mathbf K_p&=[\mathbf M_p^T\mathbf k]_1=[\mathbf M_p^T\mathbf k']_1,\\
 \mathbf c_p&=\mathbf M_p^T\mathbf k-\mathbf K_p,
 \qquad
 \mathbf c'_p=\mathbf M_p^T\mathbf k'-\mathbf K_p,
\end{aligned}
\end{equation}
with $\mathbf c_p,\mathbf c'_p\in\mathbb Z^3$. The reference states are
represented in the supercell basis as
$\tilde\psi^{(p)}_{n\mathbf k}=N_p^{-1/2}\tilde\psi_{n\mathbf k}$, normalized
in the supercell (Appendix~\ref{app:plane-wave-overlaps}). Translational symmetry of the
displaced supercell restricts Eq.~\eqref{eq:generalized-overlap} to
displaced states at $\mathbf K_p$, and we write $\mathcal H^{(p)}_x$ for
the resulting matrix of Eq.~\eqref{eq:spectral-reconstruction}.

A diagonal supercell has $\mathbf M_p=\mathrm{diag}(n_1,n_2,n_3)$ and is
commensurate with the full $n_1\times n_2\times n_3$ $\mathbf q$ mesh, so a
single supercell covers the mesh. In the nondiagonal route, a set of
commensurate matrices $\{\mathbf M_p\}$ with smaller $N_p$ is selected for
the irreducible $\mathbf q$ points following
Ref.~\cite{LloydWilliams2015}.

The self-consistent field (SCF) calculations of all supercells sample the electronic
Brillouin zone with the same primitive-cell mesh. A primitive mesh
$\mathbf k=\mathrm{diag}(\mathbf N)^{-1}(\mathbf n+\mathbf s)$ with
$\mathbf N=(N_1,N_2,N_3)$, integer $\mathbf n$, and offset $\mathbf s$ folds
exactly into supercell $p$ if $\mathrm{diag}(\mathbf N)\mathbf M_p^{-T}$ is an
integer matrix. If this condition fails for some supercell, $N_j$ is increased
to the smallest value for which it holds for all supercells; such a value
exists because any common multiple of the $N_p$ satisfies it. Each supercell
SCF calculation then samples exactly the $N_1N_2N_3/N_p$ supercell
$\mathbf k$ points onto which the primitive mesh folds, so that all undisplaced supercells are equivalent to
the primitive cell and the displaced calculations differ only by the
displacements.

In supercell $p$, atom $\kappa$ of the home cell is displaced, together with
its periodic images in the neighboring supercells, along three linearly independent vectors
$\boldsymbol\delta^{(p)}_{\kappa\xi}$, $\xi=1,2,3$, which form the columns of
$\boldsymbol\Delta^{(p)}_\kappa$. With central differences
$h^{(p)}_{\kappa\xi}=\frac12\bigl(\mathcal H^{(p)}_{\kappa\xi,+}
-\mathcal H^{(p)}_{\kappa\xi,-}\bigr)$, the spectral term is
\begin{equation}\label{eq:cartesian-vertex}
 g^{\mathrm H,(p)}_{\kappa\alpha}
 =N_p\sum_{\xi=1}^3
 \bigl[(\boldsymbol\Delta^{(p)}_\kappa)^{-1}\bigr]_{\xi\alpha}
 h^{(p)}_{\kappa\xi}.
\end{equation}
The factor $N_p$ arises because each of the $N_p$ primitive cells of the
supercell contributes equally to the lattice sum, whereas
$\mathcal H^{(p)}_x$ is evaluated with supercell-normalized states. The
analytic terms $g^{\mathrm P}$ and $g^{\mathrm R}$ are constructed from
primitive-cell states and are added once, which gives the matrix elements
evaluated in practice,
\begin{equation}\label{eq:working-vertex}
 g^{(p)}_{\kappa\alpha}=g^{\mathrm H,(p)}_{\kappa\alpha}
 +g^{\mathrm P}_{\kappa\alpha}+g^{\mathrm R}_{\kappa\alpha}.
\end{equation}
When several supercells provide matrices at the same $(\mathbf k,\mathbf q)$,
these are expressed in the same primitive-cell Bloch basis and are averaged.

\subsection{Symmetry recovery}
\label{sec:symmetry-recovery}

Matrix elements computed at symmetry-inequivalent $(\mathbf k,\mathbf q)$
are transported to the full mesh with space-group operations
$\{\mathcal R|\mathbf v\}$, with $\mathcal R$ and $\mathbf v$ in reduced
coordinates, optionally combined with time reversal. We set $\sigma=+1$ for
a spatial operation and $\sigma=-1$ for its combination with time reversal,
and write $A^{(+)}=A$ and $A^{(-)}=A^*$ for any quantity $A$. Source ($s$) and
target ($t$) momenta are related by
$\mathbf q_t=\sigma\mathcal R^{-T}\mathbf q_s+\mathbf G_q$ and
$\mathbf k_t=[\sigma\mathcal R^{-T}\mathbf k_s]_1$ with
$\mathbf G_q\in\mathbb Z^3$, and analogously for $\mathbf k'$. If
$\mathcal R\boldsymbol\tau_{\kappa'}+\mathbf v
=\boldsymbol\tau_\kappa+\mathbf L_{\kappa'}$ with
$\mathbf L_{\kappa'}\in\mathbb Z^3$, the displacement representation has the
nonzero blocks
\begin{equation}\label{eq:displacement-representation}
 \mathcal D_{\kappa\alpha,\kappa'\beta}
 =e^{-2\pi i\mathbf q_t\cdot\mathbf L_{\kappa'}}
 (\mathcal R_{\mathrm c})_{\alpha\beta},
\end{equation}
where $\mathcal R_{\mathrm c}=\mathbf A\mathcal R\mathbf A^{-1}$ is the
Cartesian rotation; $\mathcal D$ is unitary. With the operator
$\hat{\mathcal O}$ that represents $\{\mathcal R|\mathbf v\}$ on the pseudo
states, combined with complex conjugation for $\sigma=-1$, the sewing
matrices
$(W)_{mn}=\langle\tilde\psi_{m\mathbf k_t}|\tilde S_0\hat{\mathcal O}|
\tilde\psi_{n\mathbf k_s}\rangle$ and $W'$, defined analogously at
$\mathbf k'$, are unitary within band sets that contain complete degenerate
multiplets. The recovered matrix elements are
\begin{equation}\label{eq:vertex-recovery}
 g^t_{\kappa\alpha}=
 \sum_{\kappa'\beta}\mathcal D^*_{\kappa\alpha,\kappa'\beta}\,
 W'\,\bigl(g^s_{\kappa'\beta}\bigr)^{(\sigma)}\,W^\dagger,
\end{equation}
as derived in Appendix~\ref{app:symmetry}. The overlap $\tilde S_0$ in the
sewing matrices makes them the all-electron overlaps
$\langle\psi_{m\mathbf k_t}|\hat{\mathcal O}|\psi_{n\mathbf k_s}\rangle$,
because $\mathcal T_0$ commutes with the space-group operations and with
complex conjugation. To symmetrize the result, every symmetry-related source
is transported to the same target with Eq.~\eqref{eq:vertex-recovery}, and
the matrices are averaged over the space-group operations modulo lattice
translations, including those combined with time reversal, before projection
onto phonon modes.

\subsection{Wannier interpolation}
\label{sec:wannier}

Transport requires the matrix elements on much finer meshes than the
supercells provide; they are obtained by Wannier
interpolation~\cite{Giustino2007,Lee2023}. Each Bloch state is defined only
up to a phase, and states within a degenerate subspace only up to a unitary
mixing, and this gauge differs between independent DFT calculations.
Maximally localized Wannier functions~\cite{Marzari2012,Souza2001} fix a
smooth gauge: the Bloch sum of Wannier function $b$ is
$\sum_n|\psi_{n\mathbf k}\rangle V_{nb}(\mathbf k)$, where the matrices
$V(\mathbf k)$ include the disentanglement. On coarse meshes of $N_{\mathbf k}$ and $N_{\mathbf q}$
points, the Hamiltonian and the matrix elements are transformed to the
Wannier representation,
\begin{equation}\label{eq:wannier-transform}
\begin{aligned}
 H_{ab}(\mathbf n_e)&=\frac1{N_{\mathbf k}}\sum_{\mathbf k n}
   e^{-2\pi i\mathbf k\cdot\mathbf n_e}
   V^*_{na}(\mathbf k)\,\epsilon_{n\mathbf k}\,V_{nb}(\mathbf k),\\
 g_{\kappa\alpha,ab}(\mathbf n_e,\mathbf n_p)&=\frac1{N_{\mathbf k}N_{\mathbf q}}
   \sum_{\mathbf k\mathbf q mn}e^{-2\pi i(\mathbf k\cdot\mathbf n_e
   +\mathbf q\cdot\mathbf n_p)}\\
 &\quad\times V^*_{ma}(\mathbf k')\,[g_{\kappa\alpha}(\mathbf k,\mathbf q)]_{mn}\,V_{nb}(\mathbf k),
\end{aligned}
\end{equation}
where $a$ and $b$ label Wannier functions, and $\mathbf n_e$ and
$\mathbf n_p$ are lattice vectors in reduced coordinates. The inverse
transform on fine meshes is accurate only if these quantities decay quickly
with $\mathbf n_e$ and $\mathbf n_p$, which requires $V(\mathbf k)$ to refer
to the same states as $g_{\kappa\alpha}(\mathbf k,\mathbf q)$.

Here, $g_{\kappa\alpha}(\mathbf k,\mathbf q)$ is expressed in the fixed
reference states, also at the momenta reached by symmetry recovery, since
the sewing matrices of Eq.~\eqref{eq:vertex-recovery} are overlaps with
these states. We
therefore construct $V(\mathbf k)$ with Wannier90~\cite{Pizzi2020} from the
reference states themselves. When these are computed only at irreducible
momenta, the states unfolded by symmetry are used both in the matrix
elements and in Wannier90; a separate calculation on the full mesh would
introduce different phases. In PAW, $V(\mathbf k)$ acts only on the band
index and is therefore the same for the pseudo and the all-electron
states~\cite{Engel2020}.

Band velocities and effective masses converge more slowly with the coarse
$\mathbf k$ mesh than the matrix elements, because $H(\mathbf n_e)$ must
resolve the band curvature at extrema that need not lie at high-symmetry
points~\cite{Poliukhin2026}, whereas
$g_{\kappa\alpha}(\mathbf n_e,\mathbf n_p)$ decays quickly. We therefore
construct $H$ on a finer mesh, a multiple of the coarse $\mathbf k$ mesh,
from a separate primitive-cell calculation with Wannier gauge
$V_f(\mathbf k)$, and keep $g$ on the coarse meshes. At the common
$\mathbf k$ points, the states of this calculation differ from the
reference states by phases and by mixing within degenerate subspaces,
\begin{equation}\label{eq:gauge-fit}
 |\tilde f_{m\mathbf k}\rangle
 =\sum_n|\tilde\psi_{n\mathbf k}\rangle S_{nm}(\mathbf k),
\end{equation}
so that the same Wannier functions are described in the reference states by
$S(\mathbf k)V_f(\mathbf k)$. We obtain $S_{nm}(\mathbf k)=\langle\tilde\psi_{n\mathbf k}|\tilde S_0|\tilde f_{m\mathbf k}\rangle$
for the states in the disentanglement window, as in the sewing matrices, and
replace $V(\mathbf k)$ in the transform of $g$ by the matrix with
orthonormal columns closest to $S(\mathbf k)V_f(\mathbf k)$, obtained from
its singular-value decomposition. $S(\mathbf k)$ is unitary up to numerical
precision when both calculations use the same density; with hybrid
functionals, the finer calculation samples the Fock operator on a different
mesh, and the norm of the part of $|\tilde f_{m\mathbf k}\rangle$ outside the
reference states measures the resulting small change of the states.

\section{Computational workflow}
\label{sec:workflow}

\begin{figure*}[tp]
 \includegraphics[width=\textwidth]{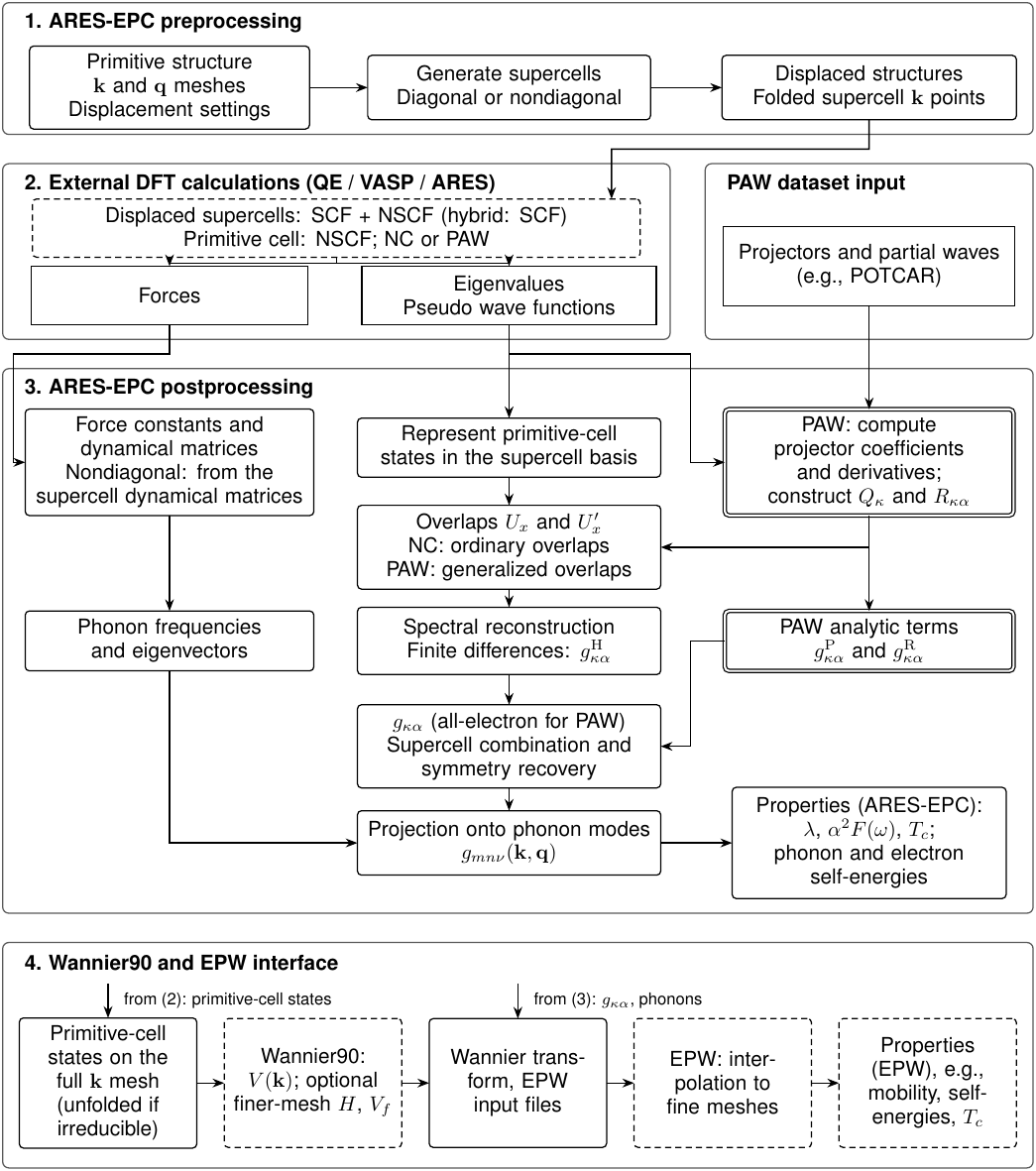}
 \caption{Computational workflow of ARES-EPC. Preprocessing (1) writes the
 inputs of the supercell calculations, external DFT calculations (2) supply
 forces, eigenvalues, and pseudo wave functions, and postprocessing (3)
 follows a phonon path and an electronic path that meet in the projection
 onto phonon modes, from which ARES-EPC computes properties. The interface to
 EPW (4) uses the primitive-cell states on the full $\mathbf k$ mesh for
 Wannier90, and the matrix elements and phonons of (3) for the input files of
 EPW. Dashed borders mark external programs, double borders the steps specific
 to PAW.
 \label{fig:workflow}}
\end{figure*}

Figure~\ref{fig:workflow} shows the implementation in ARES-EPC. The code
does not run the DFT calculations itself; preprocessing writes the inputs of
the supercell calculations, and postprocessing reads their output. The
calculations of different supercells and displacements are independent and
can run concurrently. By default, each displaced supercell is treated in two
steps (NSCF mode): an SCF calculation gives the
forces and the density, and a non-self-consistent (NSCF) calculation from
this density gives the eigenvalues and pseudo wave functions at the folded
momenta of Eq.~\eqref{eq:k-folding}. The reference states are obtained from
an NSCF calculation of the primitive cell. The supercell NSCF calculations
retain the number of bands given in Sec.~\ref{sec:computational}, whereas
the SCF calculations need only the bands that determine the density. For
hybrid functionals, whose Fock operator depends on the occupied orbitals and
not only on the density, the NSCF step is not possible, and the SCF mode is
used: each SCF calculation runs at the momenta and with the number of bands
of the NSCF calculation it replaces.

On the phonon path, the forces give the force constants of the diagonal
supercell directly. In the nondiagonal route, the dynamical matrices of each
supercell at its commensurate $\mathbf q$ points are symmetrized and Fourier
transformed into the force constants of the diagonal supercell commensurate
with the full $\mathbf q$ mesh,
from which the phonons are obtained on the full $\mathbf q$ mesh. On the
electronic path, the atomic-displacement matrix elements of each supercell
follow from
Secs.~\ref{sec:generalized-projection}--\ref{sec:supercells} and are
recovered on the full mesh by symmetry (Sec.~\ref{sec:symmetry-recovery}).
The two paths meet in the projection onto phonon modes,
Eq.~\eqref{eq:mode-vertex}, from which ARES-EPC computes properties
directly. NC and PAW calculations share the same workflow. PAW adds the projector coefficients obtained with the PAW dataset
and the analytic terms of Eq.~\eqref{eq:paw-corrections}, marked by double
borders in Fig.~\ref{fig:workflow}.

The interface to EPW~\cite{Lee2023} starts from the reference states of the
primitive cell. Wannier90 needs them on the full $\mathbf k$ mesh, so states
computed only at irreducible momenta are first unfolded by symmetry
(Sec.~\ref{sec:wannier}). Wannier90 gives the matrices $V(\mathbf k)$ and,
optionally, $H$ and $V_f(\mathbf k)$ on a finer mesh. A script then performs
the transform of Eq.~\eqref{eq:wannier-transform} on the matrix elements,
reference eigenvalues, and phonons written by ARES-EPC, and writes the result
in the format that EPW reads. We modified EPW to read $H$ and $g$ on different
electronic meshes. EPW interpolates them to fine meshes and computes the
properties it supports.

\section{Computational details}
\label{sec:computational}

Unless stated otherwise, all calculations use the PBE generalized-gradient
approximation~\cite{Perdew1996}; all are
non-spin-polarized and omit spin--orbit
coupling. NC calculations are performed with Quantum ESPRESSO (QE)
\cite{Giannozzi2009,Giannozzi2017} and scalar-relativistic optimized
norm-conserving Vanderbilt pseudopotentials of PseudoDojo
\cite{VanSetten2018,Hamann2013}, with a plane-wave cutoff of 60~Ry and an SCF
convergence threshold of $10^{-13}$~Ry. PAW calculations are performed with
VASP~\cite{KresseHafner1993,KresseFurthmueller1996} and its PBE PAW datasets
\cite{KresseJoubert1999}, with a plane-wave cutoff of 520~eV and an SCF
convergence threshold of $10^{-8}$~eV. The structures, pseudopotentials, and
smearing settings are listed in the Supplemental Material.

In ARES-EPC, all displacements have an amplitude of 0.001~\AA, and the
matrix elements are obtained by central differences. In PAW, both
displacement signs are calculated explicitly; in NC, the negative
displacement is obtained by site symmetry when possible. The acoustic sum
rule is imposed on the force constants. Space-group
operations are obtained with spglib~\cite{Togo2024}. The number of bands of
each supercell calculation is obtained from the primitive-cell eigenvalues: it
includes all states up to 3~eV above the highest reference state among the
momenta folded onto each supercell $\mathbf k$ point
[Eq.~\eqref{eq:k-folding}], plus ten bands. For the bands entering the
matrix elements, we require the projectability residual of
Sec.~\ref{sec:generalized-projection} to be below $10^{-5}$ for NC and
$10^{-4}$ for PAW.

In the diagonal-supercell benchmarks, ARES-EPC uses a diagonal supercell
$\mathbf n_{\mathrm{ref}}=(n_1,n_2,n_3)$
(Table~\ref{tab:diagonal-benchmarks}), and the electronic $\mathbf k$ mesh,
the phonon $\mathbf q$ mesh, and the primitive-cell SCF mesh are all
$\mathbf n_{\mathrm{ref}}$; the supercell SCF calculations therefore use the
$\Gamma$ point only. QE DFPT computes the matrix elements on the same
$\mathbf k$ and $\mathbf q$ meshes from a primitive-cell SCF calculation on
the $\mathbf n_{\mathrm{ref}}$ mesh, with a threshold of $10^{-18}$ on the
self-consistent potential response. VASP+phelel uses the same
$\mathbf n_{\mathrm{ref}}$ supercell, $\pm0.001$~\AA\ displacements, and
$\mathbf k$ and $\mathbf q$ meshes as ARES-EPC. Both PAW codes evaluate the
all-electron matrix element of Eq.~\eqref{eq:cartesian-definition}.
In each benchmark, ARES-EPC
and the corresponding reference code use the same structure, cutoff,
occupations, and smearing.

In the diagonal--nondiagonal comparison, both calculations use ARES-EPC with
PAW and the same $\mathbf n_{\mathrm{pair}}$ electronic $\mathbf k$ and
phonon $\mathbf q$ meshes (Table~\ref{tab:nondiagonal-benchmarks}). The
diagonal calculation uses an $\mathbf n_{\mathrm{pair}}$ supercell; the
nondiagonal calculation uses the nondiagonal supercells that are commensurate
with the $\mathbf n_{\mathrm{pair}}$ $\mathbf q$ mesh. Both use the same
primitive-cell SCF mesh, listed in the Supplemental Material, which is folded
into every supercell (Sec.~\ref{sec:supercells}).

For CaH$_6$, the Eliashberg spectral function $\alpha^2F(\omega)$ and the
EPC constant $\lambda$~\cite{Allen1983} are computed from the PAW matrix
elements, with the double delta functions replaced by Gaussians of
widths 0.05--0.50~eV in steps of 0.05~eV. The diagonal and nondiagonal
calculations on the $5^3$ $\mathbf q$ mesh use a $10^3$ $\mathbf k$ mesh, and
the nondiagonal calculation on the $8^3$ $\mathbf q$ mesh uses a $16^3$
$\mathbf k$ mesh. For each calculation, we take the width at which $\lambda$
is minimal, 0.35~eV for the $5^3$ and 0.40~eV for the $8^3$ mesh. We obtain
the critical temperature $T_c$ from the isotropic Eliashberg
equations~\cite{Eliashberg1960,Allen1983} with a Coulomb pseudopotential
$\mu^*=0.13$~\cite{Morel1962}.

For the electron mobility of Si, the PAW calculations use the PBE and the
HSE06 functional~\cite{Heyd2003,Krukau2006} (screening parameter
0.2~\AA$^{-1}$, exact-exchange fraction 0.25), each at its own relaxed
lattice constant, 5.469~\AA\ for PBE and 5.433~\AA\ for HSE06; the
SCF convergence threshold is $10^{-10}$~eV. ARES-EPC uses nondiagonal
supercells with a $12^3$ electronic $\mathbf k$ mesh and a $6^3$ phonon
$\mathbf q$ mesh. The matrix elements are computed for the lowest 16 bands,
and the wave functions only at irreducible momenta
(Sec.~\ref{sec:wannier}).

Eight Wannier functions are constructed from $sp^3$ projections on the Si
atoms, with the frozen window extending 2.5~eV (PBE) and 1.5~eV (HSE06)
above the conduction-band minimum. The Hamiltonian is constructed on a
$24^3$ mesh and the matrix elements on the $12^3\times6^3$ meshes
(Sec.~\ref{sec:wannier}). For the
NC end-to-end comparison with DFPT, ARES-EPC and QE DFPT use the PBE
lattice constant, $12^3$ $\mathbf k$ and $6^3$ $\mathbf q$ meshes, the
same Wannier projections and windows, and $H$ on the $12^3$ mesh; here the SCF threshold
is $10^{-11}$~Ry. In all mobility calculations, EPW uses $100^3$ fine
$\mathbf k$ and $\mathbf q$ meshes, adaptive smearing, an energy window of
0.2~eV around the band edges, an electron density of
$10^{13}$~cm$^{-3}$, and 300~K. The Boltzmann transport equation is solved in the self-energy
relaxation-time approximation (SERTA) and iteratively (IBTE)~\cite{Ponce2020};
dynamical quadrupoles are not included.

\section{Results}
\label{sec:results}

The seven test materials cover different bonding and electronic structures:
the simple metal Al, the heavy metal Pb with $5d$ semicore states, the
covalent semiconductors diamond and Si, the compound semiconductor GaAs with Ga
$3d$ semicore states, the layered metal MgB$_2$ with strongly coupled $E_{2g}$
modes, and the high-pressure hydride CaH$_6$, which requires the
largest supercells.

\subsection{Comparison protocol}
\label{sec:protocol}

Each calculation uses its own phonon frequencies and eigenvectors, and the
matrix elements are compared at common $(\mathbf k,\mathbf q)$ points for
complete degenerate groups (defined below) of the lowest electronic bands, listed in the Supplemental
Material. Matrix elements between degenerate states depend on the choice of basis
within the degenerate subspaces. We therefore compare the group norms
\begin{equation}\label{eq:degenerate-norm}
 \mathcal N_{IJ\Lambda}(\mathbf k,\mathbf q)=
 \Bigl[\frac{1}{|I|\,|J|\,|\Lambda|}
 \sum_{m\in I}\sum_{n\in J}\sum_{\nu\in\Lambda}
 |g_{mn\nu}(\mathbf k,\mathbf q)|^2\Bigr]^{1/2},
\end{equation}
where $I$, $J$, and $\Lambda$ are groups of degenerate final bands, initial
bands, and phonon modes, and $|I|$, $|J|$, and $|\Lambda|$ are their sizes.
The group norm is invariant under unitary rotations within
each degenerate subspace and reduces to $|g_{mn\nu}|$ for nondegenerate
states. States
are grouped if their energies differ by less than 0.05~eV for electrons and
0.1~meV for phonons in either calculation, except for the NC comparison of
MgB$_2$, which uses 0.5~meV for phonons. Phonon
groups whose lowest energy in either calculation is below 0.5~meV are
excluded; this removes the acoustic modes at $\Gamma$ and the soft modes of
CaH$_6$ and Pb in the diagonal benchmarks, whose primitive-cell SCF mesh is
only $\mathbf n_{\mathrm{ref}}$. With the denser SCF meshes of the
nondiagonal comparison and of Sec.~\ref{sec:cah6} (Supplemental Material),
these soft modes are absent. The differences between
corresponding group norms are summarized by the mean absolute error (MAE)
and the root-mean-square error (RMSE) over all compared groups, whose number ranges from
$1.6\times10^4$ to $2.6\times10^7$ per comparison.

\subsection{Comparison with DFPT and VASP+phelel}
\label{sec:diagonal-results}

Table~\ref{tab:diagonal-benchmarks} summarizes the diagonal-supercell
comparisons; the group norms of all materials are compared in the
Supplemental Material. Against QE DFPT, the MAEs lie between 0.0078 and
0.28~meV.
The largest NC differences, 90~meV (6.5\%) and 52~meV (4.3\%), occur for
MgB$_2$ at $\Gamma$, for the $E_{1u}$ and $A_{2u}$ modes between the Mg $2s$
and $2p$ semicore states. They originate from the phonons: the DFPT dynamical
matrix at $\Gamma$ violates the acoustic sum rule, with acoustic energies of
9.6--9.9~meV, whereas ARES-EPC imposes it (Sec.~\ref{sec:computational}).
As a result, the $E_{1u}$ energy is 41.9~meV in QE DFPT and 43.0~meV in
ARES-EPC, and the Mg components of the eigenvectors differ by 5.6\%; together
these account for the 6.5\%. For the $E_{2g}$ and $B_{1g}$ modes, in which Mg
does not move, the matrix elements at $\Gamma$ agree to within 0.006~meV.
For the other materials, the phonon energies of the compared modes agree to
within 0.08~meV.

\begin{figure*}[t]
\begin{minipage}[t]{0.49\textwidth}
\centering\textbf{(a)}\\[-2pt]
\includegraphics[width=\linewidth]{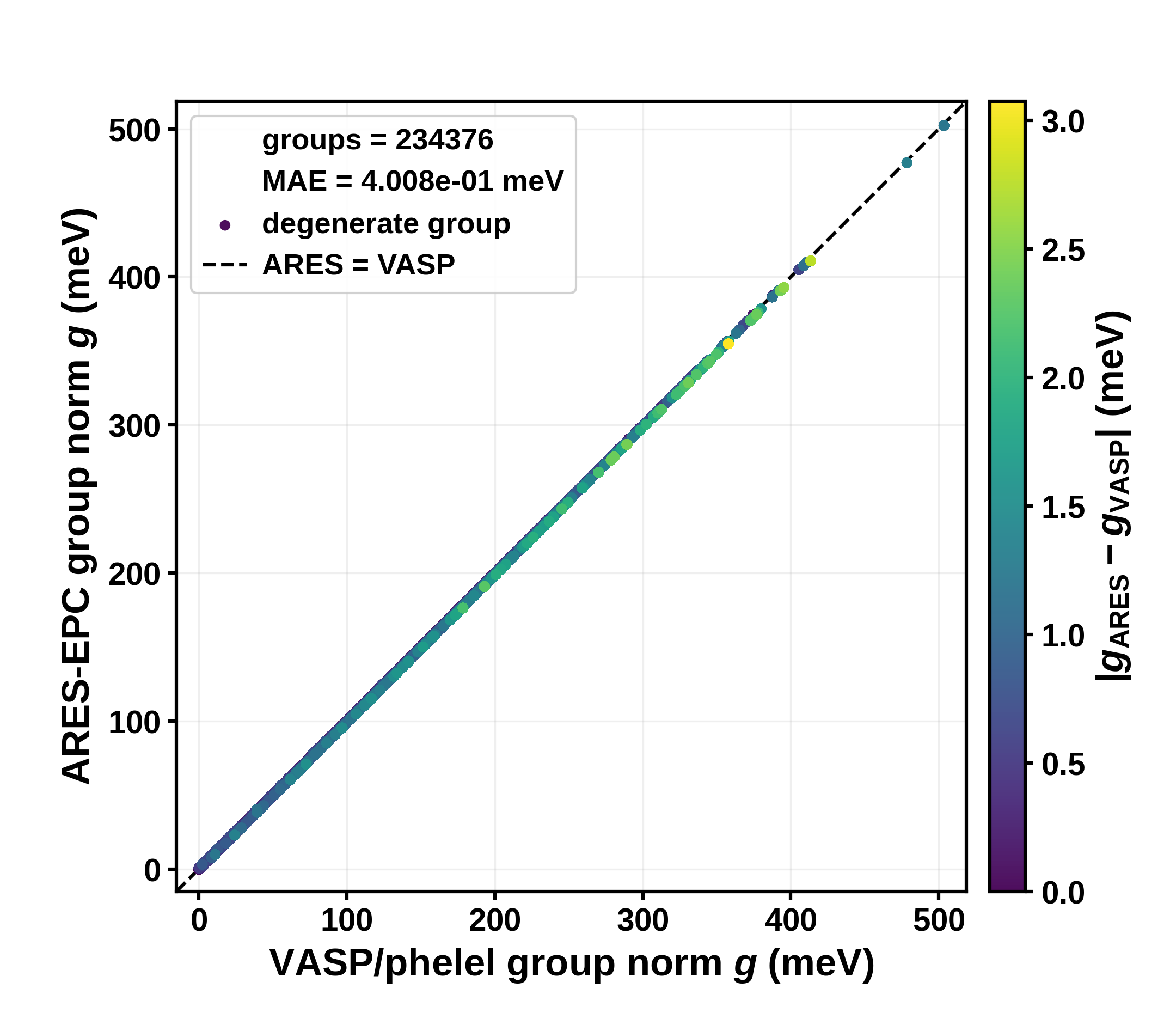}
\end{minipage}\hfill
\begin{minipage}[t]{0.49\textwidth}
\centering\textbf{(b)}\\[-2pt]
\includegraphics[width=\linewidth]{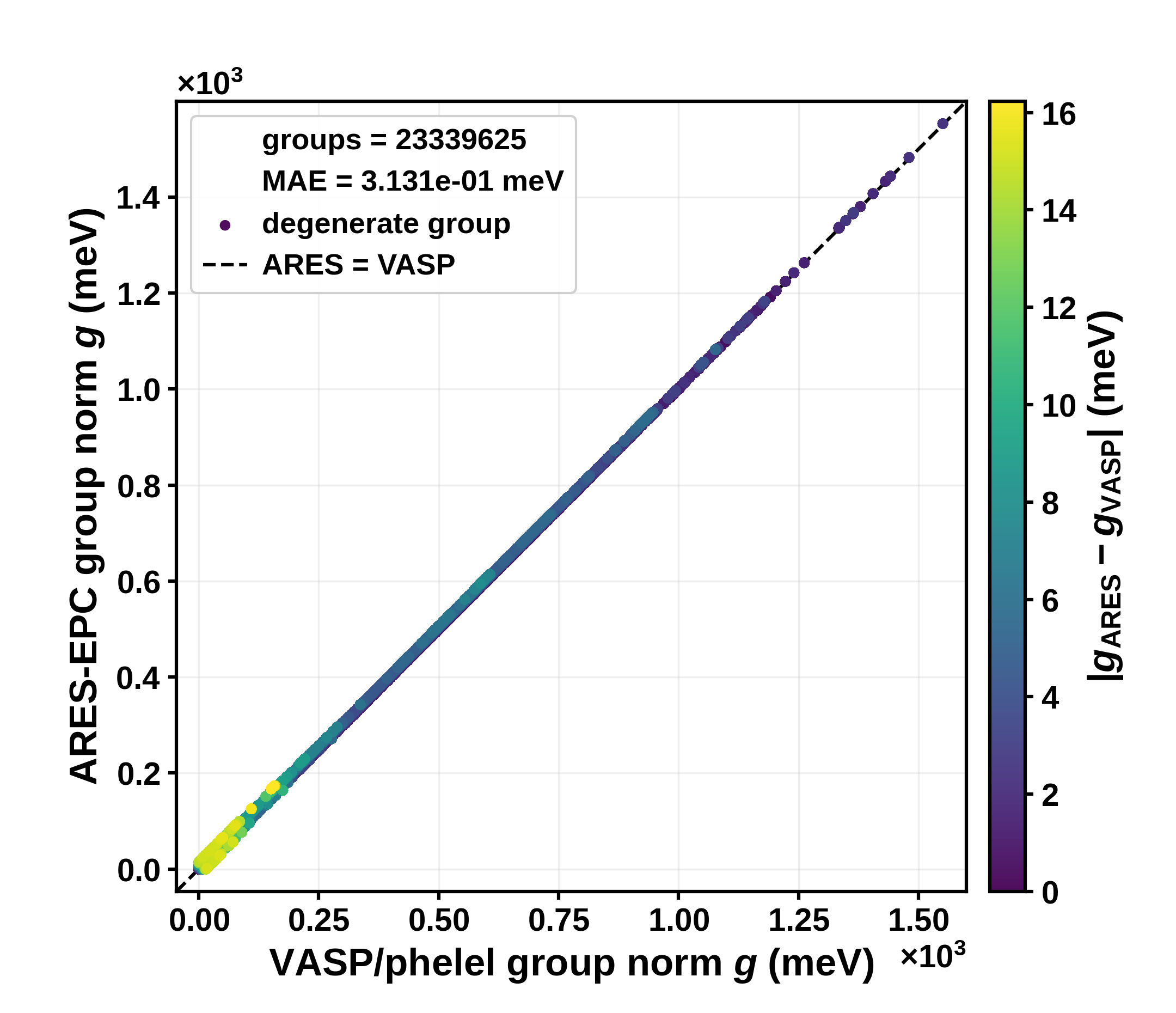}
\end{minipage}
\caption{\label{fig:paw-parity}Group norms $\mathcal N$
[Eq.~\eqref{eq:degenerate-norm}] from ARES-EPC+VASP versus VASP+phelel for
(a) diamond and (b) CaH$_6$ with diagonal supercells. Each point is one group
of electronic bands and phonon modes; the dashed line marks equality, and the
color gives the absolute difference (meV).}
\end{figure*}

Against VASP+phelel, the MAEs lie between 0.0052 and 0.40~meV.
Figure~\ref{fig:paw-parity} shows the comparison for diamond, which has
the largest PAW MAE, and for CaH$_6$, which is studied
further in Sec.~\ref{sec:cah6}. The points lie on the line of equality over the full
range of group norms. The
agreement shows that the generalized overlaps and the analytic terms of
Eq.~\eqref{eq:paw-corrections} reproduce the all-electron PAW matrix element
of VASP+phelel.

\begin{table}
\caption{\label{tab:diagonal-benchmarks}Group-norm errors (meV) of the
diagonal-supercell benchmarks. $\mathbf n_{\mathrm{ref}}$ is the diagonal
supercell expansion, which also defines the $\mathbf k$ and $\mathbf q$ meshes.
NC compares ARES-EPC+QE with QE DFPT, and PAW compares ARES-EPC+VASP with
VASP+phelel; MAE and RMSE are taken over all compared groups.}
\begin{ruledtabular}
\begin{tabular}{lccccc}
 & & \multicolumn{2}{c}{NC vs QE DFPT} & \multicolumn{2}{c}{PAW vs VASP+phelel} \\
 Material & $\mathbf n_{\mathrm{ref}}$ & MAE & RMSE & MAE & RMSE \\
\hline
 Al & $(6,6,6)$ & 0.018 & 0.19 & 0.0052 & 0.021 \\
 CaH$_6$ & $(5,5,5)$ & 0.18 & 1.1 & 0.31 & 0.69 \\
 Diamond & $(4,4,4)$ & 0.0078 & 0.018 & 0.40 & 0.55 \\
 GaAs & $(4,4,4)$ & 0.0094 & 0.032 & 0.38 & 0.64 \\
 MgB$_2$ & $(4,4,4)$ & 0.28 & 1.4 & 0.12 & 0.22 \\
 Pb & $(4,4,4)$ & 0.065 & 0.18 & 0.38 & 1.0 \\
 Si & $(4,4,4)$ & 0.0087 & 0.020 & 0.042 & 0.058
\end{tabular}
\end{ruledtabular}
\end{table}

\subsection{Nondiagonal supercells}
\label{sec:nondiagonal-results}

\begin{figure*}[t]
\begin{minipage}[t]{0.49\textwidth}
\centering\textbf{(a)}\\[-2pt]
\includegraphics[width=\linewidth]{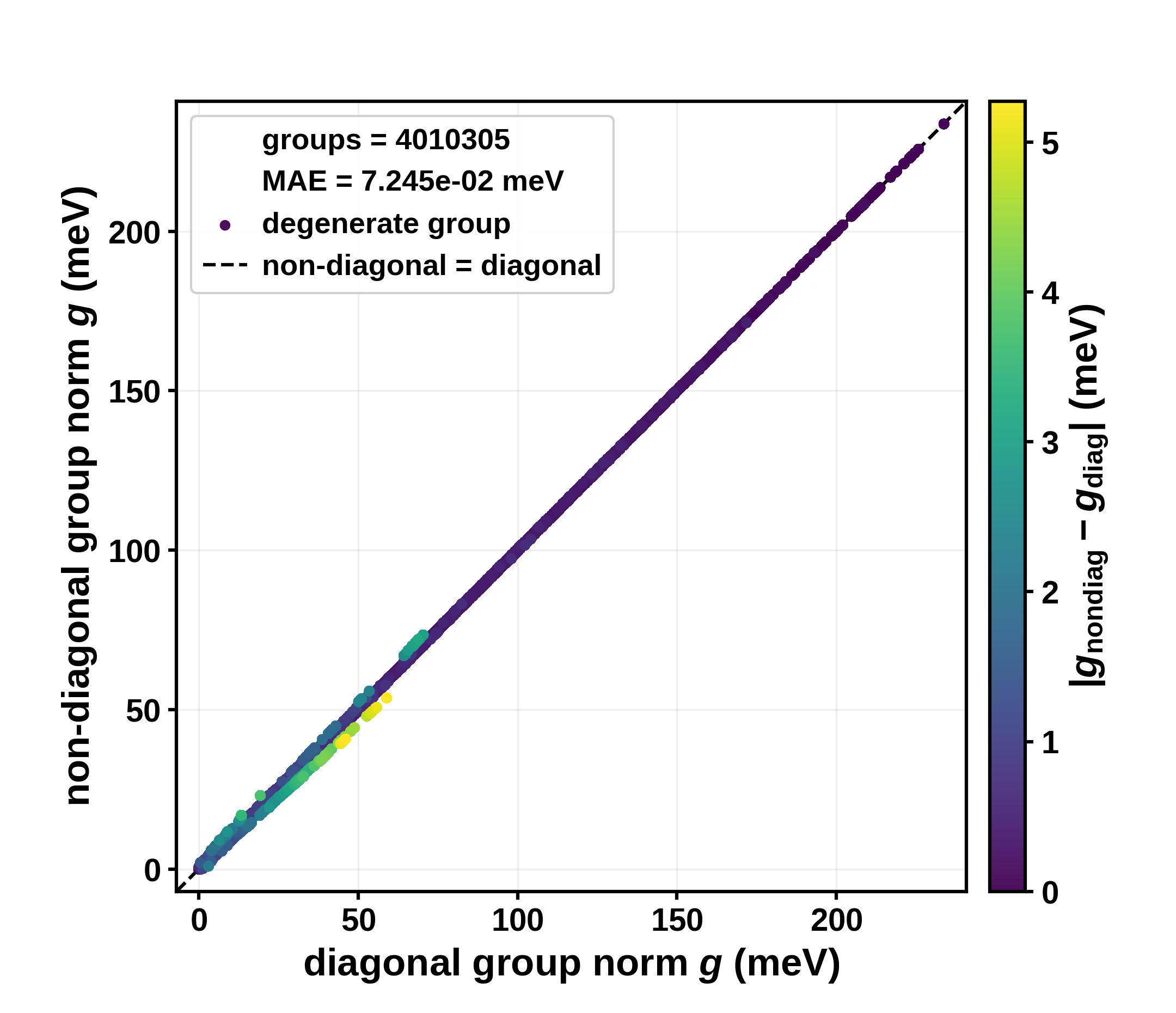}
\end{minipage}\hfill
\begin{minipage}[t]{0.49\textwidth}
\centering\textbf{(b)}\\[-2pt]
\includegraphics[width=\linewidth]{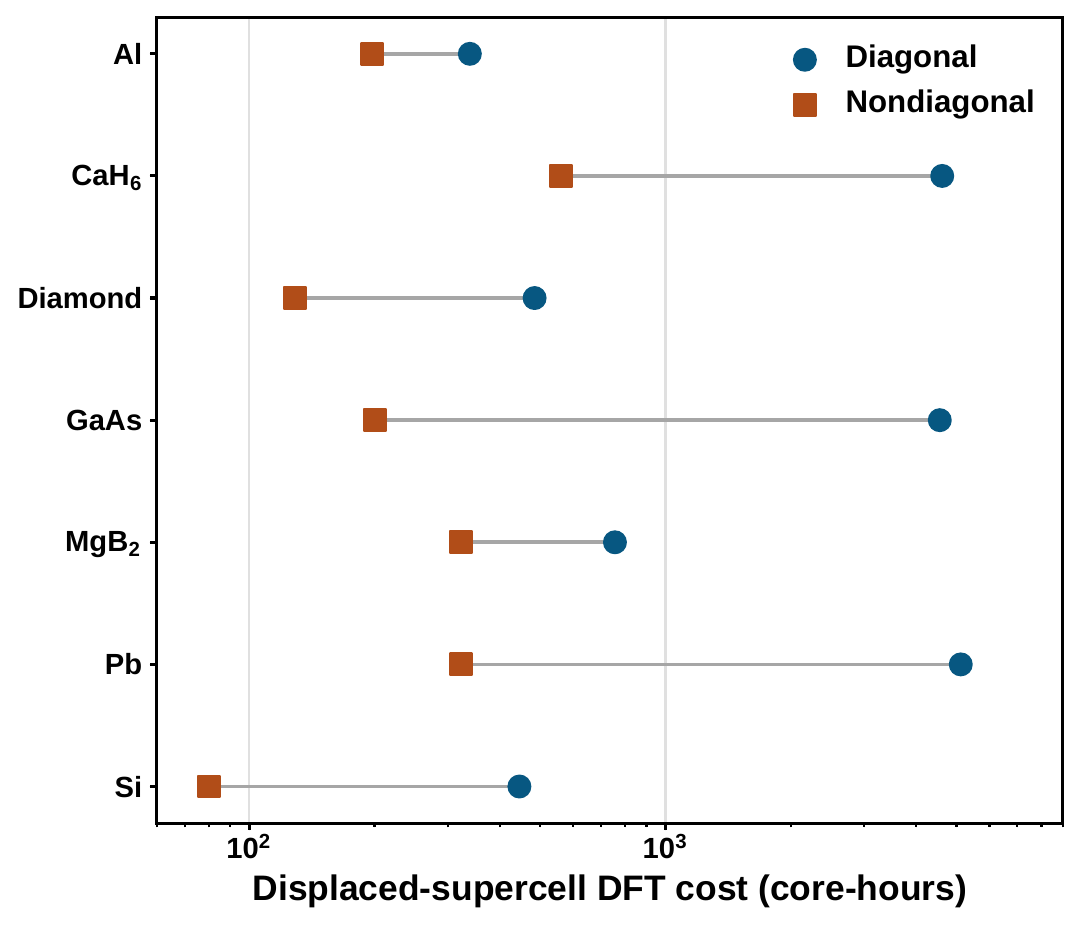}
\end{minipage}
\caption{\label{fig:nondiagonal}(a) Group norms of GaAs from nondiagonal
versus diagonal ARES-EPC+VASP calculations; the dashed line marks equality,
and the color gives the absolute difference. (b) SCF plus NSCF cost of all
displaced supercells in core-hours for the diagonal (circles) and
nondiagonal (squares) routes. Gray lines connect the two routes for each
material.}
\end{figure*}

The diagonal and nondiagonal ARES-EPC+VASP calculations use the same
$\mathbf k$ and $\mathbf q$ meshes for each material
(Table~\ref{tab:nondiagonal-benchmarks}). The phonon frequencies of the two
routes agree to within 0.1~meV, both on these $\mathbf q$ meshes and along
high-symmetry paths (Supplemental Material). The group-norm MAEs range from
0.0079 to 0.072~meV, with the largest value for GaAs
[Fig.~\ref{fig:nondiagonal}(a)]. The largest difference for an individual
group is 12.5~meV, for CaH$_6$. It belongs to a small matrix element within
the Ca $3s$ semicore band, about 40~eV below the Fermi level: 8.4~meV in the
diagonal and 20.9~meV in the nondiagonal calculation, whereas the group
norms of CaH$_6$ reach 1.4~eV (Sec.~\ref{sec:discussion}). The combination of nondiagonal supercells and the
symmetry recovery thus preserves the matrix elements of the diagonal route.

The cost is measured in core-hours of the SCF and NSCF calculations, summed
over all displaced supercells; the primitive-cell calculations and the
postprocessing are not included. The diagonal route uses a single
$\mathbf n_{\mathrm{pair}}$ supercell. Because the nondiagonal supercells
are independent and can run concurrently, we compare aggregate core-hours
rather than elapsed time. All calculations use 192 cores, at which the small
nondiagonal supercells likely run with lower parallel efficiency than the
diagonal ones, so the reported reductions are conservative. The nondiagonal
route reduces the core-hours by factors of 1.7--23 [Fig.~\ref{fig:nondiagonal}(b)]. For a given material, the
reduction is expected to grow with the size of the diagonal supercell,
because the cost of the diagonal route increases steeply with $N_p$.

\begin{table*}[tp]
\caption{\label{tab:nondiagonal-benchmarks}Comparison of diagonal and
nondiagonal ARES-EPC+VASP calculations. $\mathbf n_{\mathrm{pair}}$ is the
common $\mathbf k$ and $\mathbf q$ mesh and the diagonal supercell expansion,
$N_{\mathrm{at}}$ is the number of atoms in the diagonal supercell and in each
nondiagonal supercell, and $N_{\mathrm{disp}}$ is the total number of
displaced structures, each requiring an SCF and an NSCF calculation. $t$ is
the SCF plus NSCF cost of all displaced structures (wall time $\times$ 192
cores). Superscripts d and nd denote the diagonal and nondiagonal routes.}
\begin{ruledtabular}
\begin{tabular}{lcccccccccc}
 Material & $\mathbf n_{\mathrm{pair}}$ & $N_{\mathrm{at}}^{\mathrm d}$ &
 $N_{\mathrm{at}}^{\mathrm{nd}}$ & $N_{\mathrm{disp}}^{\mathrm d}$ &
 $N_{\mathrm{disp}}^{\mathrm{nd}}$ & MAE & RMSE & $t^{\mathrm d}$ &
 $t^{\mathrm{nd}}$ & $t^{\mathrm d}/t^{\mathrm{nd}}$ \\
 & & & & & & (meV) & (meV) & (core-h) & (core-h) & \\
\hline
 Al & $(6,6,6)$ & 216 & 6 & 2 & 32 & 0.014 & 0.047 & 339 & 198 & 1.7 \\
 CaH$_6$ & $(5,5,5)$ & 875 & 35 & 4 & 62 & 0.030 & 0.074 & 4620 & 561 & 8.2 \\
 Diamond & $(6,6,6)$ & 432 & 12 & 2 & 32 & 0.020 & 0.040 & 485 & 129 & 3.8 \\
 GaAs & $(5,5,5)$ & 250 & 10 & 4 & 32 & 0.072 & 0.21 & 4558 & 201 & 22.7 \\
 MgB$_2$ & $(4,4,4)$ & 192 & 12 & 4 & 40 & 0.0079 & 0.054 & 756 & 323 & 2.3 \\
 Pb & $(6,6,6)$ & 216 & 6 & 2 & 32 & 0.072 & 0.20 & 5119 & 322 & 15.9 \\
 Si & $(5,5,5)$ & 250 & 10 & 2 & 16 & 0.0097 & 0.017 & 446 & 80 & 5.6
\end{tabular}
\end{ruledtabular}
\end{table*}

\subsection{End-to-end comparison for Si}
\label{sec:si-validation}

\begin{figure*}[t]
\includegraphics[width=\textwidth]{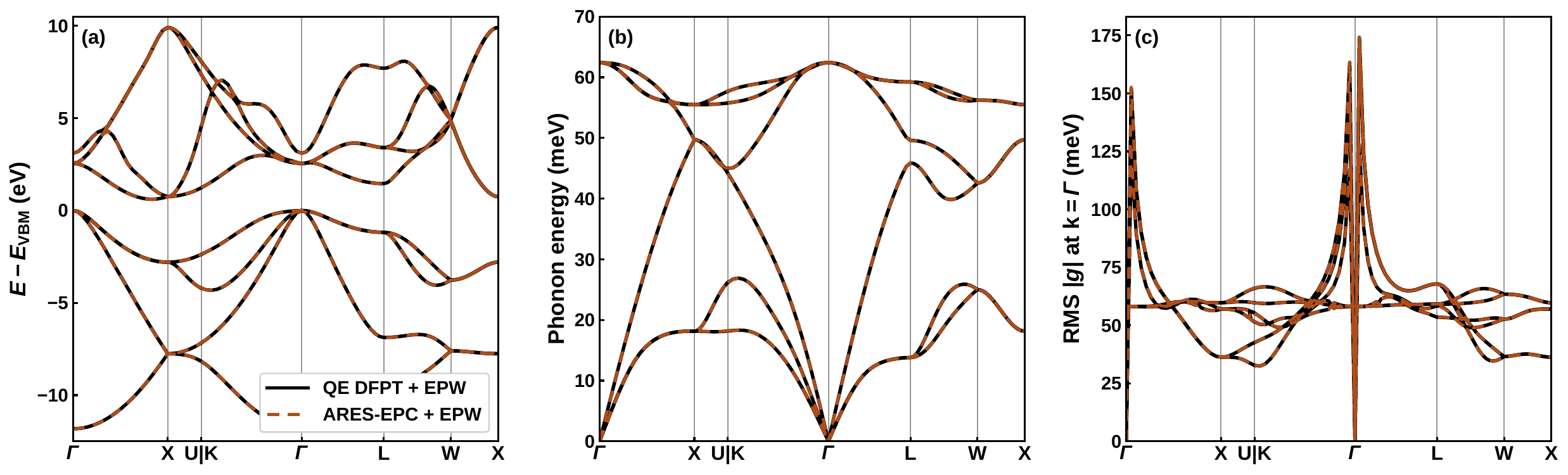}
\caption{\label{fig:si-validation}Si (NC, PBE) from QE DFPT+EPW (solid) and
ARES-EPC+EPW (dashed) along a high-symmetry path: (a) Wannier-interpolated
bands, (b) phonons, and (c) root mean square of $|g_{mn\nu}|$ over the Wannier bands $m$, $n$ at
$\mathbf k=\Gamma$ for each mode $\nu$.}
\end{figure*}

To test the matrix elements together with the interface to EPW, we compute
the electron mobility of Si from the NC phonons and matrix elements of
ARES-EPC and of QE DFPT, with all other settings shared
(Sec.~\ref{sec:computational}). Figure~\ref{fig:si-validation}
compares the interpolated bands, phonons, and matrix elements along a
high-symmetry path, and Table~\ref{tab:si-validation} summarizes the
differences. The bands agree to 0.005~meV and the phonons to 0.09~meV. On the
coarse meshes, the MAE of $|g|$ is 0.080~meV for a mean $|g|$ of 42.6~meV.
Along the path, the largest difference of $|g|$, 3.9~meV (2.9\%), occurs for
the longitudinal acoustic (LA) mode about one third of the way from $\Gamma$
to X (Sec.~\ref{sec:discussion}). The IBTE
mobilities at 300~K differ by 0.9\%. The matrix elements of ARES-EPC,
transferred to EPW through the interface, therefore reproduce the mobility
obtained from DFPT.

\begin{table}
\caption{\label{tab:si-validation}Differences between ARES-EPC+EPW and QE
DFPT+EPW for Si (NC, PBE). ``Coarse'': 40 $\mathbf k$ points of the $12^3$
mesh and the $6^3$ $\mathbf q$ mesh; ``path'': $\mathbf k=\Gamma$ and
$\mathbf q$ along the path of Fig.~\ref{fig:si-validation}. $\mu$: electron
drift mobility at 300~K (cm$^2$/V\,s).}
\begin{ruledtabular}
\begin{tabular}{lc}
 Quantity & Difference \\
\hline
 Bands, max $|\Delta\epsilon|$ (meV) & 0.0045 \\
 Phonons, max $|\Delta\omega|$ (meV) & 0.085 \\
 $|g|$ coarse, MAE / RMSE (meV) & 0.080 / 0.22 \\
 $|g|$ path, max $|\Delta g|$ (meV) & 3.9 \\
 \hline
 & QE DFPT / ARES-EPC \\
 $\mu$ IBTE & 1524 / 1537 \\
 $\mu$ SERTA & 1466 / 1482
\end{tabular}
\end{ruledtabular}
\end{table}

\subsection{Electron mobility of Si with PBE and HSE06}
\label{sec:si-hse}

\begin{figure*}[t]
\includegraphics[width=\textwidth]{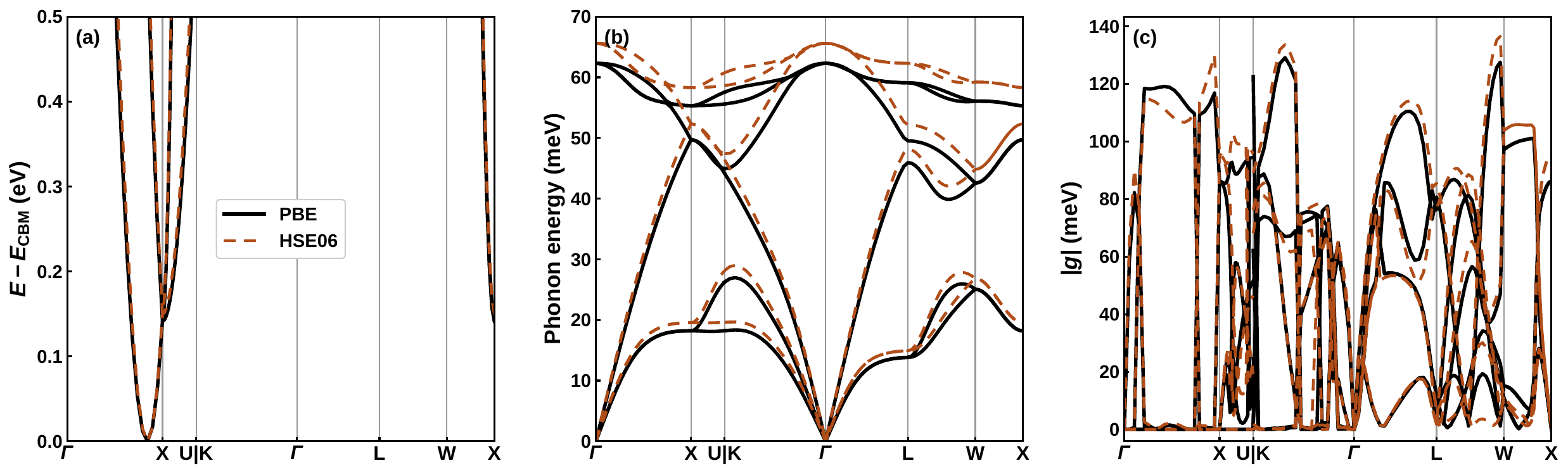}
\caption{\label{fig:si-hse}Si (PAW) with PBE (solid) and
HSE06 (dashed) along the path of Fig.~\ref{fig:si-validation}: (a)
VASP conduction bands relative to the respective conduction-band minimum, (b)
phonons from ARES-EPC, and (c) $|g_{mn\nu}|$ from ARES-EPC+EPW for each mode $\nu$ from the lowest
conduction band $n$ at the conduction-band minimum $\mathbf k$ to the lowest
conduction band $m$ at $\mathbf k+\mathbf q$.}
\end{figure*}

With PAW and the settings of Sec.~\ref{sec:computational}, we compare the
electron mobility of Si obtained with PBE and with HSE06
(Fig.~\ref{fig:si-hse}). Relative to PBE, HSE06
increases the 300~K electron mobility by 13\% in the IBTE, from 1397 to
1581~cm$^2$/V\,s, and by 12\% in the SERTA (Table~\ref{tab:si-hse}). The
SERTA mobility $\mu=\sum_\alpha\mu_{\alpha\alpha}/3$, which for cubic Si equals
each diagonal component of the mobility tensor, can be written exactly
as~\cite{Pant2026}
\begin{equation}
\begin{aligned}
\mu&=\frac{e\bar\tau}{m_c},\qquad
\frac1{m_c}=\frac{\sum_{n\mathbf k}(-\partial f/\partial\epsilon)_{n\mathbf k}
|\mathbf v_{n\mathbf k}|^2}{3\sum_{n\mathbf k}f_{n\mathbf k}},\\
\bar\tau&=\frac{\sum_{n\mathbf k}(-\partial f/\partial\epsilon)_{n\mathbf k}
|\mathbf v_{n\mathbf k}|^2\tau_{n\mathbf k}}
{\sum_{n\mathbf k}(-\partial f/\partial\epsilon)_{n\mathbf k}|\mathbf v_{n\mathbf k}|^2},
\end{aligned}
\label{eq:serta-factor}
\end{equation}
where the sums run over the conduction states, $f$ is the Fermi--Dirac
occupation,
$\mathbf v_{n\mathbf k}=\hbar^{-1}\nabla_{\mathbf k}\epsilon_{n\mathbf k}$ the
band velocity, $\tau_{n\mathbf k}$ the SERTA relaxation time, $m_c$ the
conductivity effective mass, and $\bar\tau$ the mean relaxation time.

HSE06 reduces $m_c$ by 5.3\% [Fig.~\ref{fig:si-hse}(a)] and increases $\bar\tau$ by
5.7\%, which together give the SERTA ratio of 1.116. With
$1/\bar\tau=(2\pi/\hbar)\,\bar g^2\bar\rho$, where $\bar\rho$ is an average
density of final states and $\bar g^2$ the corresponding average
coupling~\cite{Pant2026}, the longer relaxation time does not come from
weaker coupling: $\bar g^2$ increases by 13\%, whereas $\bar\rho$ decreases by
16\%. With the HSE06 bands and the PBE phonons, $\bar\rho$ is 0.92 times the
PBE value, and the harder HSE06 phonons [Fig.~\ref{fig:si-hse}(b)] give a
further factor of 0.91. Together with
$m_c$, the smaller $\bar\rho$ gives a factor of 1.26, which the stronger
coupling reduces to 1.12. The derivation of this decomposition, the
individual factors, and the convergence with the mesh of $H$ are given in the
Supplemental Material.

Table~\ref{tab:si-hse} compares the results with other calculations and with
experiment. Poliukhin \emph{et al.}~\cite{Poliukhin2026} found that HSE06
decreases the mobility, from 1553 to 1394~cm$^2$/V\,s. Their HSE06 band has a
longitudinal mass 28\% heavier than the PBE one. Ours is 7\% lighter, and the
HSE06 mass of Laurien and Rubel~\cite{Laurien2022} is also lighter than their
PBE one (Table~\ref{tab:si-hse}). The HSE06 matrix elements of Poliukhin
\emph{et al.}\ are the PBE ones scaled by a constant rescreening
factor $R=1.189$; they attribute the change of the mobility to $R$ and to the
band curvature. Figure~\ref{fig:si-hse}(c) shows that the change of the matrix
elements depends on $\mathbf q$: at the conduction-band minimum,
$\sum_\nu|g|^2$ changes by $-16\%$ to $+30\%$ along the path ($\mathbf q\neq0$), so that it
cannot be represented by a constant factor. Pant \emph{et al.}~\cite{Pant2026} obtained 1390~cm$^2$/V\,s with
$GW$ bands and $GW$ perturbation theory ($GW$PT) for the matrix elements, and found
the $GW$ corrections negligible for Si; their phonons are those of DFPT, so the
phonon hardening found here is absent, and $GW$ does not change $m_c$ of Si.

The PBE mobility lies 3.5\% and the HSE06 mobility 17\% above the experimental
value of $1350\pm100$~cm$^2$/V\,s~\cite{Ludwig1956}. Neither includes the
dynamical quadrupoles, which reduce the mobility of Si by
5--10\%~\cite{Park2020}, the electron--phonon renormalization of the
effective masses, which increases them by 3\% and reduces the mobility by
5\%~\cite{Ponce2018}, or two-phonon
scattering~\cite{Hatanpaa2023}, which reduces it further; the
closeness of the PBE value to experiment therefore does not indicate that PBE
is more accurate.

Measured as in Sec.~\ref{sec:nondiagonal-results}, the calculations of the
32 displaced structures, which give the phonons and the matrix elements of
the lowest 16 bands on the $12^3$ $\mathbf k$ and $6^3$ $\mathbf q$ meshes,
take 273~core-hours with PBE (SCF and NSCF) and $4.9\times10^4$~core-hours
with HSE06 (SCF only). The $GW$PT matrix elements of Si in
Ref.~\cite{Pant2026}, for $8^3$ $\mathbf k$ and $4^3$ $\mathbf q$ meshes and
the lowest two conduction bands, took $5.5\times10^5$~core-hours on different
hardware. Although $GW$ and HSE06 differ in the level of theory, the HSE06
matrix elements are thus obtained on 11 times as many
$(\mathbf k,\mathbf q)$ pairs at less than a tenth of the cost.

\begin{table}
\caption{\label{tab:si-hse}Electron drift mobility $\mu$ of Si at 300~K
(cm$^2$/V\,s, IBTE with SERTA in parentheses), longitudinal and transverse
effective masses $m_\parallel$ and $m_\perp$, and conductivity effective
mass $m_c$, in units of the free-electron mass $m_0$. Masses of this work are from VASP, $m_c$ from EPW;
Ref.~\cite{Laurien2022} gives VASP masses with spin--orbit coupling at the
experimental lattice constant, and the experimental masses compiled there.}
\begin{ruledtabular}
\begin{tabular}{llccc}
 & & $m_\parallel$ / $m_\perp$ & $m_c$ & $\mu$ \\
\hline
 This work & PBE & 0.958 / 0.195 & 0.292 & 1397 (1341) \\
  & HSE06 & 0.889 / 0.187 & 0.277 & 1581 (1497) \\
 Ref.~\cite{Poliukhin2026} & PBE & 0.859 / 0.202 & & 1553 \\
  & HSE06\footnotemark[1] & 1.099 / 0.189 & & 1394 \\
 Ref.~\cite{Pant2026} & $GW$\footnotemark[2] & & 0.29 & 1390 (1350) \\
 Ref.~\cite{Laurien2022} & PBE & 0.943 / 0.193 & & \\
  & HSE06 & 0.914 / 0.186 & & \\
 Expt. & & 0.916 / 0.191\footnotemark[3] & & $1350\pm100$\footnotemark[4]
\end{tabular}
\end{ruledtabular}
\footnotetext[1]{PBE matrix elements scaled by $R=1.189$.}
\footnotetext[2]{$GW$ bands, $GW$PT matrix elements, DFPT phonons; extrapolated to
infinite fine meshes and with dynamical quadrupoles.}
\footnotetext[3]{Ref.~\cite{Laurien2022}.}
\footnotetext[4]{Ref.~\cite{Ludwig1956}.}
\end{table}

\subsection{Application: CaH$_6$ at 150~GPa}
\label{sec:cah6}

\begin{figure}[tbp]
\includegraphics[width=\columnwidth]{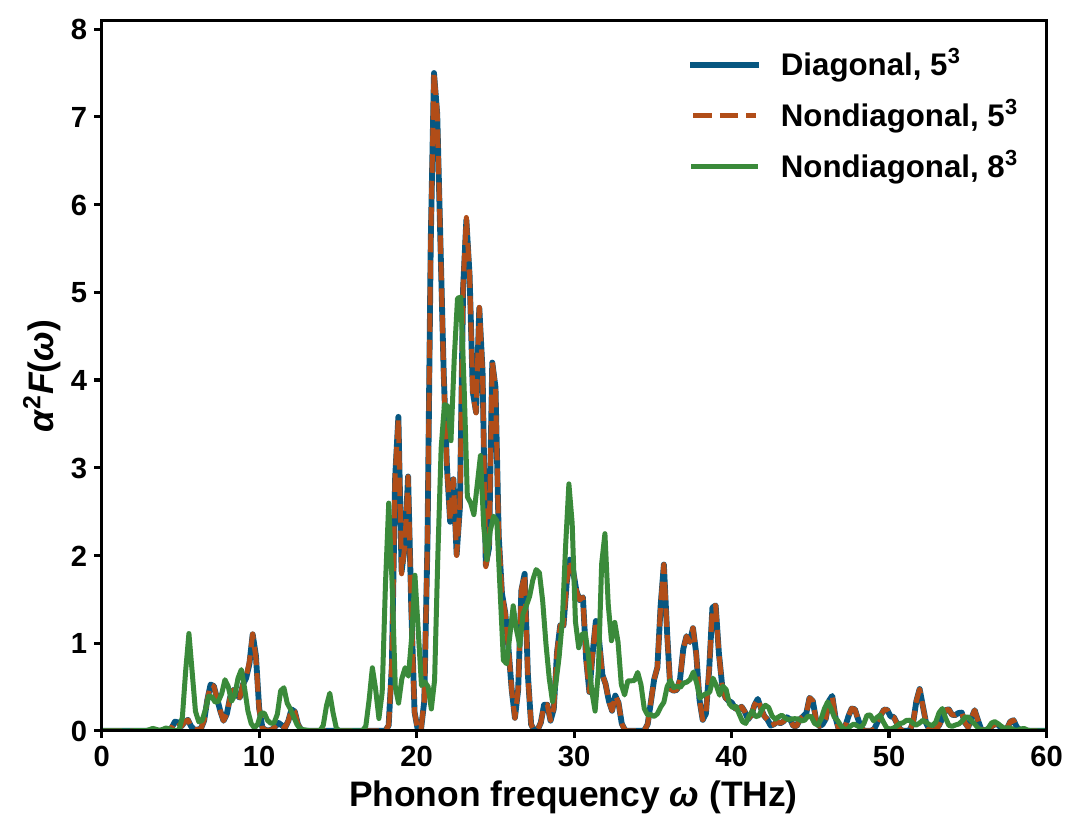}
\caption{\label{fig:cah6-spectrum}Eliashberg spectral function of CaH$_6$ at
150~GPa from ARES-EPC+VASP calculations with diagonal and nondiagonal
supercells on the $5^3$ $\mathbf q$ mesh and with nondiagonal supercells on
the $8^3$ $\mathbf q$ mesh.}
\end{figure}

CaH$_6$ is a hydride superconductor predicted and observed to have a high
$T_c$~\cite{Wang2012,Ma2022,Li2022}. We compute its Eliashberg spectral function
$\alpha^2F(\omega)$ and EPC constant $\lambda$~\cite{Allen1983} at 150~GPa
with the settings of Sec.~\ref{sec:computational}. On the $5^3$ $\mathbf q$ mesh, diagonal
and nondiagonal supercells give nearly coincident $\alpha^2F(\omega)$
[Fig.~\ref{fig:cah6-spectrum}], with $\lambda=3.245$ and 3.246, and both
give $T_c=266$~K.

\begin{table*}[tp]
\caption{\label{tab:cah6-tc}Critical temperature $T_c$ of CaH$_6$ from the
isotropic and anisotropic Eliashberg equations and from experiment, with the
$\mathbf q$ mesh, pressure $P$, Coulomb pseudopotential $\mu^*$, and EPC
constant $\lambda$.}
\begin{ruledtabular}
\begin{tabular}{llccccccc}
 & & & & & & \multicolumn{3}{c}{$T_c$ (K)} \\
\cline{7-9}
 & Method & $\mathbf q$ mesh & $P$ (GPa) & $\mu^*$ & $\lambda$ & Isotropic &
 Anisotropic & Expt. \\
\hline
 This work & Diagonal supercell & $5^3$ & 150 & 0.13 & 3.245 & 266 & & \\
  & Nondiagonal supercells & $5^3$ & 150 & 0.13 & 3.246 & 266 & & \\
  & Nondiagonal supercells & $8^3$ & 150 & 0.13 & 3.079 & 247 & & \\
 Ref.~\cite{Wang2012} & DFPT & $12^3$ & 150 & 0.13 (0.10) & 2.69 & 220 (235) & & \\
 Ref.~\cite{Jeon2022} & DFPT & $6^3$\footnotemark[1] & 150 & 0.13 & 2.75 & 238 &
 240 & \\
 Ref.~\cite{Darussalam2024} & DFPT & $8^3$ & 170 & \footnotemark[2] & 2.27 & &
 236 & \\
 Ref.~\cite{Ma2022} & Electrical resistance & & 172 & & & & & 215 \\
 Ref.~\cite{Li2022} & Electrical resistance & & 160 & & & & & 210
\end{tabular}
\end{ruledtabular}
\footnotetext[1]{Interpolated to a $24^3$ $\mathbf q$ mesh with EPW.}
\footnotetext[2]{No $\mu^*$: the gap equation contains the statically screened
Coulomb interaction from the random-phase approximation.}
\end{table*}

The nondiagonal route makes a finer $\mathbf q$ mesh accessible. An $8^3$
$\mathbf q$ mesh is covered by 18 nondiagonal supercells of 56 atoms each,
whereas the corresponding diagonal supercell would contain 3584 atoms. On
this mesh, the contribution of the modes between 20 and 25~THz to $\lambda$
decreases from 1.6 to 1.1, whereas that of the modes below 15~THz
increases from 0.48 to 0.69 [Fig.~\ref{fig:cah6-spectrum}]. The calculation gives
$\lambda=3.079$, a logarithmic average frequency $\omega_{\log}=79$~meV
compared with 86~meV on the $5^3$ mesh, and $T_c=247$~K.
Table~\ref{tab:cah6-tc} compares these results with earlier calculations and
with experiment. The $8^3$ mesh brings both $\lambda$ and $T_c$ closer to the
earlier calculations at 150~GPa, and denser $\mathbf k$ and $\mathbf q$
meshes are needed for converged results.

\section{Discussion}
\label{sec:discussion}

The generalized overlaps of Eq.~\eqref{eq:generalized-overlap} make the
spectral term the complete derivative of the PAW pseudo-Hamiltonian, so the
changes of the local potential and of the nonlocal PAW strengths need not be
differentiated separately. The analytic terms of
Eq.~\eqref{eq:paw-corrections} rely on the same approximations as the
finite-displacement PAW formulation of Chaput \emph{et al.}\
\cite{Chaput2019}, namely complete partial waves within non-overlapping
augmentation spheres and the neglect of the two Pulay-type terms of Sec.~\ref{sec:paw-ae}. The
comparison with VASP+phelel therefore tests the evaluation of $g^{\mathrm H}$
by eigenvalue projection against its evaluation from potential derivatives
within this common formulation, whereas the
comparison with DFPT tests it against an independent method.

In the nondiagonal route, the matrix elements of the individual supercells
are transformed to a common primitive-cell Bloch basis, completed by symmetry,
and averaged over the space group. The agreement with the diagonal route
shows that these steps preserve the matrix elements. The largest
individual difference occurs for a group with small matrix elements between
semicore states and likely originates from the diagonal calculation:
Eq.~\eqref{eq:cartesian-vertex} multiplies the central differences
$h^{(p)}_{\kappa\xi}$ by $N_p$, which is 125 for the diagonal supercell of
CaH$_6$ but 5 for the nondiagonal ones, so numerical noise is amplified 25
times more. It does not affect $\lambda$, which agrees between the two routes
(Sec.~\ref{sec:cah6}). In the end-to-end comparison for Si, the largest difference, for the LA
matrix elements along $\Gamma$--X, may originate from the SCF convergence of the
supercell calculations: in a test with a diagonal supercell, a similar
difference disappeared when the SCF threshold was tightened.

The cost of the diagonal route has two sources. The SCF calculations of the
large supercells dominate it and grow rapidly with the number of atoms. In
addition, the spectral representation of Eq.~\eqref{eq:spectral-reconstruction} requires the NSCF calculations to
retain the bands of Sec.~\ref{sec:computational}, whose number grows linearly
with $N_p$ because they include the reference states at all folded
momenta. Nondiagonal
supercells reduce $N_p$ and thereby both contributions. The nondiagonal route
requires more displaced calculations, but they are independent and smaller.
For hybrid functionals, which require the SCF mode with all these bands
(Sec.~\ref{sec:workflow}), nondiagonal supercells make phonon and
matrix-element calculations practical: for the $6^3$ $\mathbf q$ mesh of Si,
the HSE06 calculations of Sec.~\ref{sec:si-hse} use supercells of 12 atoms,
whereas the diagonal supercell would contain 432 atoms.

Because the displaced calculations supply only eigenvalues and pseudo wave
functions, the procedure used here for HSE06 applies in the same way to
DFT$+U$, meta-GGA functionals, and other methods that provide PAW
eigenstates. The present calculations have two restrictions that are not
inherent in the method. First, the long-range part of the matrix elements
decays slowly in the Wannier representation and is therefore subtracted
before the interpolation and restored afterwards. The interface to EPW does
this for the dipole part~\cite{Verdi2015}, with Born charges and the
dielectric tensor as input, but not yet for the quadrupole
part~\cite{Park2020}, which also affects Si (Sec.~\ref{sec:si-hse}).
Second, the calculations omit spin--orbit coupling. The formulation extends
to it and to noncollinear magnetism: the overlaps and projector coefficients become sums
over spinor components, the spin--orbit contribution to the PAW strengths is
contained in the displaced eigenvalues and eigenstates, and the symmetry
operations act on spinors. We leave these extensions to future work.

\section{Conclusions}
\label{sec:conclusions}

We have extended eigenvalue projection for electron--phonon matrix elements
to the PAW method. With generalized overlaps in the PAW metric of each
displaced structure and analytic terms constructed from the PAW dataset, the
all-electron matrix element is obtained from the eigenvalues and pseudo wave
functions of the displaced calculations alone, without derivatives of the
local potential or of the PAW strengths. We further combined the method with
nondiagonal supercells, whose SCF calculations sample the Brillouin zone
equivalently to the primitive cell, and recovered the full matrix elements by
crystal symmetry. Both are implemented in ARES-EPC, together with an
interface to EPW.

For seven materials, the NC matrix elements agree with QE DFPT to MAEs of
0.0078--0.28~meV, the PAW matrix elements agree with VASP+phelel to MAEs of
0.0052--0.40~meV, and nondiagonal supercells reproduce the diagonal-supercell
results to MAEs of at most 0.072~meV, while reducing the aggregate DFT cost
of the displaced calculations by factors of 1.7--23. For CaH$_6$, diagonal
and nondiagonal supercells give the same $\lambda$ and $T_c$, and the
nondiagonal route makes a $\mathbf q$ mesh accessible that would otherwise
require a diagonal supercell of 3584 atoms.

For Si, the electron mobility obtained with EPW from the NC matrix elements
of ARES-EPC agrees with that from QE DFPT to 0.9\%. The HSE06 matrix
elements are obtained from nondiagonal supercells of 12 atoms, whereas the
diagonal supercell for the same $6^3$ $\mathbf q$ mesh would contain 432
atoms. HSE06 increases the mobility by 13\%, because the
lighter conduction band and the harder phonons outweigh the stronger
coupling. Other methods available in
PAW, such as DFT$+U$ and meta-GGA functionals, can be treated in the same
way.

\makeatletter
\let\si@hangfrom@section\@hangfrom@section
\let\si@sectioncntformat\@sectioncntformat
\makeatother

\appendix
\section{Generalized overlaps and completeness}
\label{app:plane-wave-overlaps}

\subsection{Folding and ordinary overlaps}

Let $\Omega$ and $\Omega_p=N_p\Omega$ be the primitive- and supercell
volumes. Reference and displaced pseudo states are expanded in
cell-normalized plane waves,
\begin{equation}\label{eq:app-plane-waves}
\begin{aligned}
 \tilde\psi_{n\mathbf k}(\mathbf r)
 &=\Omega^{-1/2}\sum_{\mathbf G}c_{n\mathbf k}(\mathbf G)
   e^{i\mathbf B(\mathbf k+\mathbf G)\cdot\mathbf r},\\
 \tilde\Psi^x_j(\mathbf r)
 &=\Omega_p^{-1/2}\sum_{\boldsymbol\Gamma}c^x_j(\boldsymbol\Gamma)
   e^{i\mathbf B_p(\mathbf K_p+\boldsymbol\Gamma)\cdot\mathbf r},
\end{aligned}
\end{equation}
where $\mathbf G$ and $\boldsymbol\Gamma$ are primitive and supercell
reciprocal-lattice vectors in reduced coordinates. With
$\boldsymbol\Gamma_p(\mathbf G)=\mathbf M_p^T\mathbf G+\mathbf c_p$,
Eq.~\eqref{eq:k-folding} gives
$\mathbf B_p[\mathbf K_p+\boldsymbol\Gamma_p(\mathbf G)]
=\mathbf B(\mathbf k+\mathbf G)$. The supercell-normalized reference state
$\tilde\psi^{(p)}_{n\mathbf k}=N_p^{-1/2}\tilde\psi_{n\mathbf k}$ therefore has
the coefficient $c_{n\mathbf k}(\mathbf G)$ at $\boldsymbol\Gamma_p(\mathbf G)$
and zero elsewhere; the factor $N_p^{-1/2}$ is absorbed by
$\Omega_p^{-1/2}=N_p^{-1/2}\Omega^{-1/2}$. The ordinary overlap is
\begin{equation}\label{eq:app-pw-overlap}
 U^{\mathrm{PW}}_x(j,n)
 =\sum_{\mathbf G}\bigl[c^x_j(\boldsymbol\Gamma_p(\mathbf G))\bigr]^*
 c_{n\mathbf k}(\mathbf G),
\end{equation}
with the sum restricted to $\mathbf G$ whose folded vectors $\boldsymbol\Gamma_p(\mathbf G)$ lie in the plane-wave set
of the displaced supercell; for equal cutoffs, this includes all $\mathbf G$.
The final-state overlap $U'^{\mathrm{PW}}_x$ follows with $\mathbf k'$ and
$\mathbf c'_p$.

\subsection{Projector overlaps}

For a channel $\mu=(\iota,\ell,m_\ell)$ with radial index $\iota$, let
$\mathcal P_{\kappa\iota\ell}(|\mathbf v|)$ be the reciprocal-space radial
projector of the PAW dataset, evaluated by cubic interpolation. With the
Cartesian momentum $\mathbf v=\mathbf B_p(\mathbf K_p+\boldsymbol\Gamma)$, the
position-independent projector factor is
\begin{equation}\label{eq:app-projector-fourier-factor}
 F^{(p)}_{\kappa\mu}(\mathbf v)
 =\Omega_p^{-1/2}\,i^\ell\,
 Y_{\ell m_\ell}(\hat{\mathbf v})\,\mathcal P_{\kappa\iota\ell}(|\mathbf v|),
\end{equation}
where $Y_{\ell m_\ell}$ are orthonormal real spherical harmonics,
\begin{equation*}
\begin{aligned}
 Y_{\ell m_\ell}(\theta,\varphi)
 &=\left[\frac{(2-\delta_{m_\ell0})(2\ell+1)}{4\pi}
   \frac{(\ell-|m_\ell|)!}{(\ell+|m_\ell|)!}\right]^{1/2}\\
 &\quad\times P_\ell^{|m_\ell|}(\cos\theta)\,f_{m_\ell}(\varphi),
\end{aligned}
\end{equation*}
with $f_{m_\ell}=\sin(|m_\ell|\varphi)$, $1$, and $\cos(m_\ell\varphi)$ for
$m_\ell<0$, $m_\ell=0$, and $m_\ell>0$, and associated Legendre functions
$P_\ell^{|m_\ell|}$ without the Condon--Shortley phase. The PAW channel
matrices $Q_\kappa$ and $R_{\kappa\alpha}$ use the same basis.

For a projector centered at the reduced supercell position
$\mathbf t^x_{l\kappa}$ and any state $w$ at $\mathbf K_p$ with coefficients
$c_w(\boldsymbol\Gamma)$, we write the projector coefficient with its
atom-centered Bloch phase factored out,
\begin{equation}\label{eq:app-projector-coefficient}
\begin{aligned}
 \bar\Pi^x_{l\kappa\mu}(w)
 &\equiv e^{-2\pi i\mathbf K_p\cdot\mathbf t^x_{l\kappa}}
   \langle\tilde p^x_{l\kappa\mu}|w\rangle\\
 &=\sum_{\boldsymbol\Gamma}
   F^{(p)}_{\kappa\mu}(\mathbf v)\,
   e^{2\pi i\boldsymbol\Gamma\cdot\mathbf t^x_{l\kappa}}\,
   c_w(\boldsymbol\Gamma).
\end{aligned}
\end{equation}
Each coefficient uses the full plane-wave set of its own state, and both
coefficients in an overlap are evaluated at the displaced positions. Because
both states have momentum $\mathbf K_p$, the phases cancel, and the generalized
overlap of Eq.~\eqref{eq:generalized-overlap} is
\begin{equation}\label{eq:app-paw-overlap}
\begin{aligned}
 U_x(j,n)&=U^{\mathrm{PW}}_x(j,n)\\
 &\quad+\sum_{l\kappa\mu\eta}
 \bigl[\bar\Pi^x_{l\kappa\mu}(\tilde\Psi^x_j)\bigr]^*
 (Q_\kappa)_{\mu\eta}\,
 \bar\Pi^x_{l\kappa\eta}(\tilde\psi^{(p)}_{n\mathbf k}).
\end{aligned}
\end{equation}

\subsection{Completeness of the displaced states}
\label{app:completeness}

Multiplying the generalized completeness relation by $\tilde S_x^{-1}$ gives
$\sum_j|\tilde\Psi^x_j\rangle\langle\tilde\Psi^x_j|=\tilde S_x^{-1}$, where
$\tilde S_x^{-1}$ is understood within the plane-wave space of the displaced
supercell. For reference states that fold to $\mathbf K_p$,
\begin{equation}\label{eq:app-completeness-proof}
\begin{aligned}
 {}[U_x'^{\dagger}U_x]_{mn}
 &=\langle\tilde\psi^{(p)}_{m\mathbf k'}|\tilde S_x
   \Bigl(\sum_j|\tilde\Psi^x_j\rangle\langle\tilde\Psi^x_j|\Bigr)
   \tilde S_x|\tilde\psi^{(p)}_{n\mathbf k}\rangle\\
 &=\langle\tilde\psi^{(p)}_{m\mathbf k'}|\tilde S_x|
   \tilde\psi^{(p)}_{n\mathbf k}\rangle
 =[\Sigma_x]_{mn},
\end{aligned}
\end{equation}
for both $\mathbf k'=\mathbf k$ and $\mathbf k'\neq\mathbf k$, provided that
the restriction in Eq.~\eqref{eq:app-pw-overlap} removes no plane waves. $\Sigma_x$ is evaluated independently as
\begin{equation}\label{eq:app-completeness-target}
\begin{aligned}
 {}[\Sigma_x]_{mn}
 &=\langle\tilde\psi^{(p)}_{m\mathbf k'}|\tilde\psi^{(p)}_{n\mathbf k}\rangle\\
 &\quad+\sum_{l\kappa\mu\eta}
 \bigl[\bar\Pi^x_{l\kappa\mu}(\tilde\psi^{(p)}_{m\mathbf k'})\bigr]^*
 (Q_\kappa)_{\mu\eta}\,
 \bar\Pi^x_{l\kappa\eta}(\tilde\psi^{(p)}_{n\mathbf k}),
\end{aligned}
\end{equation}
and the truncation error entering the projectability residual is
$U_{x,N_s}'^{\dagger}U_{x,N_s}-\Sigma_x=-\sum_{j>N_s}[U_x'(j,\cdot)]^\dagger U_x(j,\cdot)$.
In NC, $\Sigma_x$ reduces to
$\delta_{\mathbf k\mathbf k'}\delta_{mn}$.

\section{Plane-wave evaluation of the analytic PAW terms}
\label{app:paw-corrections}

The coefficients of Eq.~\eqref{eq:projector-coefficients} carry the atomic
Bloch phase $e^{2\pi i\mathbf k\cdot\boldsymbol\tau_\kappa}$; we write
$C_\kappa=e^{2\pi i\mathbf k\cdot\boldsymbol\tau_\kappa}\bar C_\kappa$ and
$D_{\kappa\alpha}=e^{2\pi i\mathbf k\cdot\boldsymbol\tau_\kappa}\bar D_{\kappa\alpha}$.
Let $F_{\kappa\mu}$ denote Eq.~\eqref{eq:app-projector-fourier-factor} with
$\Omega_p$ and $\mathbf B_p$ replaced by $\Omega$ and $\mathbf B$. Then
\begin{equation}\label{eq:app-primitive-projector}
 \bar C_\kappa(\mu,n)
 =\sum_{\mathbf G}F_{\kappa\mu}\bigl(\mathbf B(\mathbf k+\mathbf G)\bigr)\,
 e^{2\pi i\mathbf G\cdot\boldsymbol\tau_\kappa}\,c_{n\mathbf k}(\mathbf G).
\end{equation}
A rigid translation of the projector changes only the phase
$e^{i\mathbf B(\mathbf k+\mathbf G)\cdot\mathbf r_\kappa}$ of each plane-wave
component, where $\mathbf r_\kappa$ is the Cartesian atomic position, so
\begin{equation}\label{eq:app-primitive-projector-derivative}
\begin{aligned}
 \bar D_{\kappa\alpha}(\mu,n)
 &=\sum_{\mathbf G}i[\mathbf B(\mathbf k+\mathbf G)]_\alpha\,
 F_{\kappa\mu}\bigl(\mathbf B(\mathbf k+\mathbf G)\bigr)\\
 &\quad\times e^{2\pi i\mathbf G\cdot\boldsymbol\tau_\kappa}\,
 c_{n\mathbf k}(\mathbf G),
\end{aligned}
\end{equation}
which retains the $\mathbf k$ contribution to the derivative. The final-state
coefficients follow with $\mathbf k'$. The products
$P_{\kappa\alpha}$, $\bar P_{\kappa\alpha}$, and $Z_{\kappa\alpha}$ of
Eq.~\eqref{eq:x-decomposition} are formed from $\bar C$ and $\bar D$ and
multiplied by $e^{-2\pi i(\mathbf k'-\mathbf k)\cdot\boldsymbol\tau_\kappa}$.
The reduced difference $\mathbf k'-\mathbf k$ differs from $\mathbf q$ by a
reciprocal-lattice vector, which changes this phase when
$\boldsymbol\tau_\kappa\neq0$; the phase must therefore be evaluated with
$\mathbf k'-\mathbf k$ rather than $\mathbf q$.

\section{Symmetry transformation of the matrix elements}
\label{app:symmetry}

Consider a complex displacement $u^s_b$, $b=(\kappa,\alpha)$, at
$\mathbf q_s$, which induces
$\delta H^s=\sum_bu^s_b\,\partial_{b,\mathbf q_s}H^{\mathrm{AE}}$, whose
matrix between the states at $\mathbf k'_s$ and $\mathbf k_s$ is
$\sum_bu^s_b\,g^s_b$. The symmetry-related displacement at $\mathbf q_t$ is
$\mathbf u^t=\mathcal D\,(\mathbf u^s)^{(\sigma)}$ with $\mathcal D$ of
Eq.~\eqref{eq:displacement-representation}, and it induces
$\delta H^t=\hat{\mathcal O}\,\delta H^s\,\hat{\mathcal O}^{-1}$. Expanding the
transformed source states in the target bands with the sewing matrices,
\begin{equation}\label{eq:app-sewing}
 W'^{\dagger}\Bigl(\sum_a u^t_a\,g^t_a\Bigr)W
 =\Bigl(\sum_b u^s_b\,g^s_b\Bigr)^{(\sigma)},
\end{equation}
where the complex conjugation for $\sigma=-1$ follows from the antiunitarity
of $\hat{\mathcal O}$. Since $W$ and $W'$ are unitary within closed
electronic subspaces,
$\sum_a u^t_ag^t_a=W'\bigl(\sum_bu^s_bg^s_b\bigr)^{(\sigma)}W^\dagger$.
Inserting $\mathbf u^t=\mathcal D(\mathbf u^s)^{(\sigma)}$ and comparing the
coefficients of $(u^s_b)^{(\sigma)}$ gives
$\sum_a\mathcal D_{ab}\,g^t_a=W'(g^s_b)^{(\sigma)}W^\dagger$. Because
$\mathcal D$ is unitary, $(\mathcal D^T)^{-1}=\mathcal D^*$, which yields
Eq.~\eqref{eq:vertex-recovery}.

\vspace*{\fill}

\bibliography{references}

\clearpage
\onecolumngrid
\makeatletter
\let\@hangfrom@section\si@hangfrom@section
\let\@sectioncntformat\si@sectioncntformat
\makeatother
\setcounter{section}{0}
\setcounter{equation}{0}
\setcounter{figure}{0}
\setcounter{table}{0}
\renewcommand{\thesection}{S\arabic{section}}
\renewcommand{\theequation}{S\arabic{equation}}
\renewcommand{\thefigure}{S\arabic{figure}}
\renewcommand{\thetable}{S\arabic{table}}

\begin{center}
\textbf{\large Supplemental Material for ``Electron--phonon matrix elements
for hybrid functionals in the projector augmented-wave method: Eigenvalue
projection, nondiagonal supercells, and Wannier interpolation''}
\end{center}
\vspace{1em}

\section{Structures and pseudopotentials}

Table~\ref{tab:si-structures} lists the primitive structures. Within each
comparison of the main text, both calculation routes use the listed structure
and atom order. Table~\ref{tab:si-pseudopotentials} lists the
pseudopotentials and the number of valence electrons.

\begin{table}[ht]
\caption{\label{tab:si-structures}Primitive structures. $a$ and $c$ are the
conventional lattice parameters; $\Omega$ is the primitive-cell volume.
Reduced atomic positions are given with respect to the primitive lattice
vectors used in the calculations. The Si mobility calculations use the lattice
constants given in Sec.~\ref{sec:computational} of the main text.}
\begin{ruledtabular}
\begin{tabular}{llcccl}
 Material & Space group & $a$ (\AA) & $c$ (\AA) & $\Omega$ (\AA$^3$) &
 Atomic positions \\
\hline
 Al & $Fm\bar3m$ & 4.0389 & & 16.472 & Al $(0,0,0)$ \\
 CaH$_6$ & $Im\bar3m$ & 3.5075 & & 21.575 &
 Ca $(0,0,0)$; H $(\frac12,\frac34,\frac14)$ and equivalents \\
 Diamond & $Fd\bar3m$ & 3.5607 & & 11.287 &
 C $(\frac12,\frac12,\frac12)$, $(\frac34,\frac34,\frac34)$ \\
 GaAs & $F\bar43m$ & 5.7502 & & 47.532 &
 Ga $(0,0,0)$; As $(\frac14,\frac14,\frac14)$ \\
 MgB$_2$ & $P6/mmm$ & 3.0746 & 3.5269 & 28.873 &
 Mg $(0,0,0)$; B $(\frac23,\frac13,\frac12)$, $(\frac13,\frac23,\frac12)$ \\
 Pb & $Fm\bar3m$ & 4.9895 & & 31.054 & Pb $(0,0,0)$ \\
 Si & $Fd\bar3m$ & 5.4305 & & 40.037 &
 Si $(\frac14,\frac14,\frac14)$, $(0,0,0)$
\end{tabular}
\end{ruledtabular}
\end{table}

\begin{table}[ht]
\caption{\label{tab:si-pseudopotentials}Pseudopotentials. NC: PseudoDojo
v0.5 scalar-relativistic PBE pseudopotentials (standard accuracy)
\cite{VanSetten2018}. PAW: VASP PBE datasets (potpaw\_PBE.64).
$Z_v$ is the number of valence electrons.}
\begin{ruledtabular}
\begin{tabular}{lccc}
 Element & NC $Z_v$ & PAW dataset & PAW $Z_v$ \\
\hline
 Al & 3 & Al & 3 \\
 Ca & 10 & Ca\_sv & 10 \\
 H & 1 & H & 1 \\
 C & 4 & C & 4 \\
 Ga & 13 & Ga\_d & 13 \\
 As & 15 & As & 5 \\
 Mg & 10 & Mg & 2 \\
 B & 3 & B & 3 \\
 Pb & 14 & Pb\_d & 14 \\
 Si & 4 & Si & 4
\end{tabular}
\end{ruledtabular}
\end{table}

\section{Calculation settings}

Table~\ref{tab:si-settings} lists the occupation and smearing settings and
the compared bands. The diagonal benchmarks use the meshes
$\mathbf n_{\mathrm{ref}}$ of Table~\ref{tab:diagonal-benchmarks} of the main text; the
diagonal--nondiagonal comparison uses $\mathbf n_{\mathrm{pair}}$ of
Table~\ref{tab:nondiagonal-benchmarks}, with the primitive-cell SCF meshes $\mathbf n_{\mathrm{SCF}}$ given
here. The CaH$_6$ spectral-function calculations use primitive-cell SCF
meshes of $20^3$ for the $5^3$ and $24^3$ for the $8^3$ $\mathbf q$ mesh;
within $\pm0.05$~eV of the selected Gaussian widths, $\lambda$
changes by less than 0.6\%. For the Si mobility, the largest norm of the part
of $|\tilde f_{m\mathbf k}\rangle$ of Eq.~\eqref{eq:gauge-fit} of the main text
outside the reference states is $2.6\times10^{-6}$ for PBE and
$2.9\times10^{-5}$ for HSE06.

\begin{table}[ht]
\caption{\label{tab:si-settings}Smearing and comparison settings. QE uses
Methfessel--Paxton (MP)~\cite{Methfessel1989} or Gaussian (G) smearing with width $\sigma$ in Ry;
VASP uses first-order MP (ISMEAR = 1) or Gaussian (ISMEAR = 0) smearing with
$\sigma$ in eV. Bands are counted from the lowest valence band.
$\mathbf n_{\mathrm{SCF}}$ is the primitive-cell SCF mesh of the
diagonal--nondiagonal comparison.}
\begin{ruledtabular}
\begin{tabular}{lccccc}
 Material & QE smearing ($\sigma$, Ry) & VASP smearing ($\sigma$, eV) &
 Bands (NC) & Bands (PAW) & $\mathbf n_{\mathrm{SCF}}$ \\
\hline
 Al & MP, 0.02 & MP, 0.2 & 1--4 & 1--4 & $24^3$ \\
 CaH$_6$ & MP, 0.02 & MP, 0.2 & 1--10 & 1--10 & $20^3$ \\
 Diamond & fixed occupations & G, 0.05 & 1--4 & 1--4 & $18^3$ \\
 GaAs & G, 0.001 & G, 0.05 & 1--10 & 1--10 & $15^3$ \\
 MgB$_2$ & MP, 0.02 & MP, 0.2 & 1--10 & 1--6 & $24^3$ \\
 Pb & MP, 0.0147 & MP, 0.2 & 1--10 & 1--10 & $24^3$ \\
 Si & fixed occupations & G, 0.05 & 1--4 & 1--4 & $15^3$
\end{tabular}
\end{ruledtabular}
\end{table}

\section{Si mobility: mesh of the Hamiltonian}

Table~\ref{tab:si-hgrid} shows the effect of the electronic mesh on which
$H$ is constructed, with the matrix elements on the $12^3\times6^3$ meshes in
both cases. The HSE06 calculation on the $12^3$ mesh uses a frozen window
extending 2.6~eV above the conduction-band minimum. On the $12^3$ mesh, the
longitudinal mass is 22\% (PBE) and
16\% (HSE06) lighter than the VASP one, whereas on the $24^3$ mesh the masses
agree with VASP to about 1\%. Compared with the $24^3$ mesh, $m_c$ on the
$12^3$ mesh is about 3\% lighter with both functionals, and the mobilities
are 7\% (PBE) and 17\% (HSE06) higher.

\begin{table}[ht]
\caption{\label{tab:si-hgrid}Longitudinal and transverse effective masses
$m_\parallel$ and $m_\perp$ at the conduction-band minimum, conductivity
effective mass $m_c$ ($m_0$), and IBTE electron mobility $\mu$ at 300~K
(cm$^2$/V\,s) of Si for $H$ on the $12^3$ and $24^3$ meshes, with the VASP
masses of Table~\ref{tab:si-hse} of the main text.}
\begin{ruledtabular}
\begin{tabular}{llccc}
 & Mesh of $H$ & $m_\parallel$ / $m_\perp$ & $m_c$ & $\mu$ \\
\hline
 PBE & $12^3$ & 0.749 / 0.188 & 0.284 & 1500 \\
  & $24^3$ & 0.967 / 0.194 & 0.292 & 1397 \\
  & VASP & 0.958 / 0.195 & & \\
\hline
 HSE06 & $12^3$ & 0.743 / 0.181 & 0.268 & 1850 \\
  & $24^3$ & 0.887 / 0.189 & 0.277 & 1581 \\
  & VASP & 0.889 / 0.187 & &
\end{tabular}
\end{ruledtabular}
\end{table}

\section{Decomposition of the SERTA mobility of Si}

Equation~\eqref{eq:serta-factor} of the main text holds for any occupation.
The relaxation time is given by~\cite{Pant2026}
\begin{equation}
\begin{aligned}
\frac1{\tau_{n\mathbf k}}=\frac{2\pi}{\hbar}\sum_{m\nu\mathbf q}
|g_{mn\nu}(\mathbf k,\mathbf q)|^2
\big[&(n_{\mathbf q\nu}+1-f')\,\delta(\epsilon-\epsilon'-\hbar\omega_{\mathbf q\nu})\\
+&(n_{\mathbf q\nu}+f')\,\delta(\epsilon-\epsilon'+\hbar\omega_{\mathbf q\nu})\big],
\end{aligned}
\label{eq:si-serta-rate}
\end{equation}
with $\epsilon=\epsilon_{n\mathbf k}$, $\epsilon'=\epsilon_{m\mathbf k+\mathbf q}$,
$f'=f_{m\mathbf k+\mathbf q}$, and the Bose--Einstein occupation
$n_{\mathbf q\nu}$. Throughout this section, sums over $\mathbf k$ and
$\mathbf q$ are normalized to the Brillouin zone, and energies are measured
from the conduction-band minimum.

\emph{Average coupling and density of final states.} Following
Ref.~\cite{Pant2026}, we write
\begin{equation}
\frac1{\bar\tau}=\frac{2\pi}{\hbar}\,\bar g^2\bar\rho,\qquad
\mu=\frac{e}{m_c}\Big(\frac{2\pi}{\hbar}\,\bar g^2\bar\rho\Big)^{-1},
\label{eq:si-gbar}
\end{equation}
where $\bar\rho$ is an average density of final states and $\bar g^2$ is
defined by Eq.~\eqref{eq:si-gbar}, so that the second relation holds exactly
for any choice of $\bar\rho$. As in Ref.~\cite{Pant2026}, we take
\begin{equation}
\rho_{n\mathbf k}=\sum_{m\mathbf q\nu}\Big[
(n_{\mathbf q\nu}+f_{m\mathbf k+\mathbf q})\,\delta(\epsilon_{n\mathbf k}-\epsilon_{m\mathbf k+\mathbf q}+\hbar\omega_{\mathbf q\nu})
+(n_{\mathbf q\nu}+1-f_{m\mathbf k+\mathbf q})\,\delta(\epsilon_{n\mathbf k}-\epsilon_{m\mathbf k+\mathbf q}-\hbar\omega_{\mathbf q\nu})\Big],
\label{eq:si-rho}
\end{equation}
the scattering rate of Eq.~\eqref{eq:si-serta-rate} with
$(2\pi/\hbar)|g_{mn\nu}|^2$ replaced by unity, and average it as
\begin{equation}
\frac1{\bar\rho}=\frac{\sum_{n\mathbf k}(-\partial f/\partial\epsilon)_{n\mathbf k}
|\mathbf v_{n\mathbf k}|^2\rho_{n\mathbf k}^{-1}}
{\sum_{n\mathbf k}(-\partial f/\partial\epsilon)_{n\mathbf k}|\mathbf v_{n\mathbf k}|^2}.
\label{eq:si-rhobar}
\end{equation}
$\bar g^2$ then contains only the matrix elements. The $\delta$ functions are
replaced by the adaptive Gaussian of the EPW calculation. The HSE06/PBE
ratio of $\bar\rho$ is 0.837 and that of $\bar g^2$ is 1.130
(Table~\ref{tab:si-decomposition}). Evaluating
Eq.~\eqref{eq:si-rho} with the HSE06 bands and the PBE phonons gives 0.922 times
the PBE value of $\bar\rho$, and then replacing also the phonons gives a further
factor of 0.908; in the reverse order, the factors are 0.907 and 0.923. The
band factor reflects the lower density of states of the lighter HSE06 band,
whose partition function
$Z=2\sum_{n\mathbf k}e^{-\epsilon_{n\mathbf k}/k_BT}$ over the conduction
states, with the factor 2 for spin, has the ratio 0.906.

\begin{table}[ht]
\caption{\label{tab:si-decomposition}Factors of the SERTA electron mobility
of Si at 300~K from Eq.~\eqref{eq:serta-factor} of the main text and
Eq.~\eqref{eq:si-gbar}, with the velocities and relaxation times of EPW and
$\bar\rho$ from the adaptive Gaussian of the EPW calculation. $Z$ is given
per unit cell.}
\begin{ruledtabular}
\begin{tabular}{lcc}
 Factor & PBE / HSE06 & HSE06/PBE \\
\hline
 $m_c$ ($m_0$) & 0.2923 / 0.2767 & 0.947 \\
 $\bar\tau$ (ps) & 0.2229 / 0.2356 & 1.057 \\
 $\bar\rho$ (eV$^{-1}$) & 0.600 / 0.502 & 0.837 \\
 $\bar g^2$ (meV$^2$) & 784 / 885 & 1.130 \\
 $Z$ ($10^{-3}$) & 1.238 / 1.122 & 0.906 \\
 $\mu_{\mathrm{SERTA}}$ (cm$^2$/V\,s) & 1341 / 1497 & 1.116
\end{tabular}
\end{ruledtabular}
\end{table}

\section{Group-norm comparisons for all materials}

Figures~\ref{fig:si-nc}--\ref{fig:si-nd} show the group-norm comparisons for
all seven materials: NC versus QE DFPT, PAW versus VASP+phelel, and
nondiagonal versus diagonal PAW supercells. Each point is one group of
electronic bands and phonon modes [Eq.~\eqref{eq:degenerate-norm} of the main text]; the dashed
line marks equality, and the color gives the absolute difference.

\newcommand{\sipanel}[2]{%
 \begin{minipage}[t]{0.32\textwidth}\centering
 \textbf{#2}\\[-1pt]\includegraphics[width=\linewidth]{#1}\end{minipage}}

\begin{figure}[ht]
\sipanel{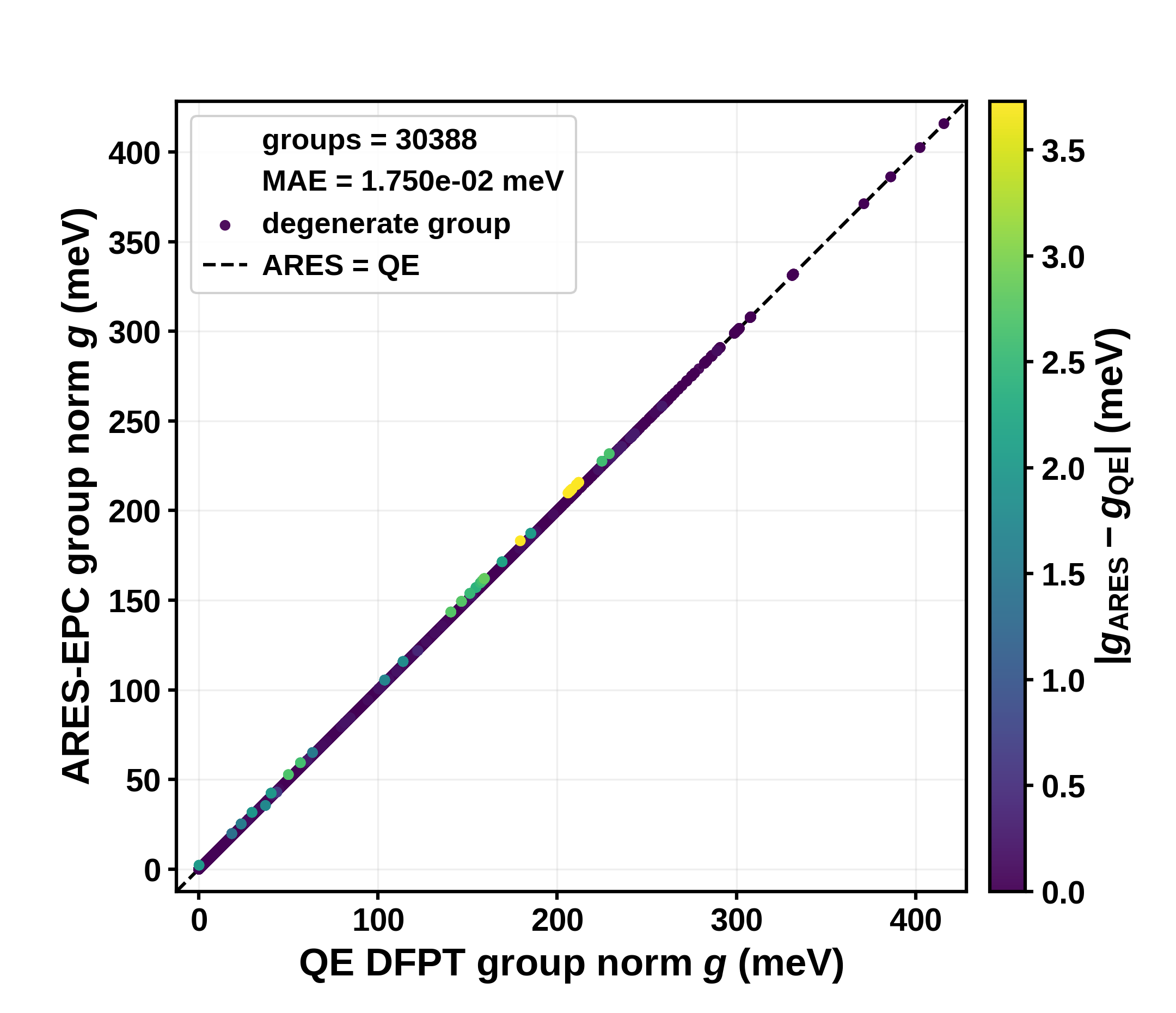}{(a) Al}\hfill
\sipanel{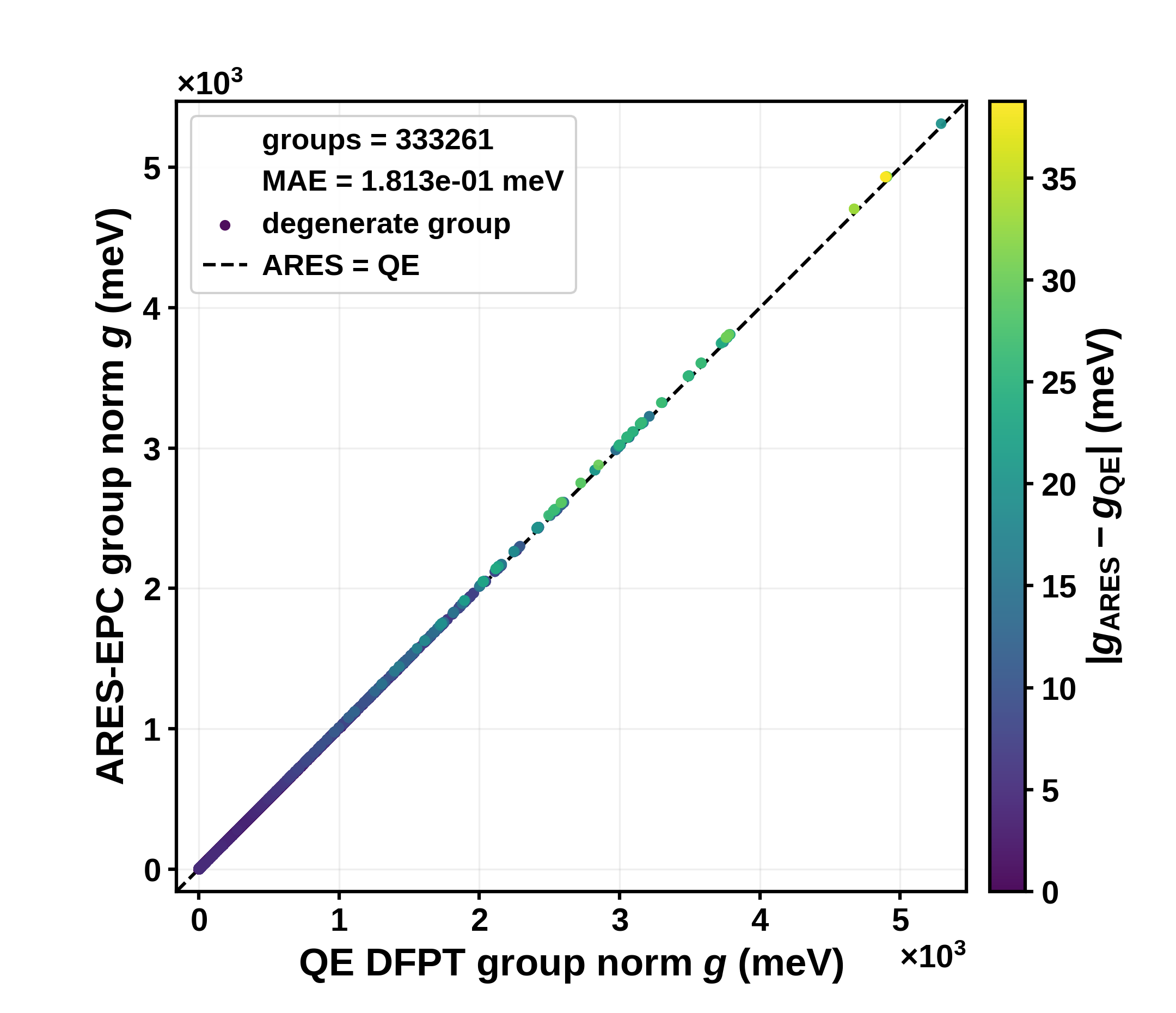}{(b) CaH$_6$}\hfill
\sipanel{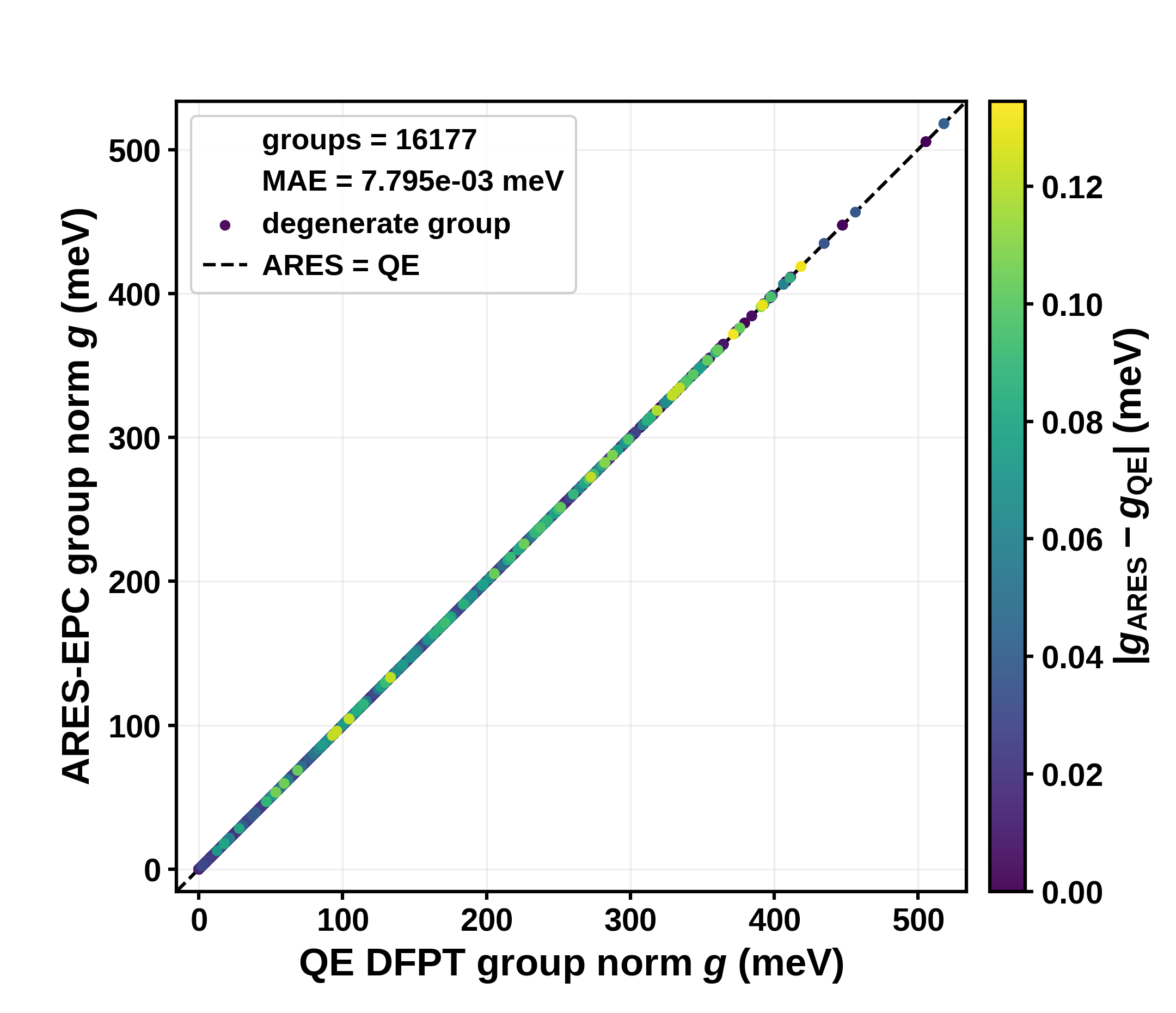}{(c) Diamond}\\[4pt]
\sipanel{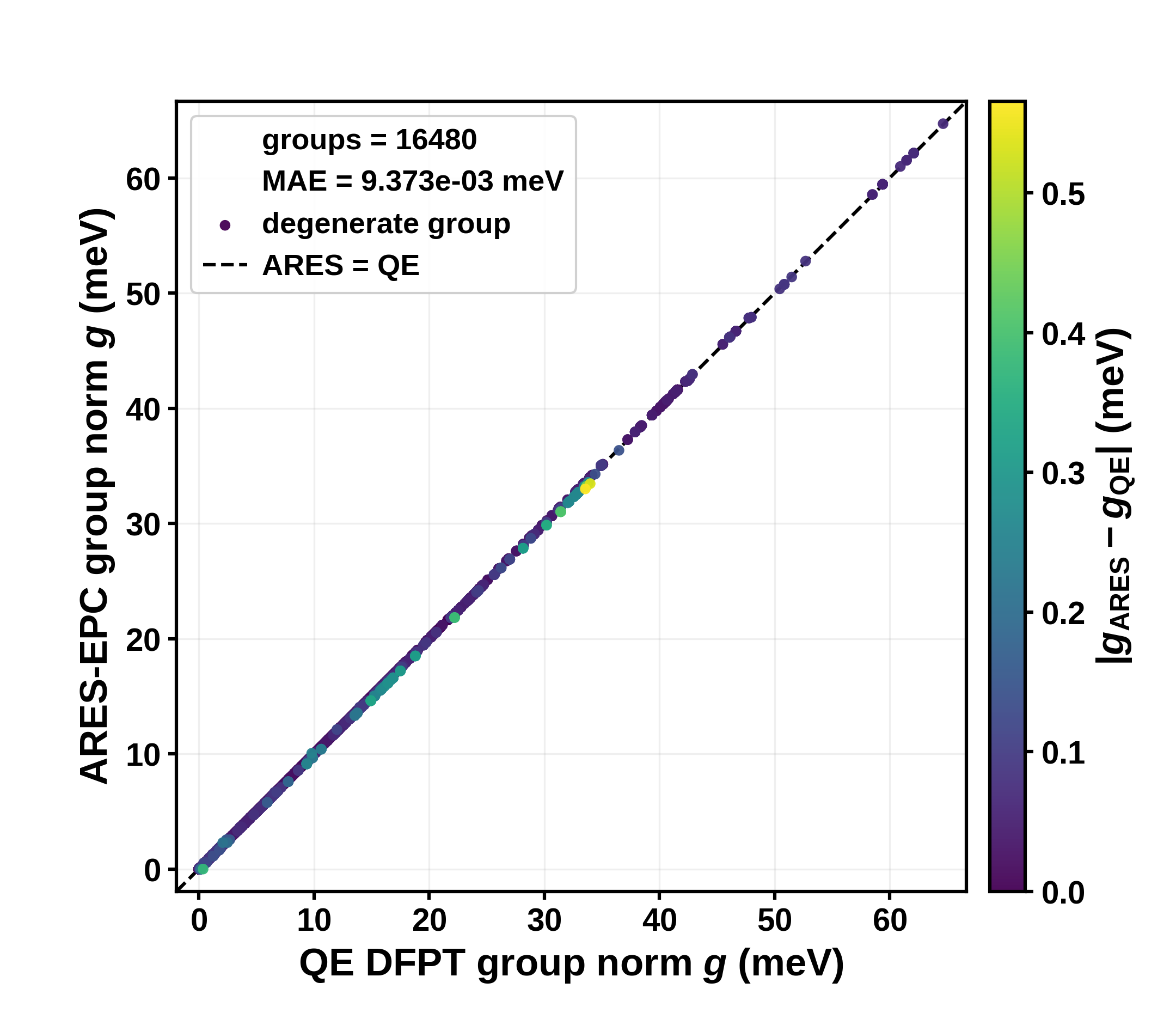}{(d) GaAs}\hfill
\sipanel{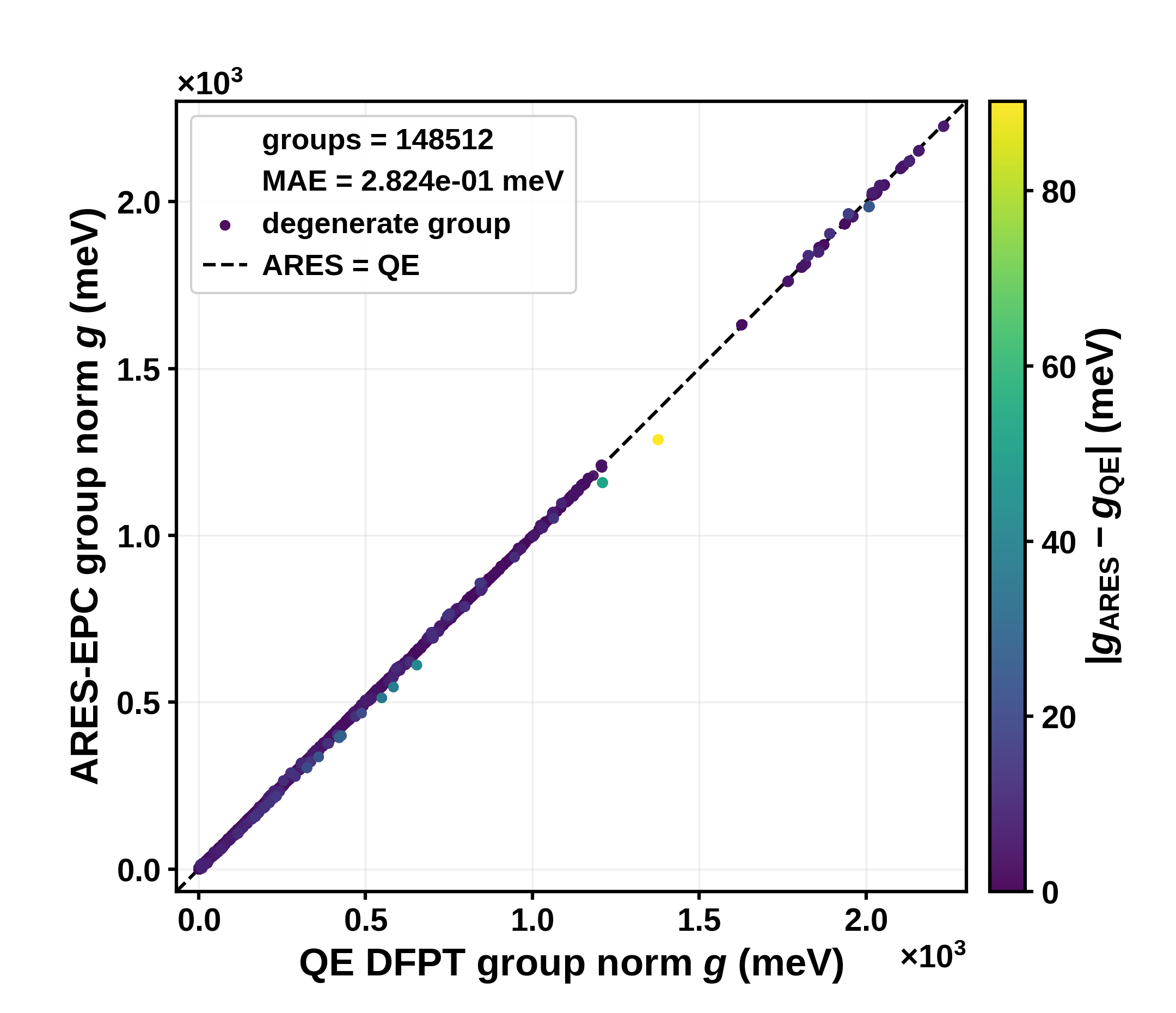}{(e) MgB$_2$}\hfill
\sipanel{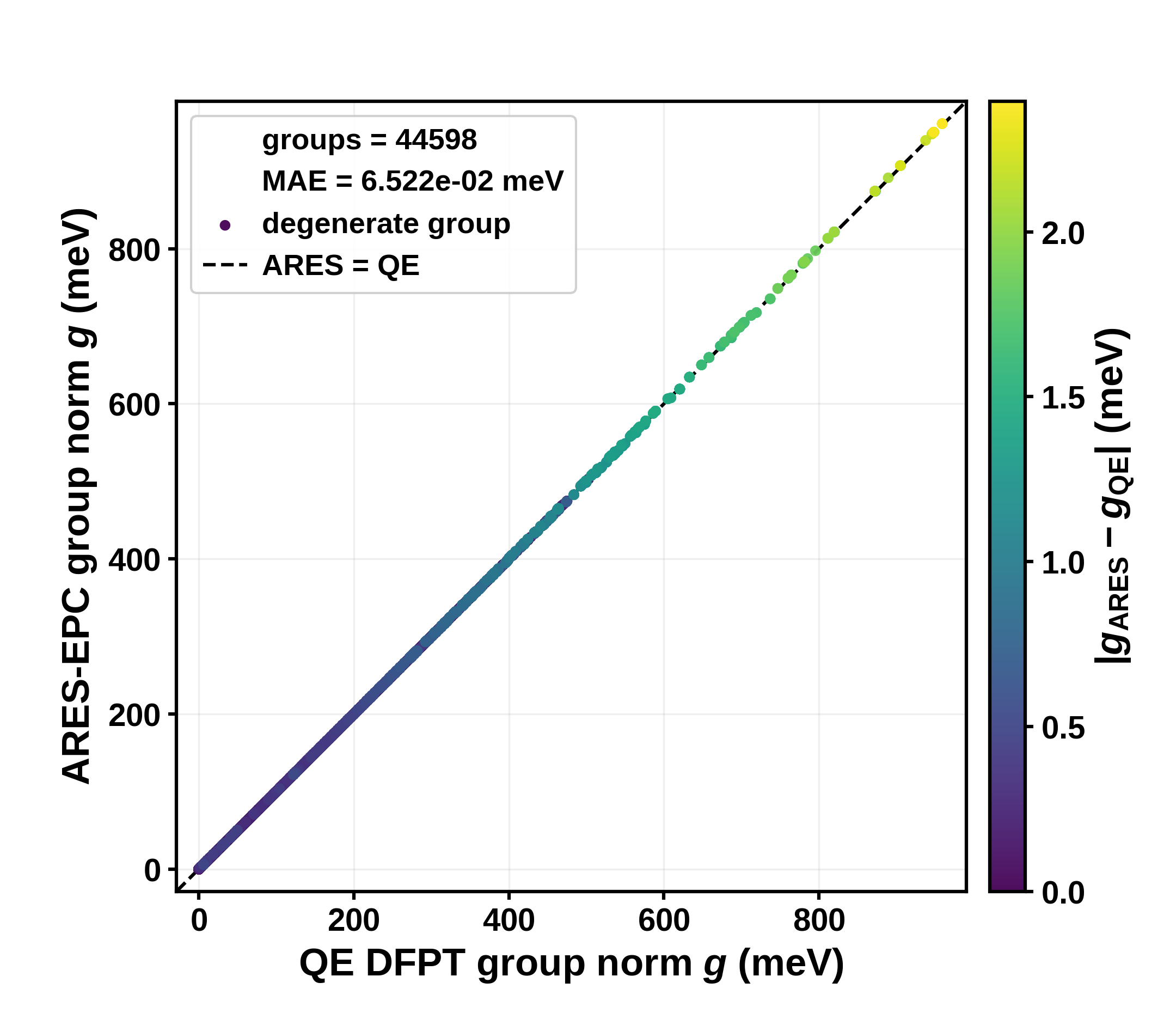}{(f) Pb}\\[4pt]
\sipanel{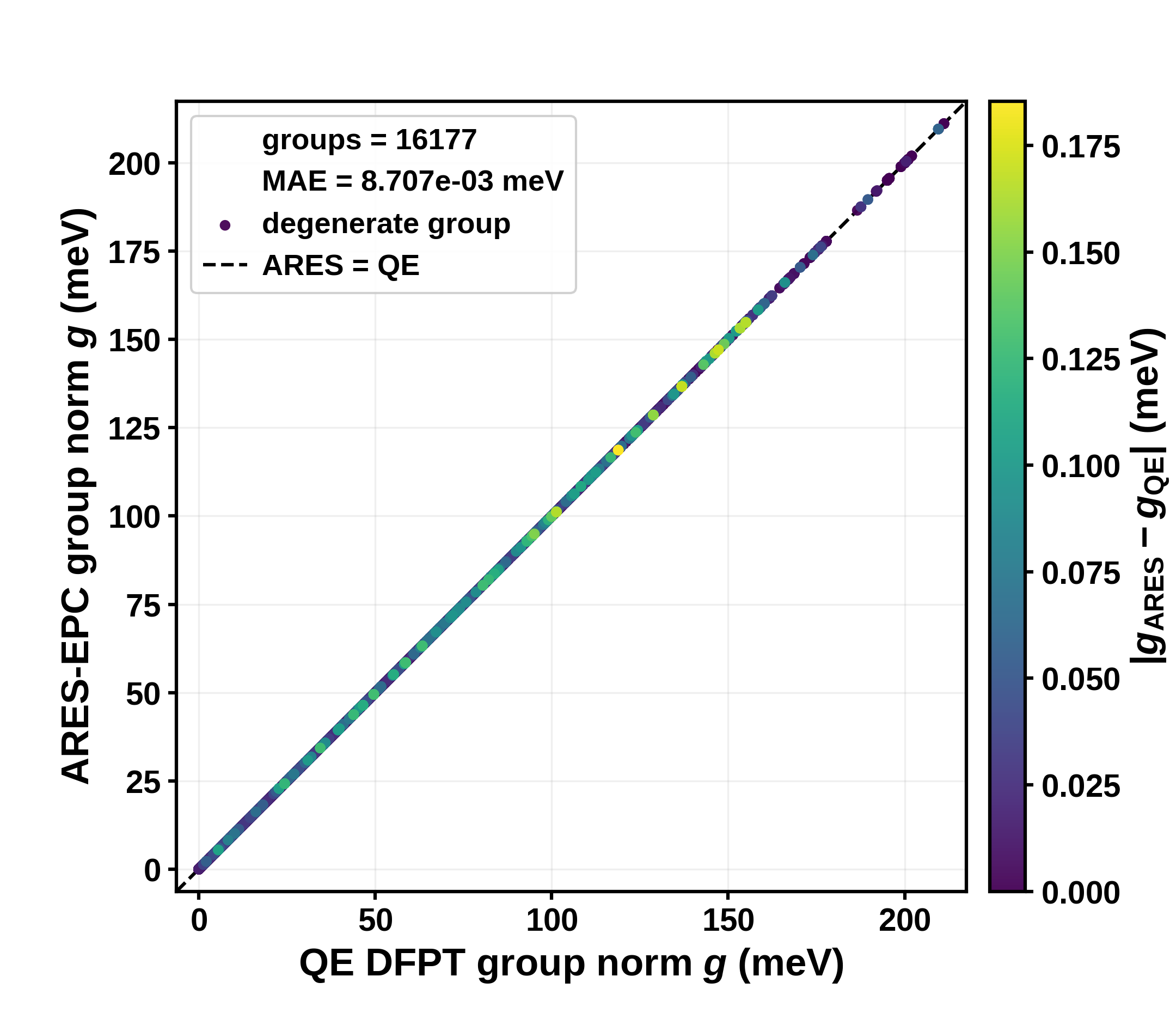}{(g) Si}\hfill
\begin{minipage}[t]{0.64\textwidth}\end{minipage}
\caption{\label{fig:si-nc}Group norms from ARES-EPC+QE versus QE DFPT with
diagonal supercells.}
\end{figure}

\begin{figure}[ht]
\sipanel{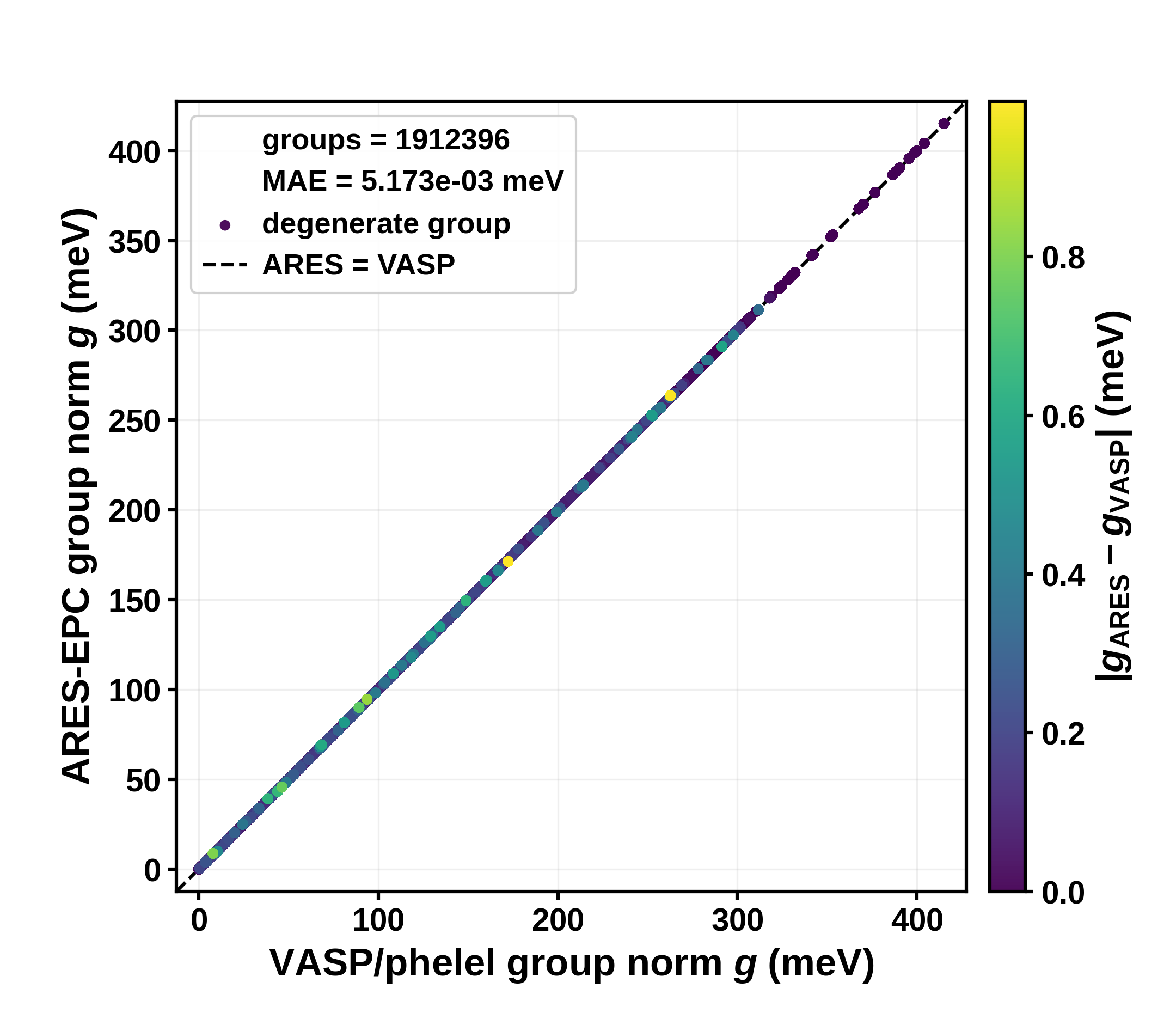}{(a) Al}\hfill
\sipanel{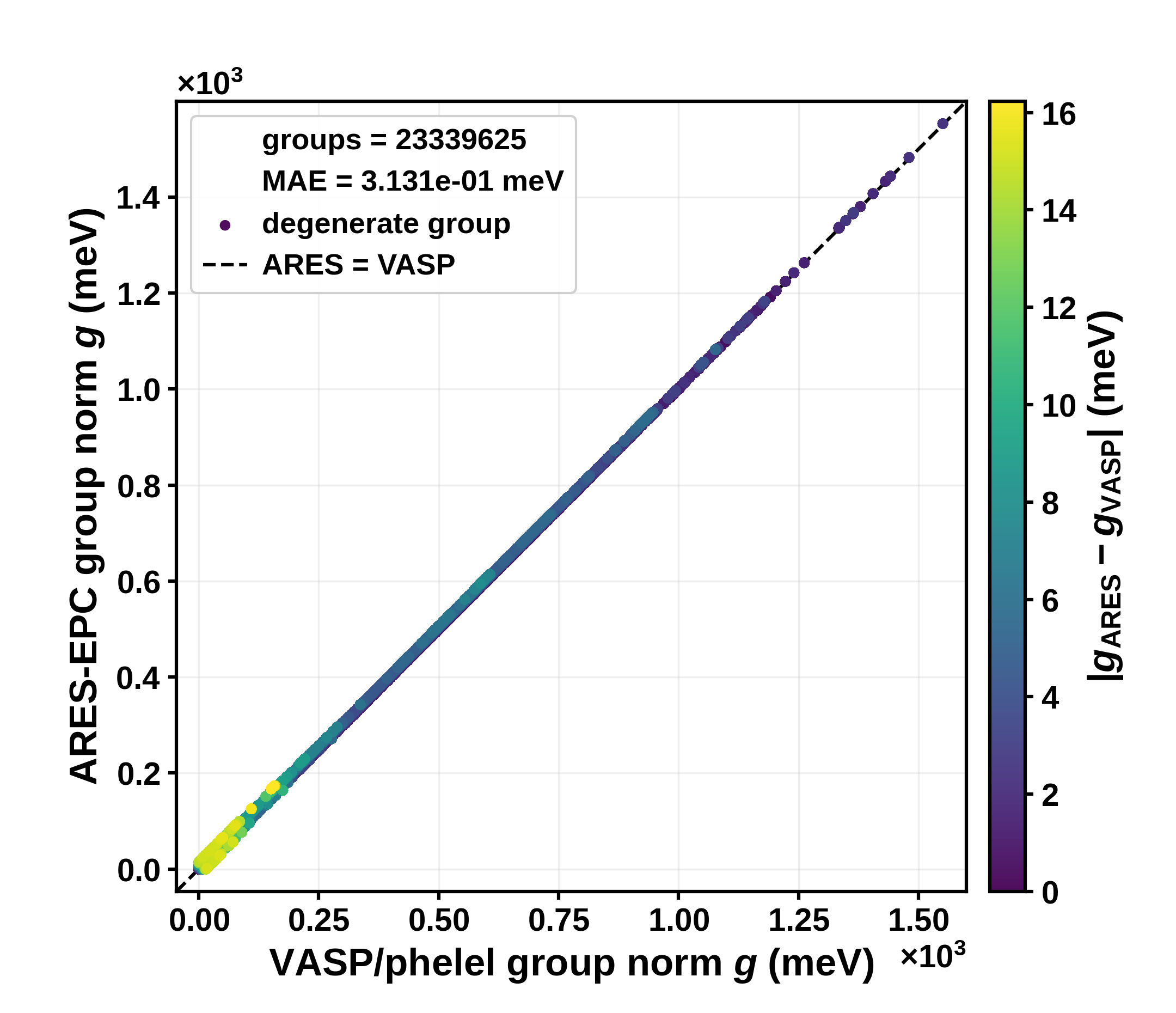}{(b) CaH$_6$}\hfill
\sipanel{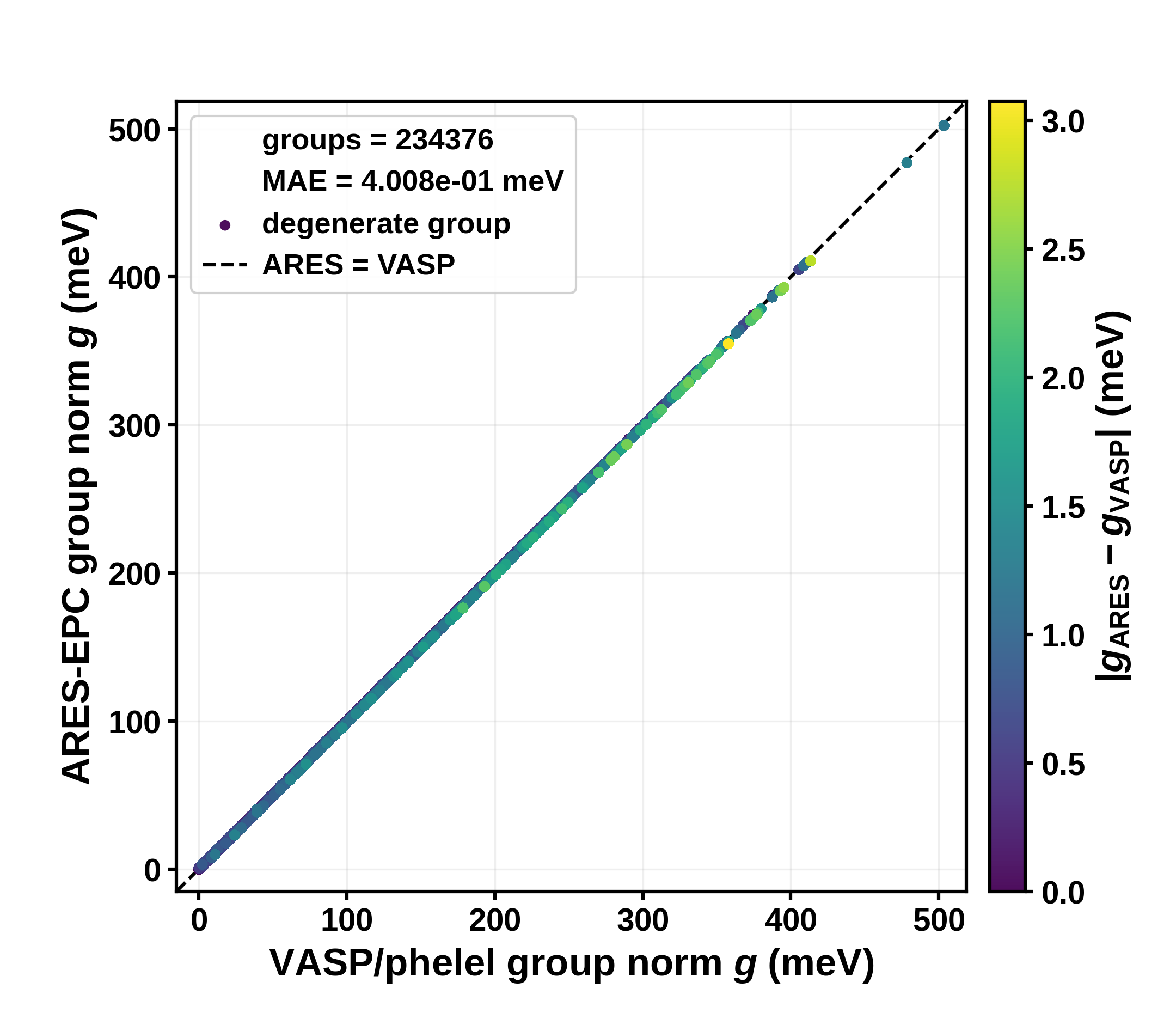}{(c) Diamond}\\[4pt]
\sipanel{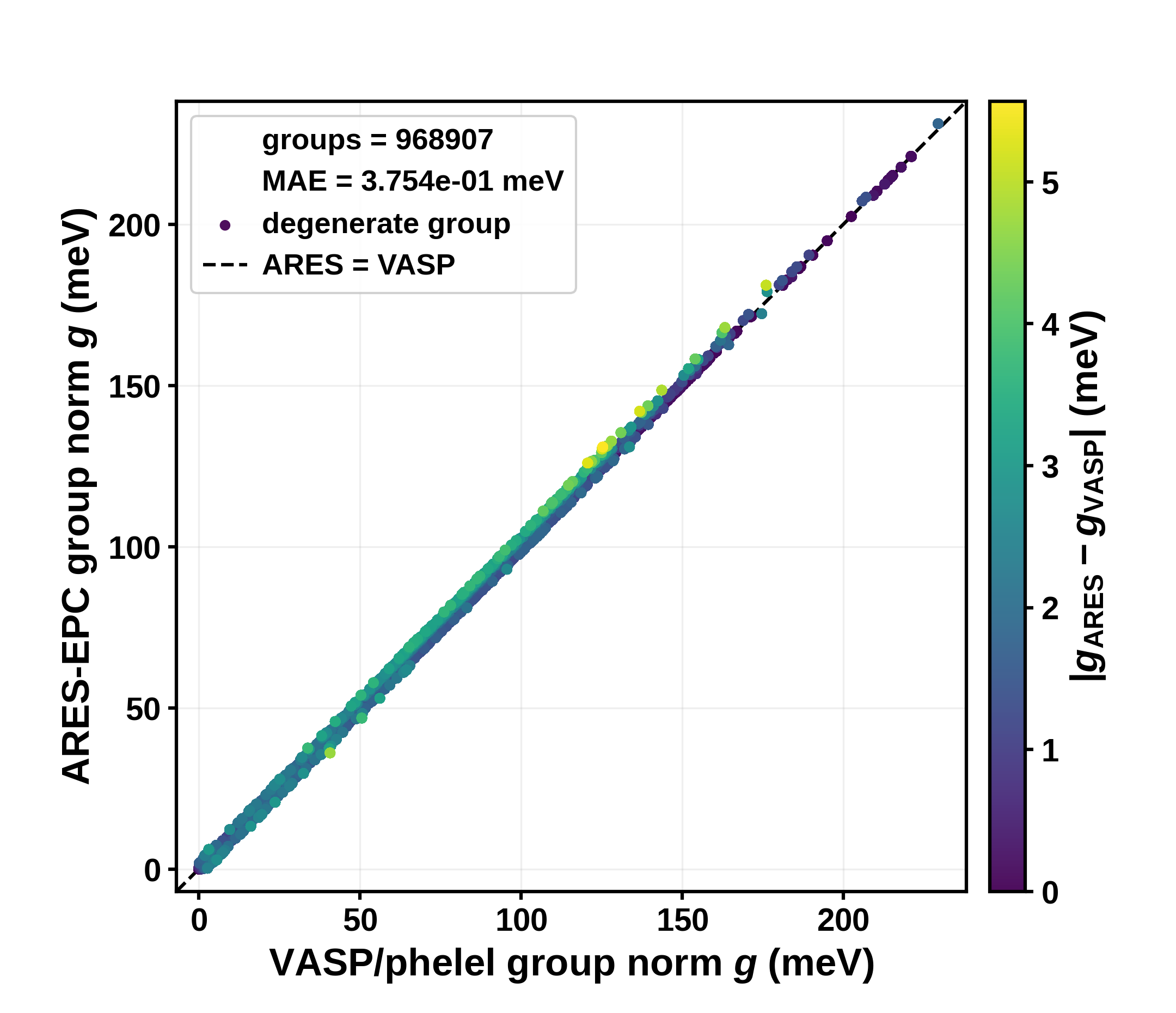}{(d) GaAs}\hfill
\sipanel{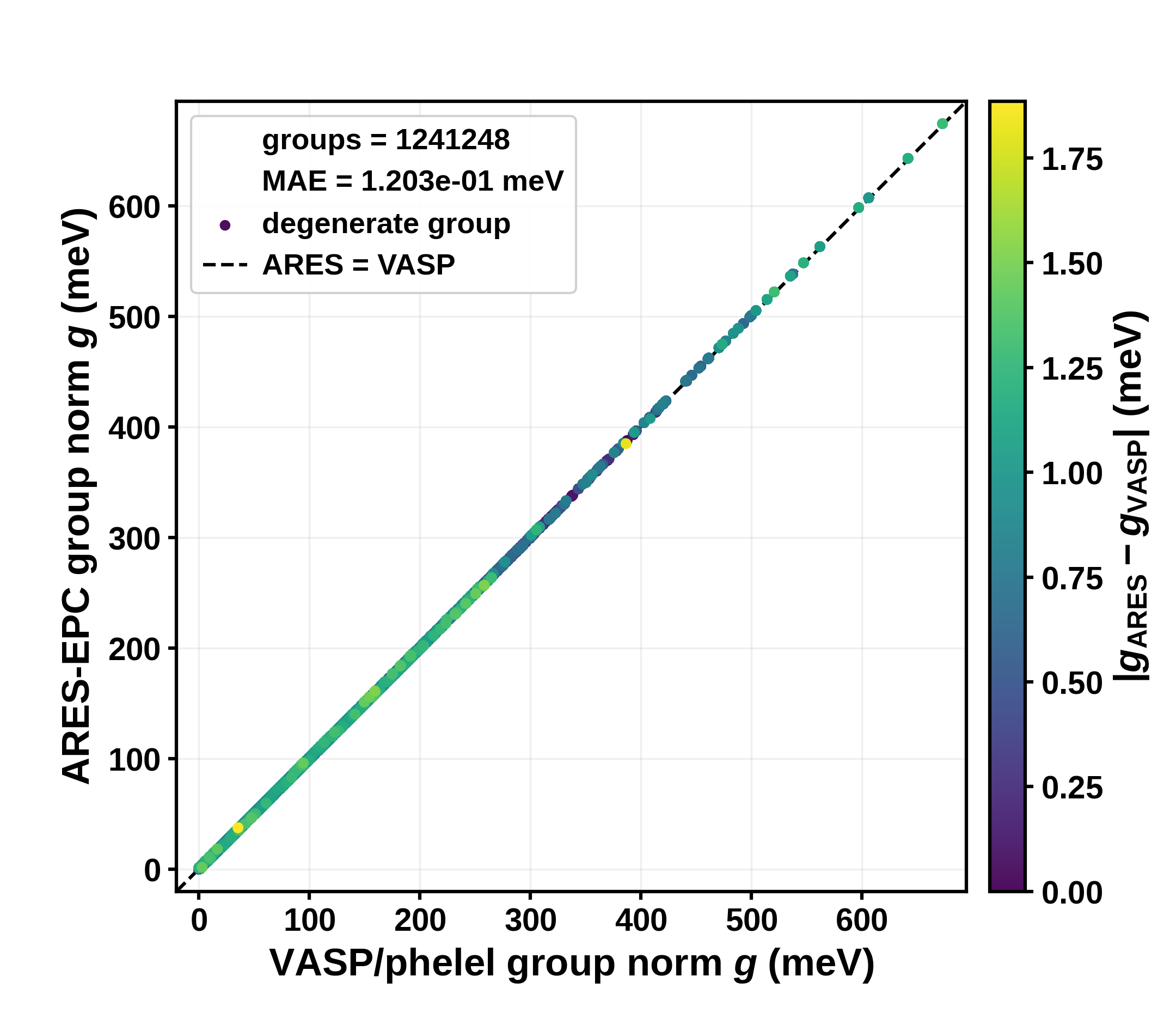}{(e) MgB$_2$}\hfill
\sipanel{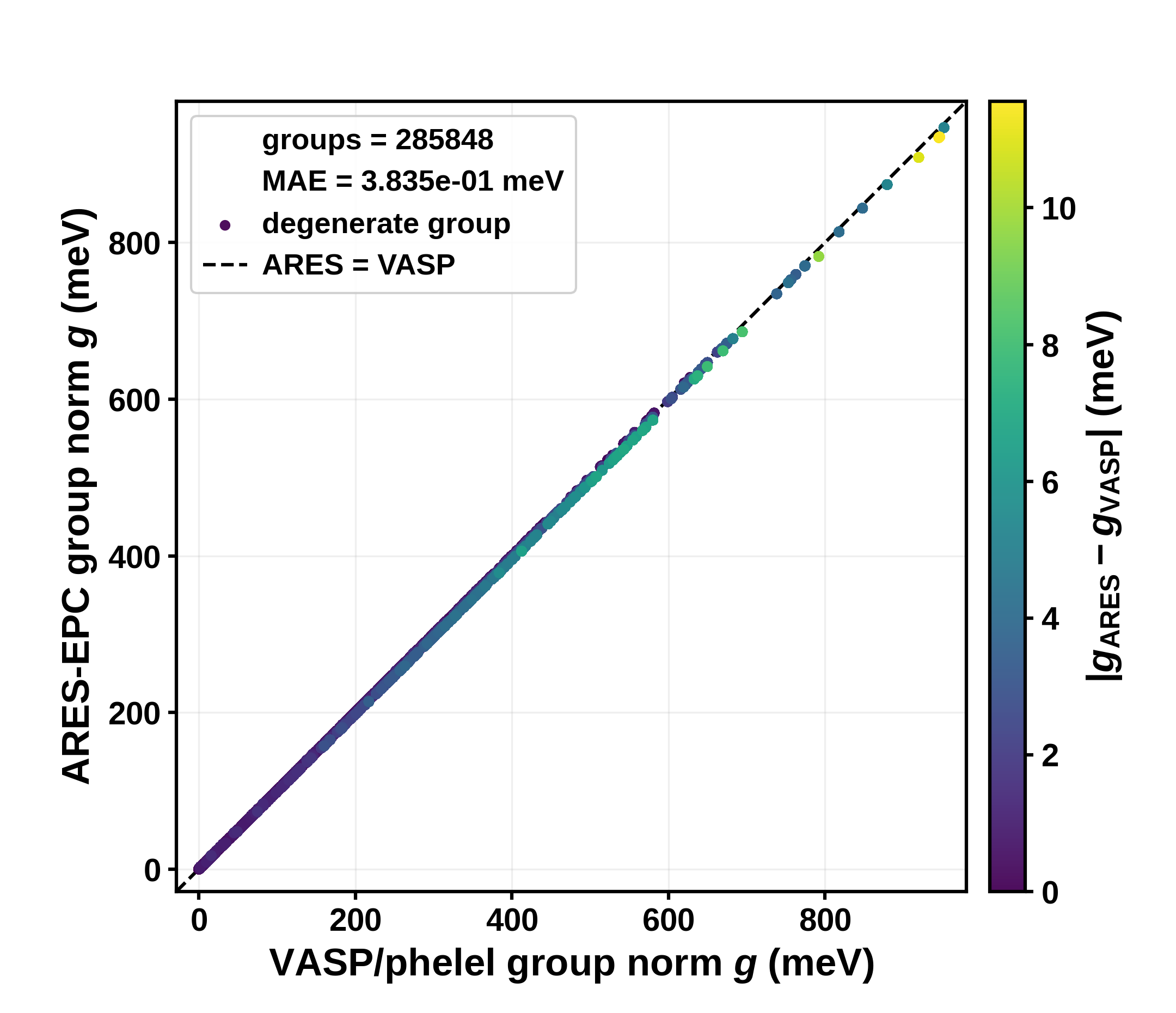}{(f) Pb}\\[4pt]
\sipanel{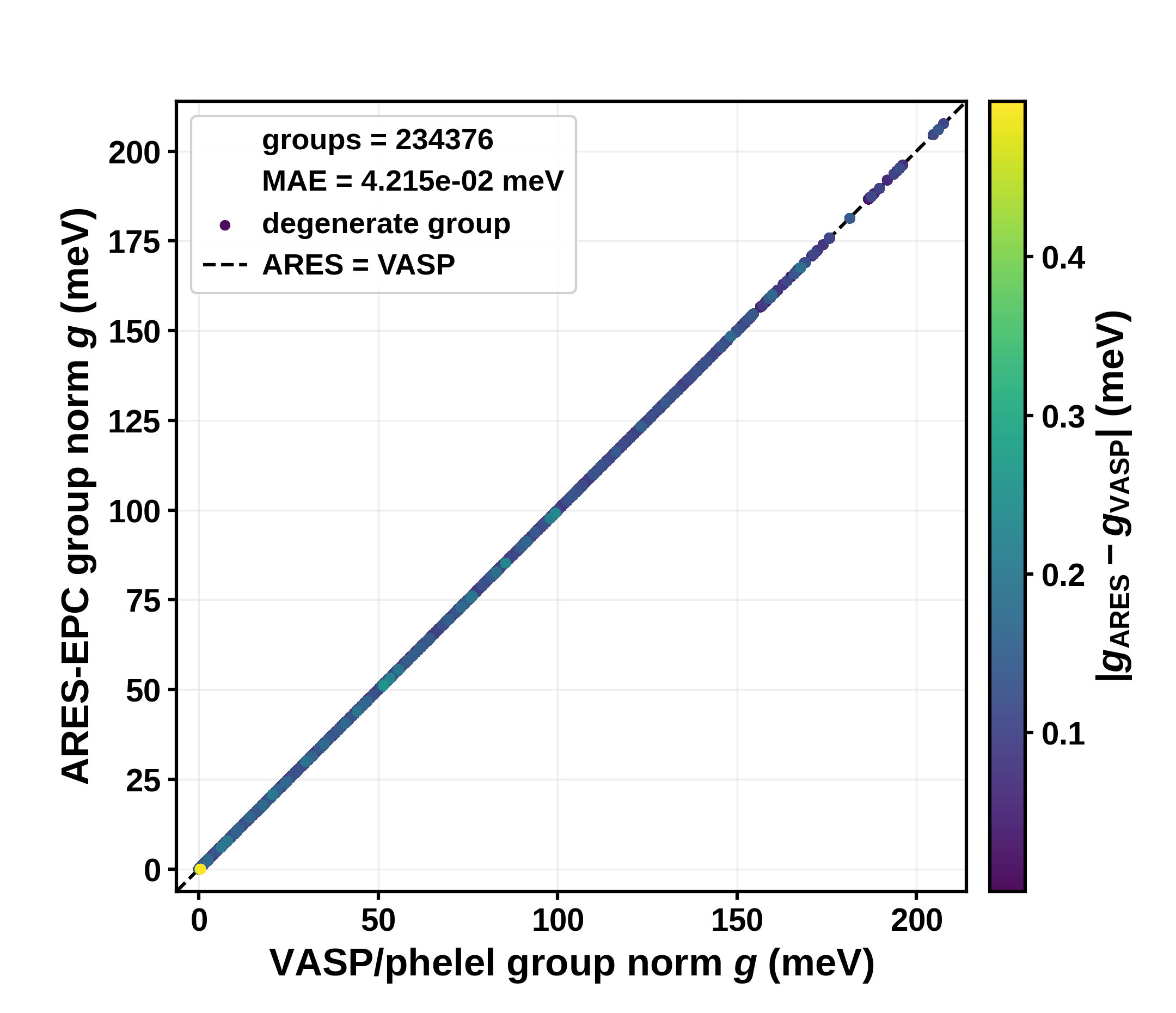}{(g) Si}\hfill
\begin{minipage}[t]{0.64\textwidth}\end{minipage}
\caption{\label{fig:si-paw}Group norms from ARES-EPC+VASP versus VASP+phelel
with diagonal supercells.}
\end{figure}

\begin{figure}[ht]
\sipanel{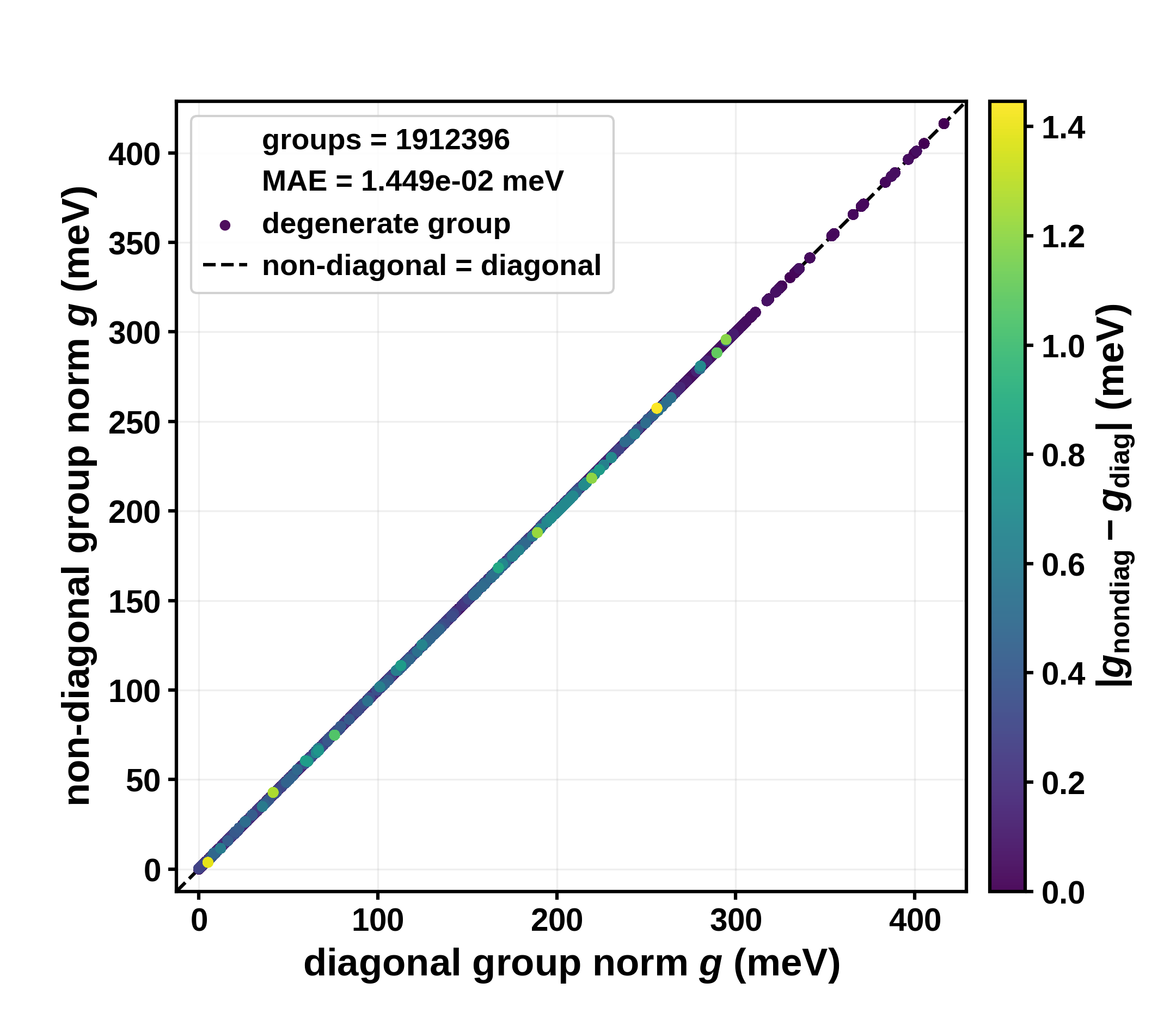}{(a) Al}\hfill
\sipanel{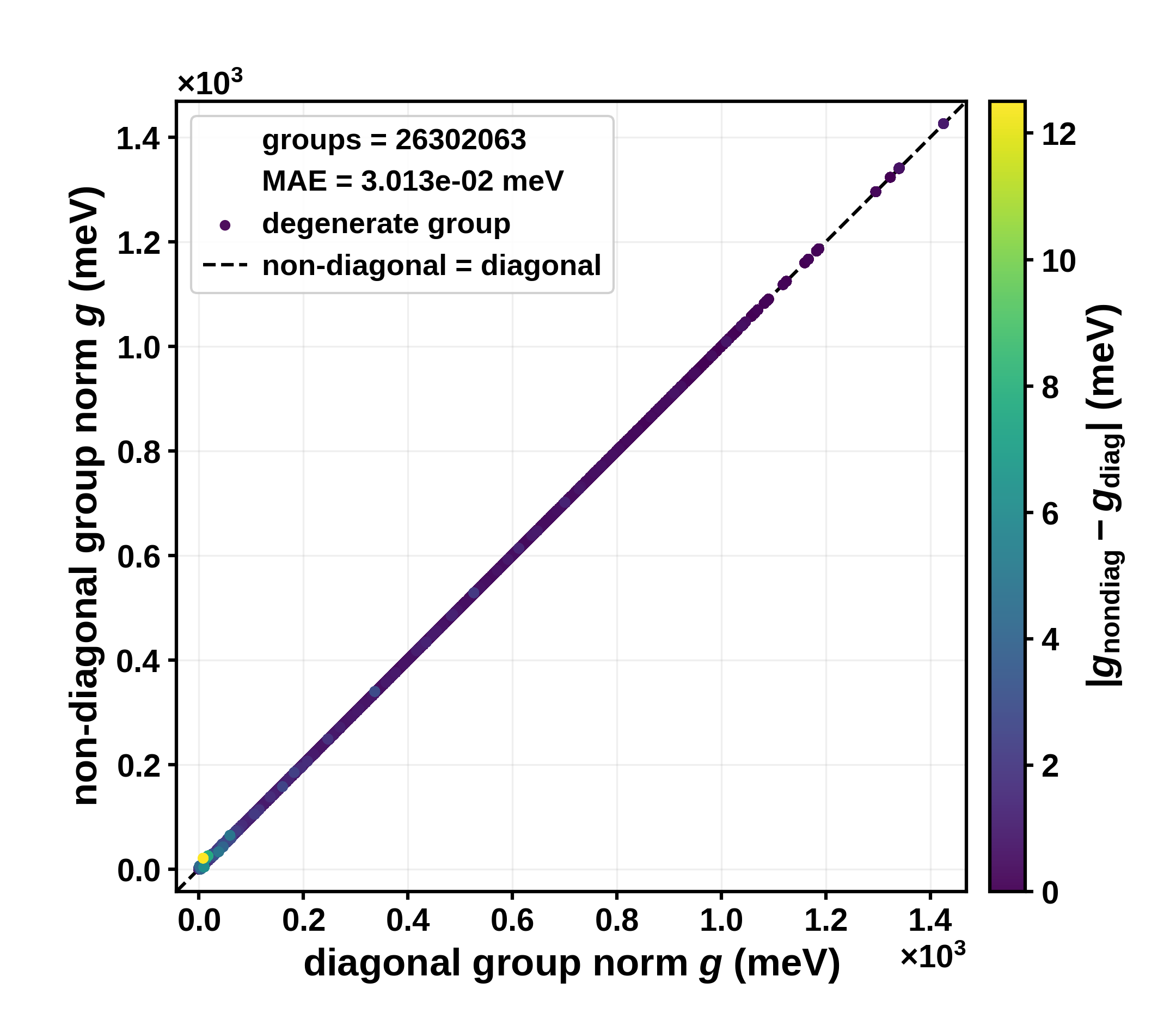}{(b) CaH$_6$}\hfill
\sipanel{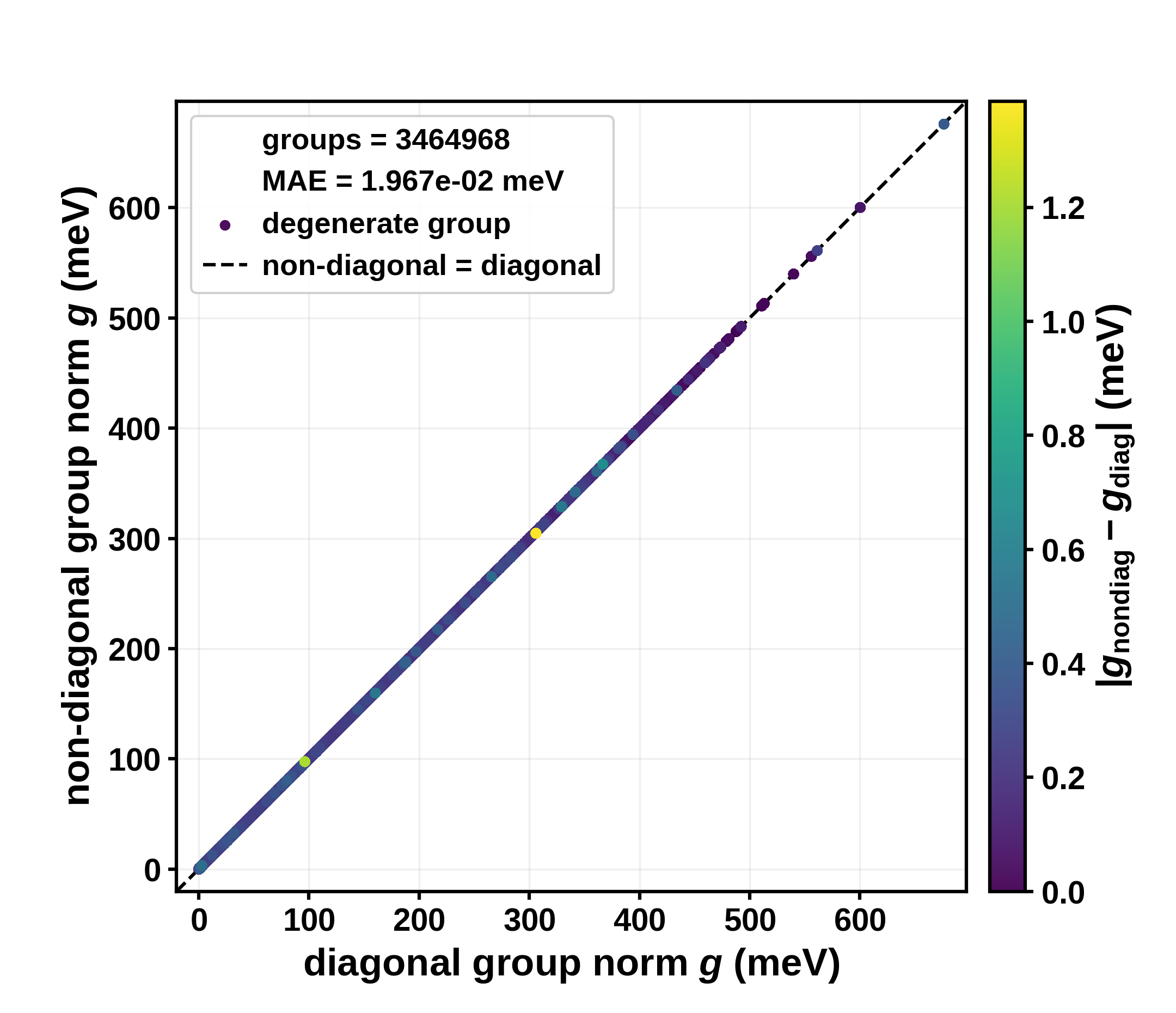}{(c) Diamond}\\[4pt]
\sipanel{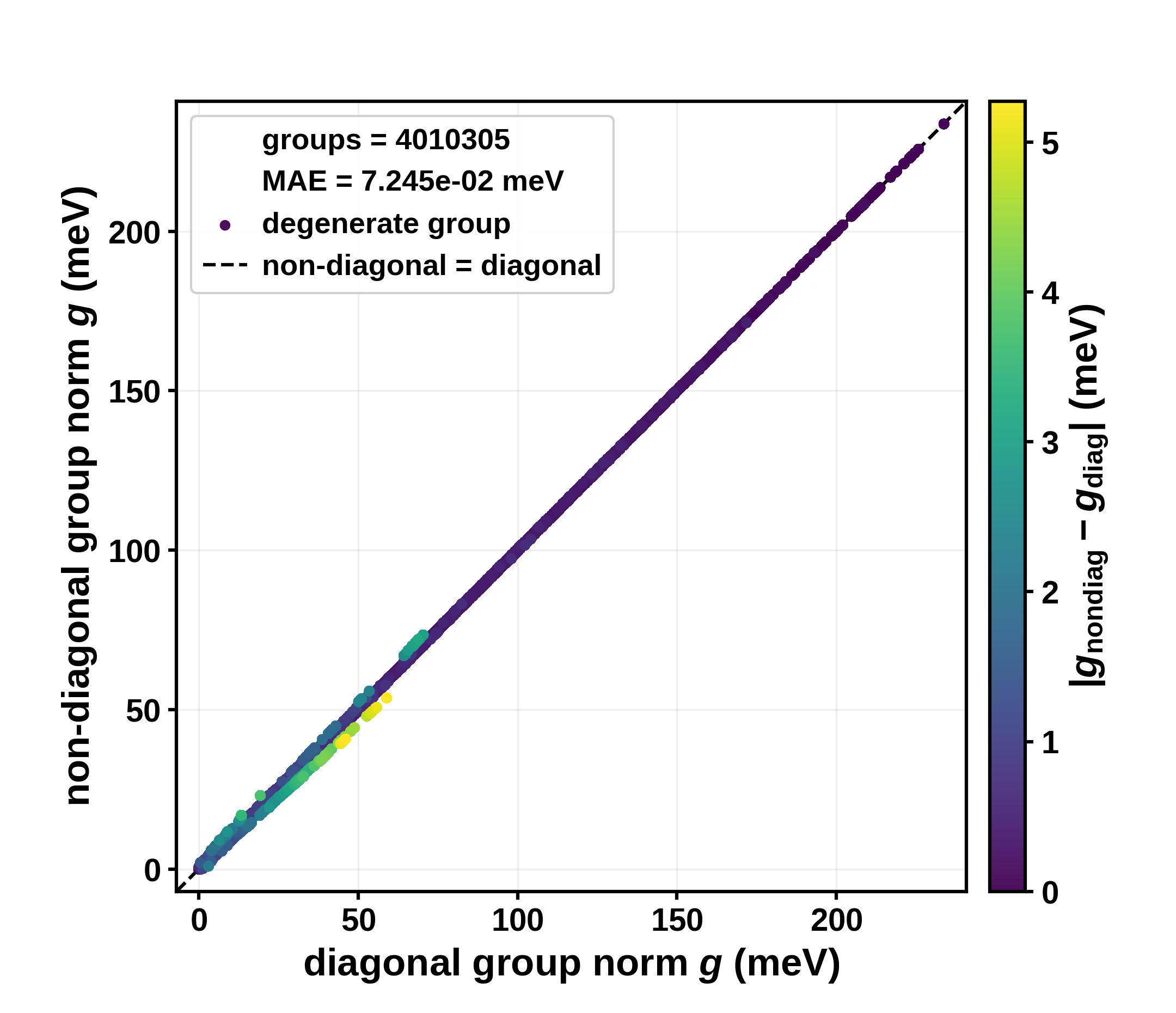}{(d) GaAs}\hfill
\sipanel{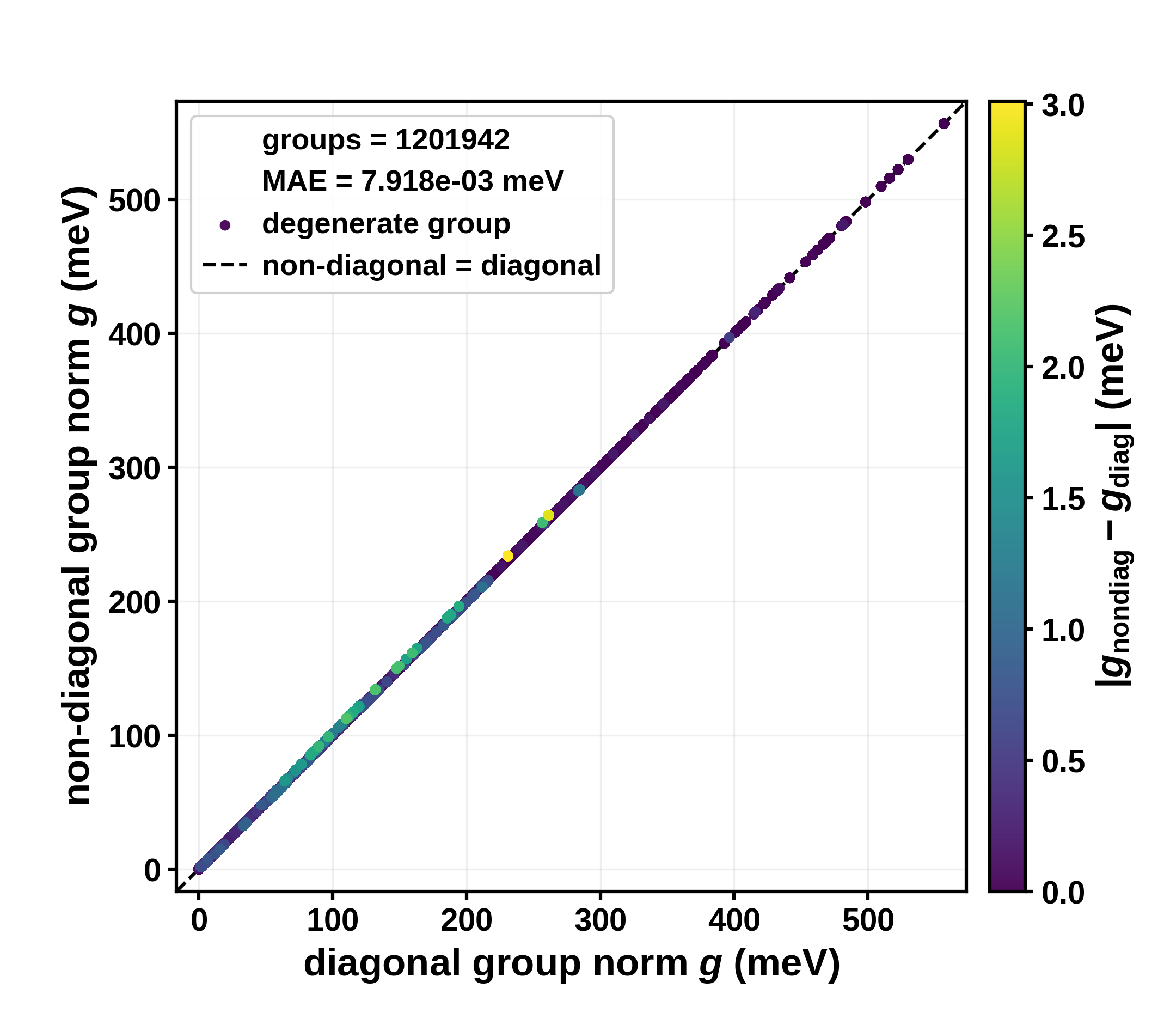}{(e) MgB$_2$}\hfill
\sipanel{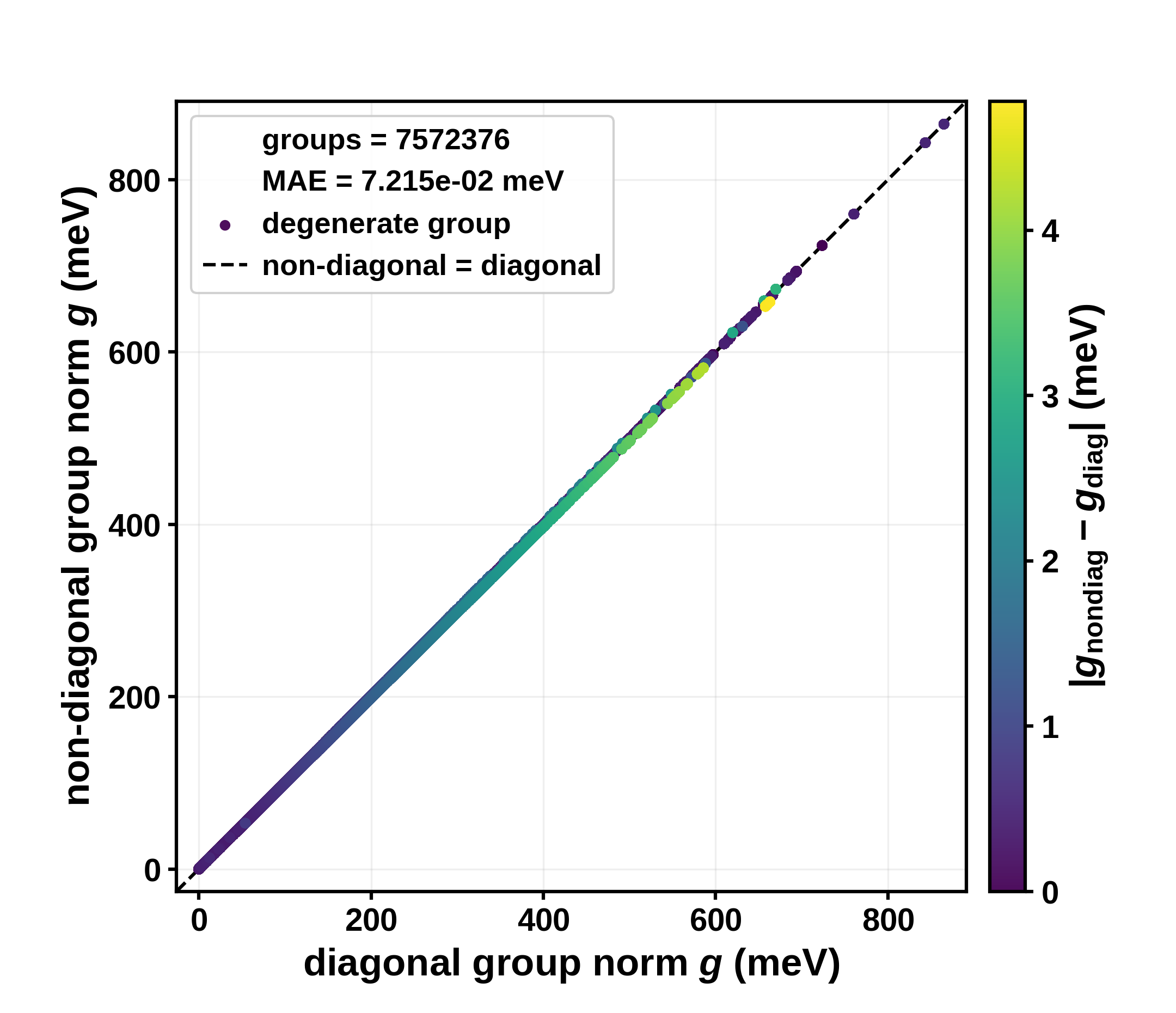}{(f) Pb}\\[4pt]
\sipanel{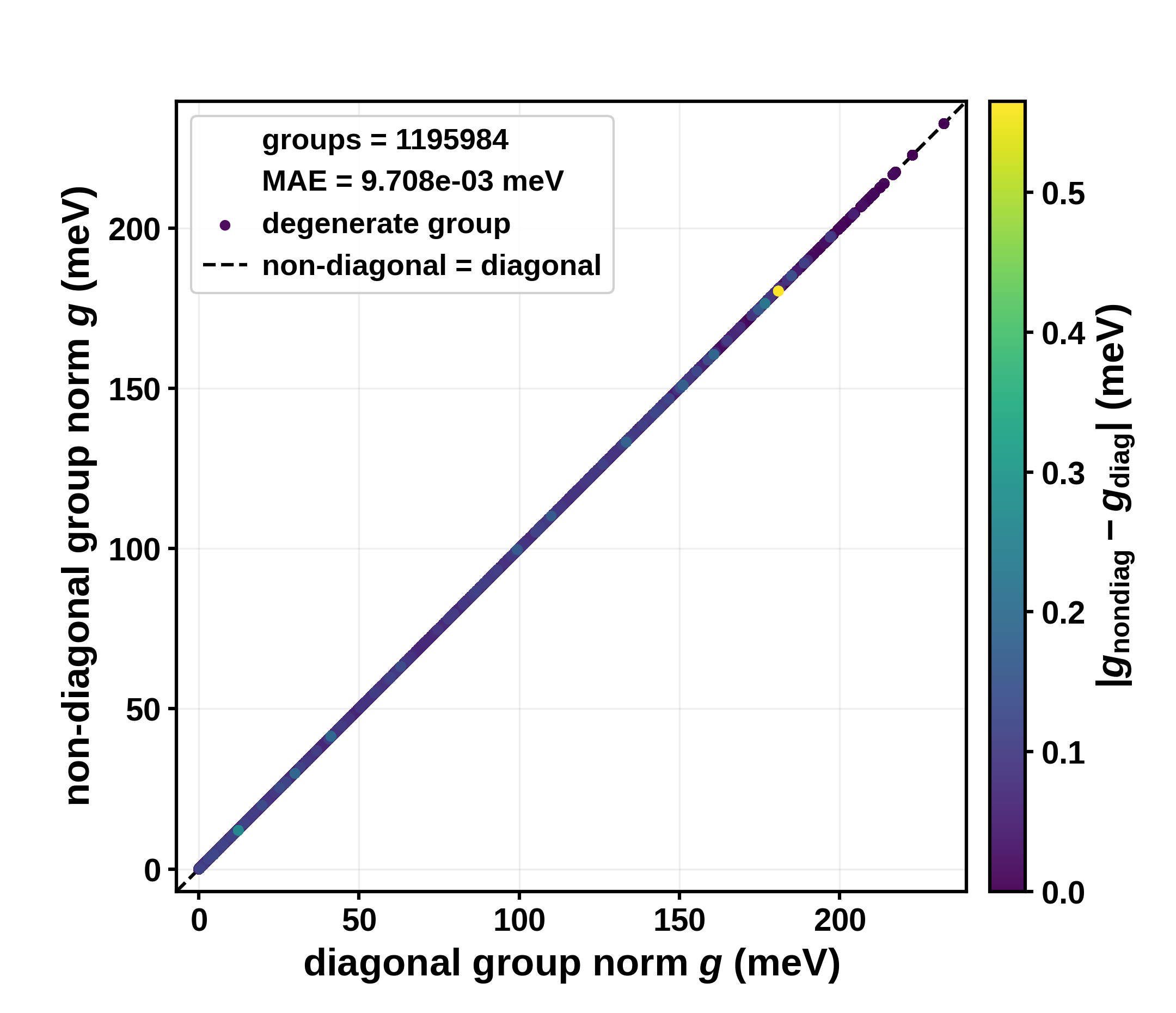}{(g) Si}\hfill
\begin{minipage}[t]{0.64\textwidth}\end{minipage}
\caption{\label{fig:si-nd}Group norms from nondiagonal versus diagonal
ARES-EPC+VASP calculations.}
\end{figure}

\section{Phonon dispersions from diagonal and nondiagonal supercells}

\begin{figure}[ht]
\sipanel{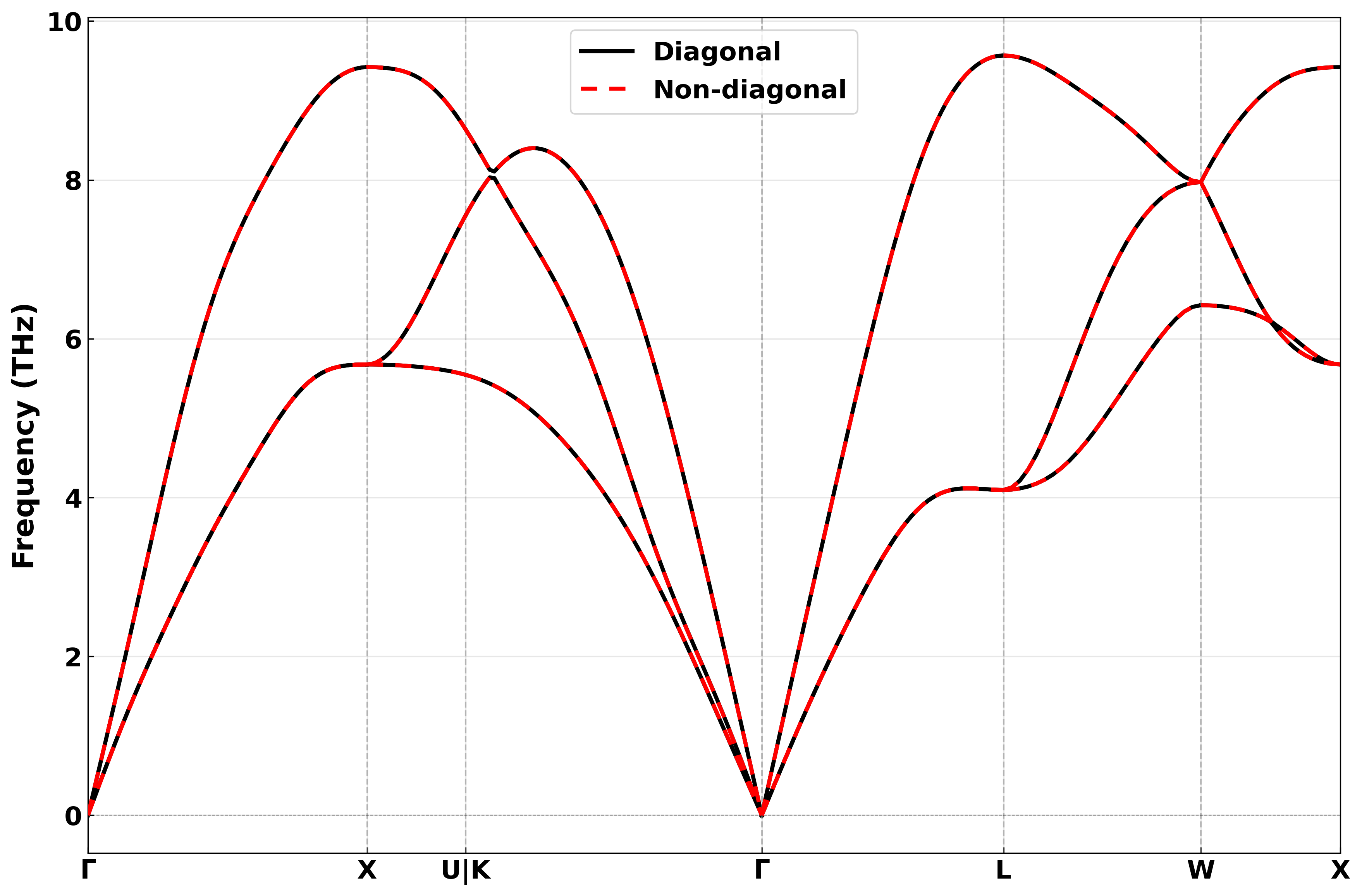}{(a) Al}\hfill
\sipanel{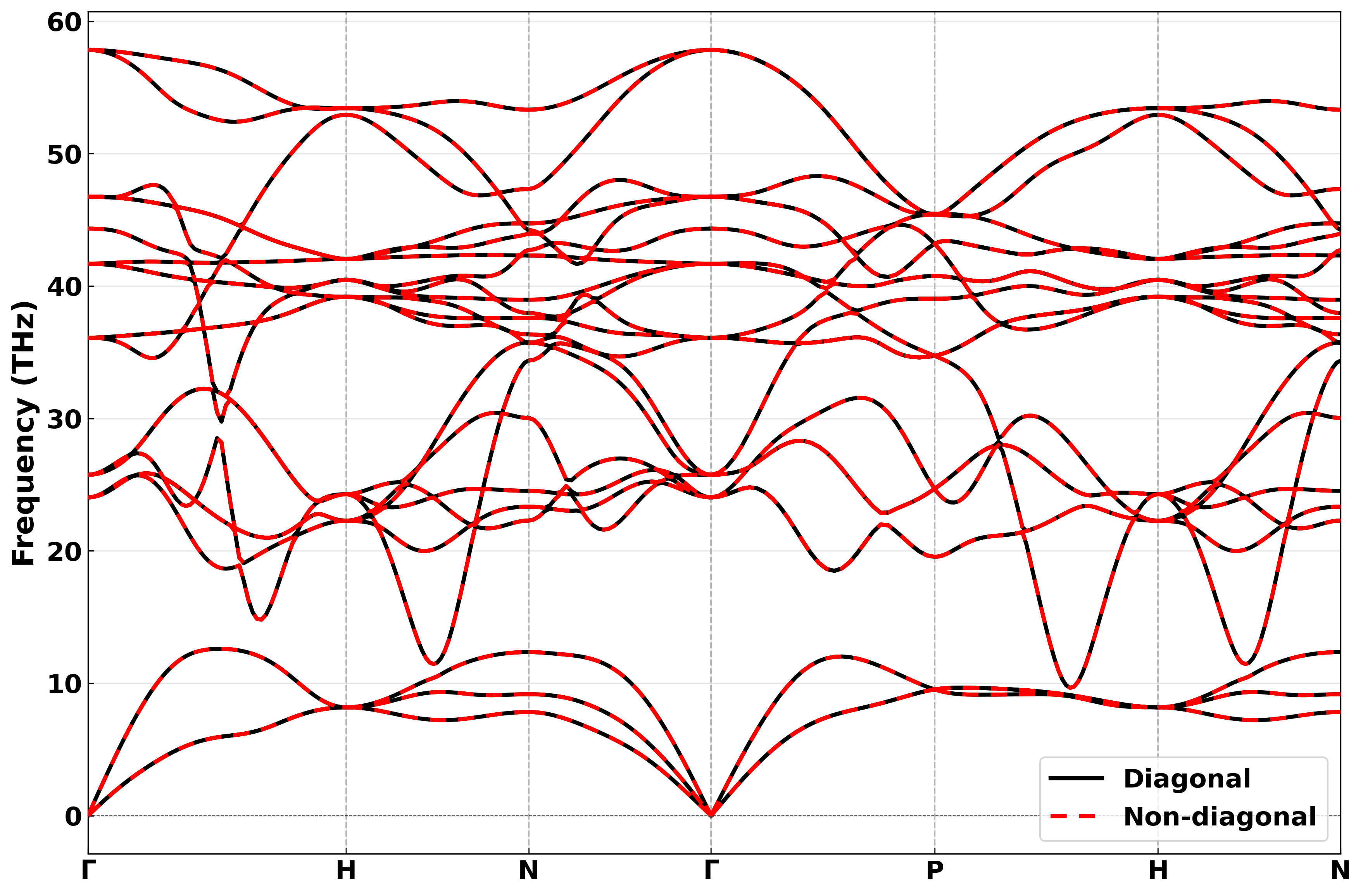}{(b) CaH$_6$}\hfill
\sipanel{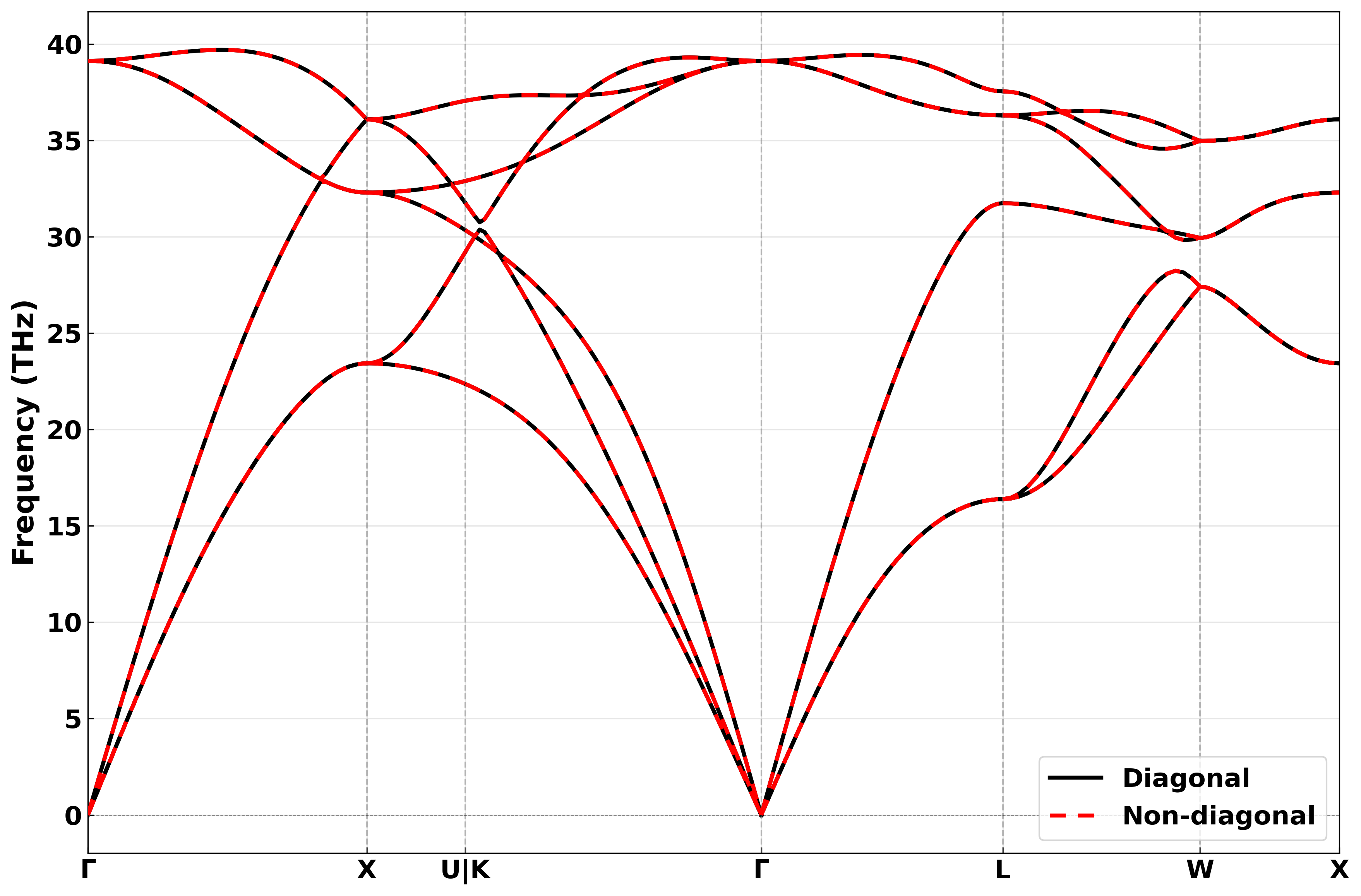}{(c) Diamond}\\[4pt]
\sipanel{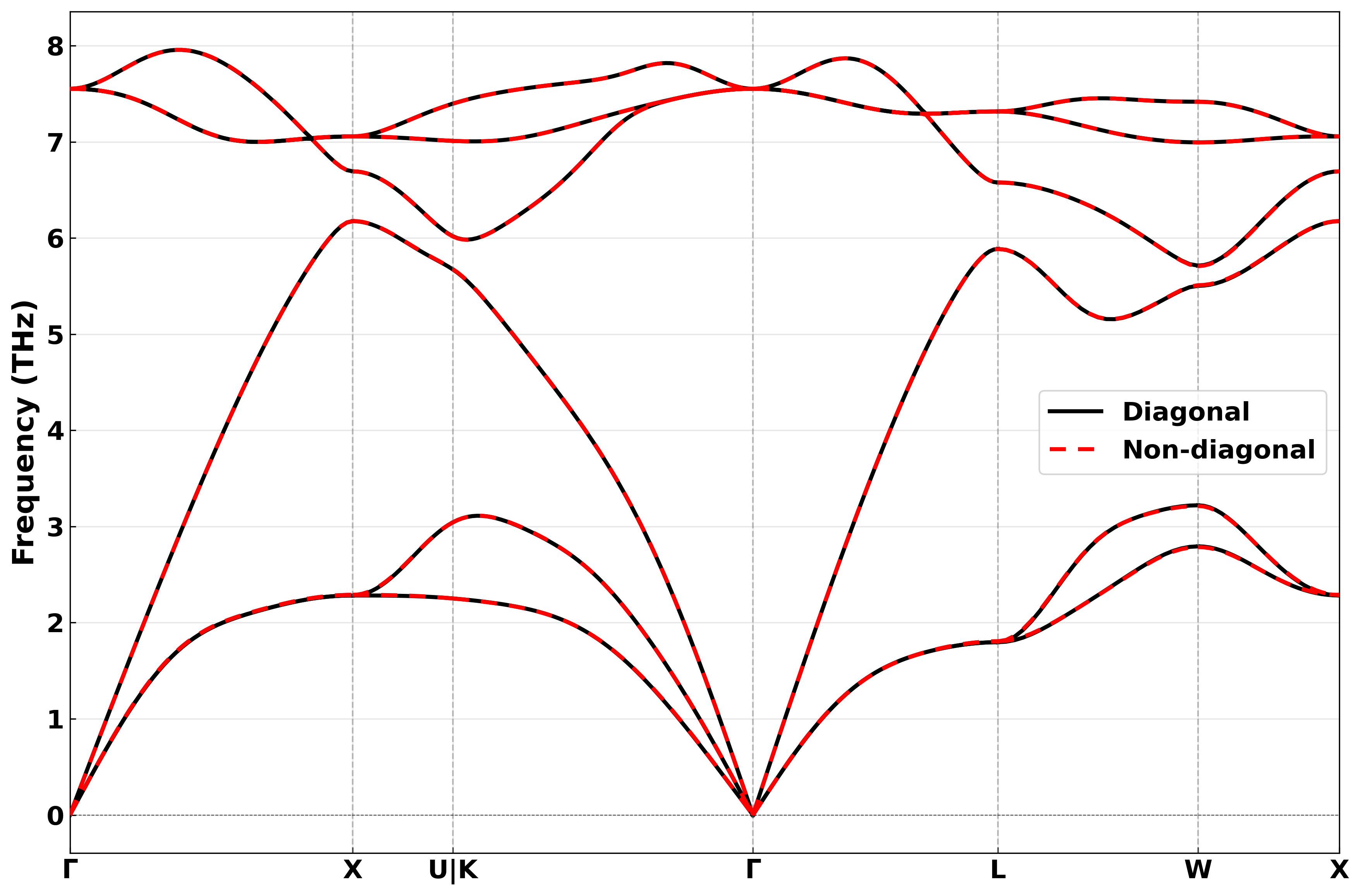}{(d) GaAs}\hfill
\sipanel{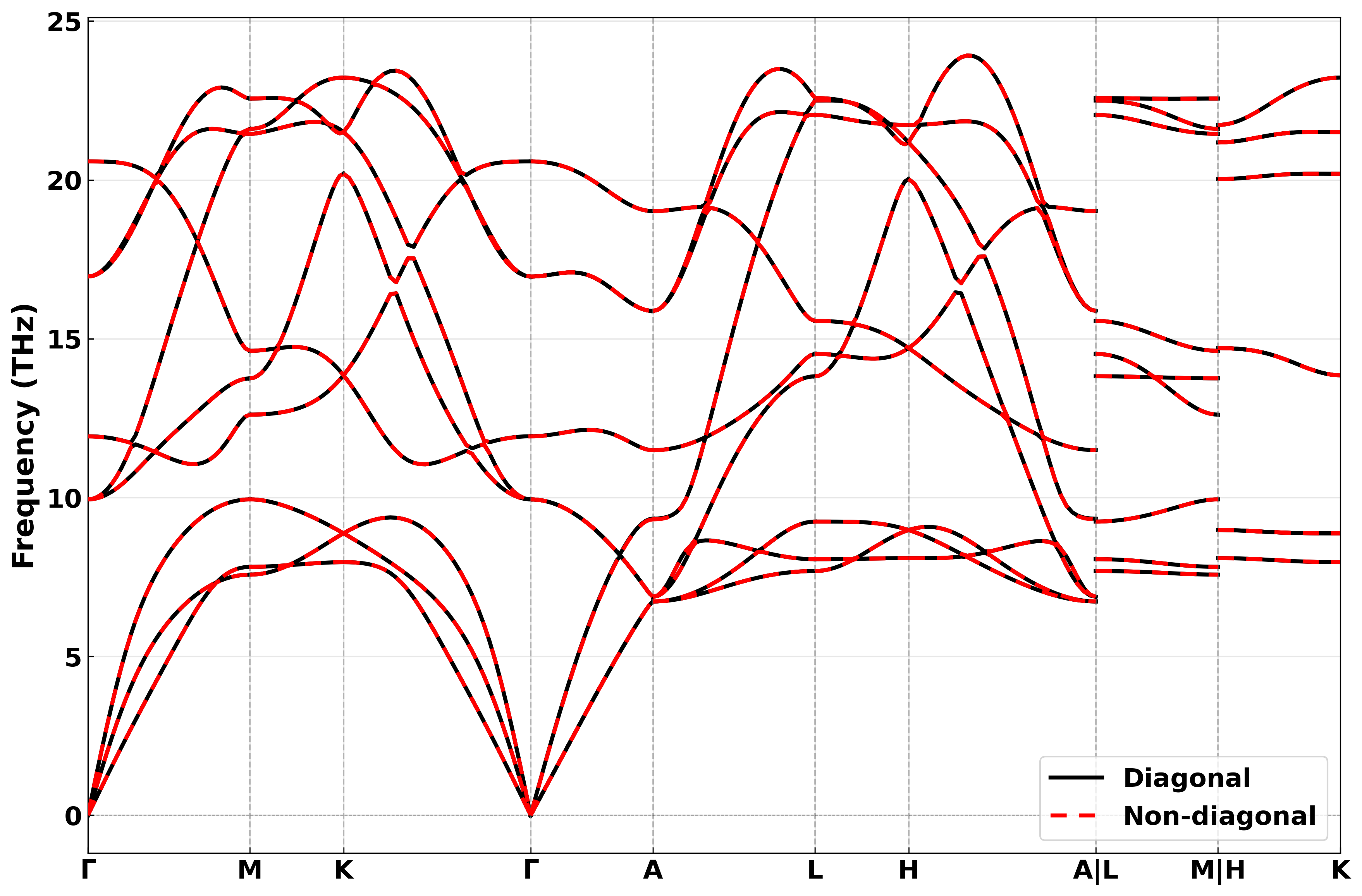}{(e) MgB$_2$}\hfill
\sipanel{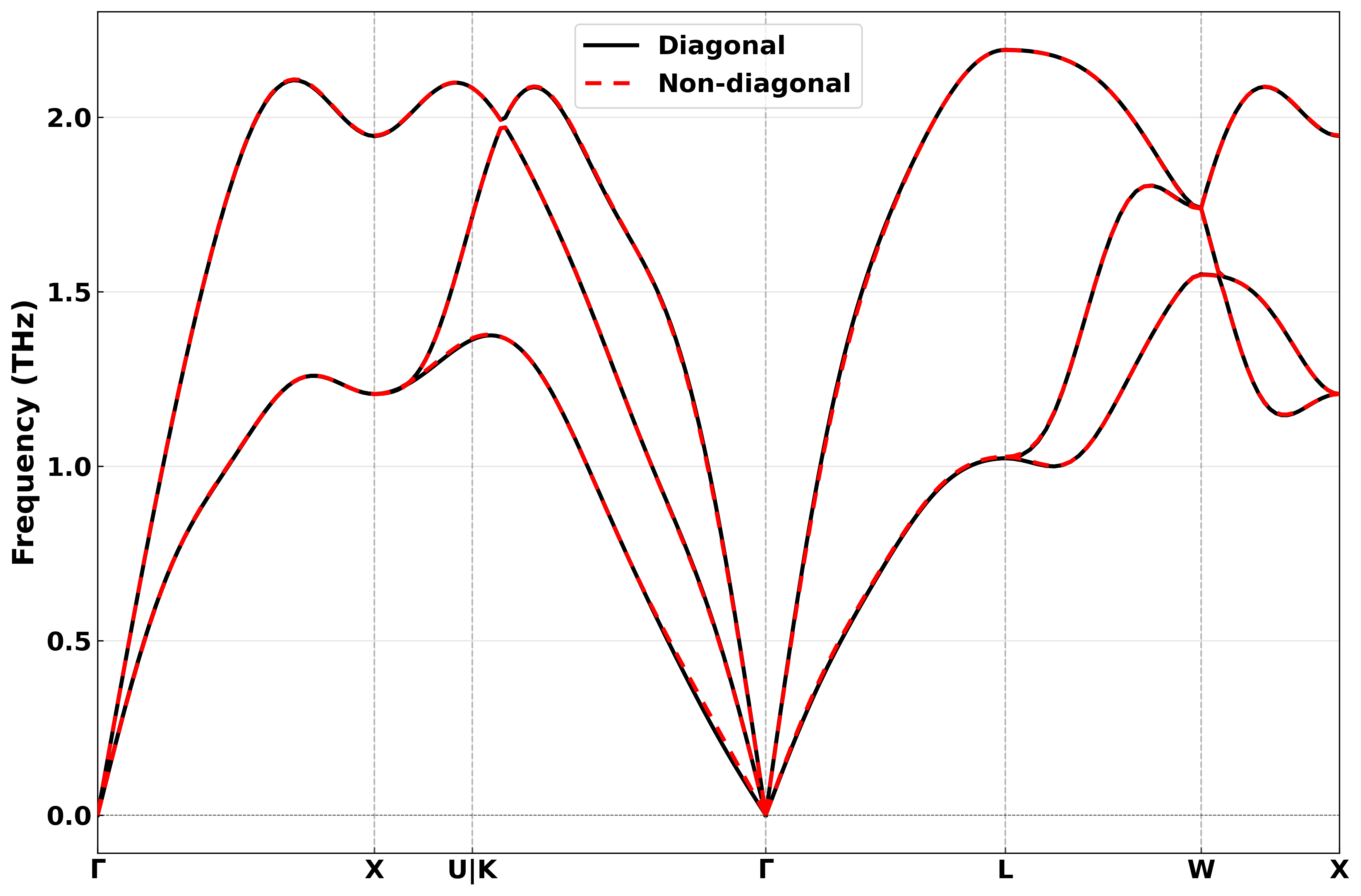}{(f) Pb}\\[4pt]
\sipanel{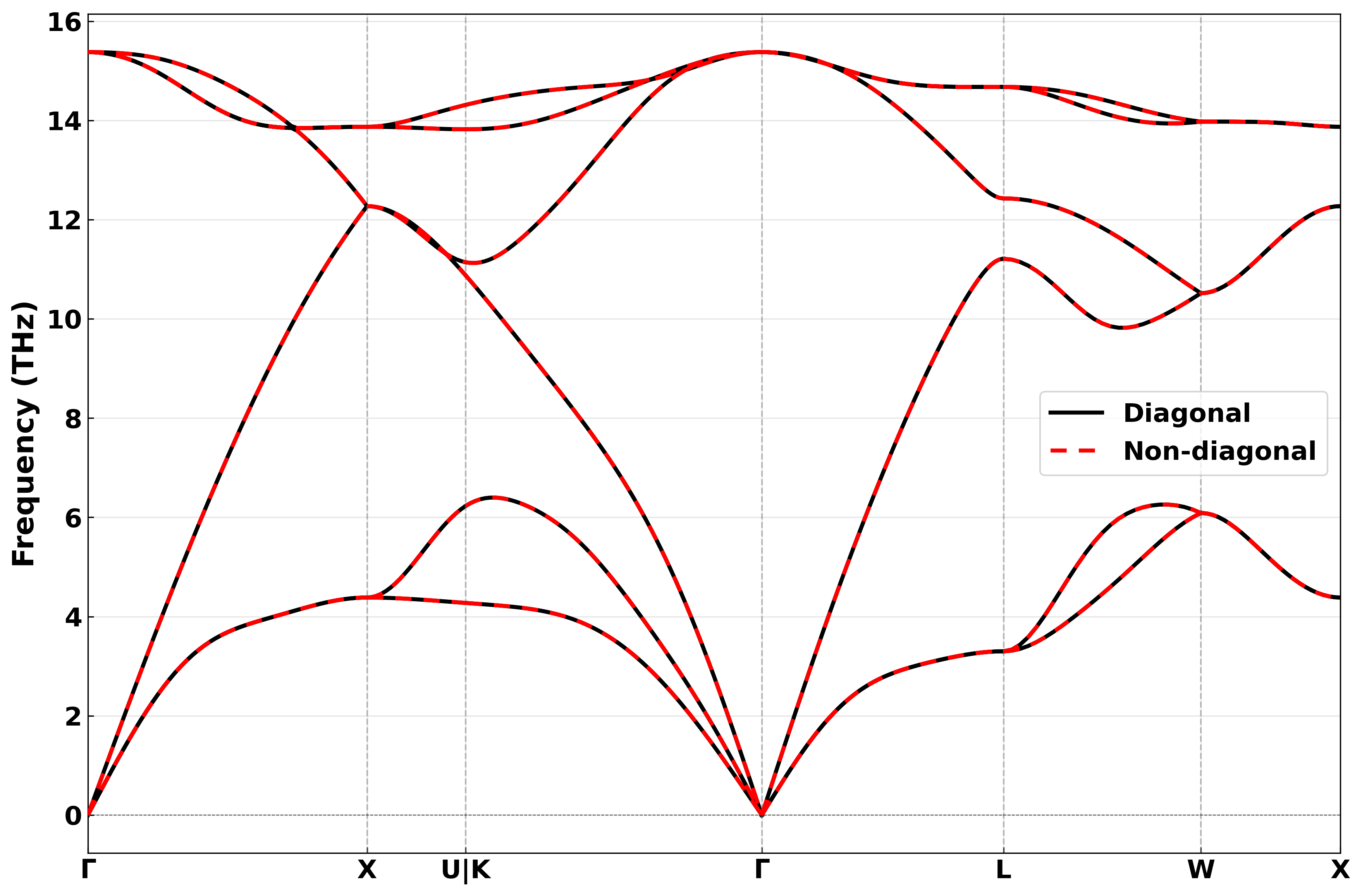}{(g) Si}\hfill
\begin{minipage}[t]{0.64\textwidth}\end{minipage}
\caption{\label{fig:si-phonon}Phonon dispersions from diagonal (solid) and
nondiagonal (dashed) ARES-EPC+VASP calculations.}
\end{figure}

Figure~\ref{fig:si-phonon} compares the phonon dispersions obtained from the
diagonal supercells and from the nondiagonal supercells of
Table~\ref{tab:nondiagonal-benchmarks} of the main text. In the nondiagonal
route, the dynamical matrices of the supercells are transformed into the
force constants of the equivalent diagonal supercell, from which the
dispersion is interpolated along the high-symmetry path. No nonanalytic
LO--TO correction is applied in either route. Table~\ref{tab:si-phonon}
lists the largest and mean frequency differences along the path and the
largest difference on the $\mathbf n_{\mathrm{pair}}$ $\mathbf q$ mesh.

\begin{table}[ht]
\caption{\label{tab:si-phonon}Phonon-frequency differences (meV) between the
nondiagonal and diagonal routes: largest and mean differences along the
high-symmetry path of Fig.~\ref{fig:si-phonon}, and largest difference on the
$\mathbf n_{\mathrm{pair}}$ $\mathbf q$ mesh.}
\begin{ruledtabular}
\begin{tabular}{lccc}
 Material & Path, max & Path, mean & Mesh, max \\
\hline
 Al & 0.024 & 0.0041 & 0.023 \\
 CaH$_6$ & 0.051 & 0.0079 & 0.048 \\
 Diamond & 0.035 & 0.0044 & 0.034 \\
 GaAs & 0.039 & 0.0085 & 0.032 \\
 MgB$_2$ & 0.099 & 0.0039 & 0.096 \\
 Pb & 0.089 & 0.012 & 0.051 \\
 Si & 0.0052 & 0.00098 & 0.0042
\end{tabular}
\end{ruledtabular}
\end{table}

\end{document}